\PassOptionsToPackage{pdfpagelabels=false}{hyperref} 
\documentclass[fleqn]{mnras}

\usepackage{graphicx}	
\usepackage{amsmath}	
\usepackage{amssymb}	
\usepackage{color}
\usepackage{float}

\usepackage{ulem}
\usepackage{comment}
\usepackage[utf8]{inputenc}
\usepackage[english]{babel}
\usepackage{xcolor}
\usepackage{rotating}
\usepackage{natbib}
\usepackage{booktabs}
\usepackage{tabularx}

\def\eq{\begin{equation}}
\def\en{\end{equation}}

\def\etal{{\it et al}\thinspace}

\def\apj{{\it Ap.J.}\thinspace}
\def\apjl{{\it Ap.J. Lett.}\thinspace}
\def\apjs{{\it Ap.J. Suppl.}\thinspace}
\def\aj{{\it A.J.}\thinspace}
\def\jaa{{\it J.Ap.A.}\thinspace}
\def\aap{{\it A\&A}\thinspace}
\def\aaps{{\it A\&A Suppl.}\thinspace}
\def\mnras{{\it MNRAS}\thinspace}
\def\P3hat{{\mathaccent 94 P}_3}

\def\etal{{\it et al.}\thinspace}

\def\eg{{\it e.g.,}\thinspace}

\title[Arecibo Pulsar Beam Geometry at Lower Frequencies]{Radio Pulsar Beam Geometry Down to the 100-MHz Band:  76 Additional Sources Within the Arecibo Sky\thanks{This paper is dedicated to our colleagues at the Institute for Astronomy, Kharkiv, Ukraine}}

\author[Rankin, Wahl, Venkataraman, \& Olszanski]
{Joanna Rankin$^{1,4}$\thanks{E-mail: Joanna.Rankin@uvm.edu}, Haley Wahl$^{1, 2, 3}$, Arun Venkataraman{$^5$}, Timothy Olszanski$^{1, 2, 3}$\\
$^{1}$Physics Department, University of Vermont, Burlington, VT 05405, USA \\
$^{2}$Department of Physics and Astronomy, West Virginia University, P.O. Box 6315, Morgantown, WV 26505\\
$^{3}$Center for Gravitational Waves and Cosmology, West Virginia University, Morgantown, WV 26505\\
$^{4}$Anton Pannekoek Institute for Astronomy, University of Amsterdam, Science Park 904, 1098 XH Amsterdam \\
$^{5}$Arecibo Observatory, bo. La Esperanza, P.P. Box 53995, Arecibo, Puerto Rico, 0612}

\date{Accepted XXX. Received XXX; in original form XXX}

\pubyear{XX}

\begin{document}
\label{firstpage}
\pagerange{\pageref{firstpage}--\pageref{lastpage}}
\maketitle

\begin{abstract}
This paper provides analyses of the emission beam structure of 76 ``B''-named pulsars within the Arecibo sky.  Most of these objects are included in both the Gould \& Lyne and LOFAR High Band surveys and thus complement our other works treating various parts of these populations.  These comprise a further group of mostly well studied pulsars within the Arecibo sky that we here treat similarly to those in Olszanski \etal---and extend our overall efforts to study all of the pulsars in both surveys.  The analyses are based on observations made with the Arecibo Telescope at 327 MHz and 1.4 GHz.  Many have been observed at frequencies down to 100 MHz using either LOFAR or the Pushchino Radio Astronomy Observatory as well as a few with the Long Wavelength Array at lower frequencies. This work uses the Arecibo observations as a foundation for interpreting the low frequency profiles and emission-beam geometries.  We attempt to build quantitative geometric emission-beam models using the core/double-cone topology, while reviewing the evidence of previous studies and arguments for previous classifications on these sources.
These efforts were successful for all but two pulsars, and interesting new subpulse modulation patterns were identified in a number of the objects.  We interpret the Arecibo pulsar population in the context of the entire population of ``B'' pulsars. 
\end{abstract}




\section{Introduction}
In astrophysics we have only the radiation from celestial sources to study, and we can only regard such sources as understood when we manage to comprehend the physical processes responsible for their emission.  Radio pulsars provide unique challenges because their radiation is highly beamed, and we usually have no direct way of knowing just what part of the entire beam crosses our sightline on each rotation.\footnote{Pulsars J1141--6545 and J1906+0746 provide interesting exceptions when precession of the magnetic axis is significant enough to allow observers to map the pulsar's emission geometry as the pulsar processes \citep{Manchester+10,Desvignes19}.} Attempts to decipher the topology of pulsar beams began shortly after the discovery of pulsars, and this history is reviewed in recent publications both by \citet{olszankski+22} and \citet{paperiv}.  

Our purpose here is two-fold: first, to assemble and complete publication\footnote{A group of the brightest Arecibo pulsars were studied similarly in \citet{olszankski+22}.} of the Arecibo 1.4-GHz and 327-MHz polarimetry we have carried out on the ``B''-named pulsar population within the Arecibo sky (declinations between about --1.\degr\ and 38\degr) over the last two decades. And, second, to extend study of the spectral behavior of their radio emission beams down to the 100-MHz band or below.  The pulsars of this ``B'' population were discovered prior to the mid-1990s and studied by either the Lovell 75-m at Jodrell Bank in England---most in the course of the \citet{GL98} survey---or the Parkes 70-m telescope in Australia.  This population thus includes the great majority of objects bright enough to be studied over a broad frequency band, some down to low frequencies---the 100-MHz band---and a few into the decameter band below.  These all then complement the high quality Arecibo observations we now have now available.  

The Pushchino Radio Astronomy Observatory (PRAO) has long pioneered 103/111-MHz studies of pulsar emission using their Large Phased Array (LPA).  Recent surveys by \citet[hereafter KL99]{kuzmin1999} and \citet[MM10]{malov2010} provide a foundation for this work.  More recently, the Low Frequency Array (LOFAR) in the Netherlands has produced an abundance of high-quality profiles with their High Band Survey \cite[hereafter BKK+, PHS+]{bilous2016, pilia2016} in the 100-200 MHz band, and the Long Wavelength Array is beginning to produce quality profiles in the decameter band \citet{KTSD23}.  

A radio pulsar emission-beam model with a central ``core'' pencil beam and two concentric conal beams has proven useful and largely successful both qualitatively and quantitatively to model the beam geometry at frequencies around 1 GHz (\citet{rankin1993a} and its Appendix \citet{rankin1993b}; together hereafter ET VI); see \S~\ref{sec:ccbeams} below.  Few attempts, however, have been made to explore the systematics of pulsar beam geometry over the entire radio spectrum.\footnote{\citet{Olszanski2019} studied the beam geometries of a group of Arecibo pulsars from 327 MHz up to the 4.5 GHz band.}  Here, we present analyses aimed at elucidating the multiband beam geometry of a particular group of ``B'' pulsars within the Arecibo sky, many of which were first studied systematically by Gould \& Lyne's \citeyear{GL98} survey using the Jodrell Bank Lovell telescope.  For a few we are also able to conduct single-pulse analyses that can assist in elucidating the beam geometries.  

In what follows we provide analyses intended to assess the efficacy of the core/double-cone beam model at frequencies down to 100 MHz or below, and to compare this geometry with new and existing 1-GHz models from ET Vi and elsewhere. For a few, we are also able to conduct single-pulse analyses which assist in elucidating the beam geometries.  Our overarching goal in these works is to identify the physical implications of pulsar beamform variations with radio frequency.  Here we present our analyses of the emission beam geometry of a group of older, less-studied pulsars observed using Arecibo at 1.4 GHz and/or 327 MHz. These are all represented in the PRAO and LOFAR surveys and some as well in earlier surveys such as \citet[W99, W04]{W99, Weisberg2004}, and \citet[HR10]{hankins2010}.  

In this work, \S\ref{sec:obs} describes the Arecibo observations, \S\ref{sec:ccbeams} reviews the geometry and theory of core and conal beams, \S\ref{sec:models} describes how our beaming models are computed and displayed,  \S\ref{sec:scattering} discusses scattering and its effects at low frequencies, \S\ref{sec:discussion} the analysis and discussion, and \S\ref{sec:summary} gives a short summary. The main text of the paper introduces our analyses while the tables, model plots, and the detailed discussions are given in the Appendix.  In this Appendix, we discuss the interpretation and beam geometry of each pulsar and Figures~A1--A6 and A31--A42 show the results of other analyses clarifying the beam configurations.  Figs.~A7--A30 then give the beam-model plots and (mostly) Arecibo profiles on which they are based.  The supplementary material provides the three tables in ascii format.

\section{Observations} 
\label{sec:obs}
We present observations carried out using the upgraded Arecibo Telescope\footnote{https://www.naic.edu/ao/telescope-description} in Puerto Rico with its Gregorian feed system, 327-MHz (``P-band") or 1100-1700-MHz (``L-band") receivers, with either Wideband Arecibo Pulsar Processors (WAPPs\footnote{http://www.naic.edu/˜wapp}) or Mock spectrometer\footnote{http://www.naic.edu/ao/scientist-user-portal/astronomy/mock-spectrometer} backends.  At P-band, four 12.5-MHz bands were used across the 50 MHz available.  Four nominally 100-MHz bands centered at 1170, 1420, 1520 and 1620 MHz were used at L-band, and the lower three were usually free enough of radio frequency interference (RFI) such that they could be added together to give approximately 300-MHz bandwidth nominally at 1400 MHz.  The four Stokes parameters were calibrated from the auto- and cross-voltage correlations computed by the spectrometers, corrected for interstellar Faraday rotation, various instrumental polarization effects, and dispersion.  The resolution of each observation is usually about a milliperiod, and the sample numbers in Table~A1 reflect resampling modulo the pulsar period per a current timing solution from the ATNF pulsar Catalog\footnote{https://www.atnf.csiro.au/research/pulsar/psrcat/}.  For the 100-MHz observations, please consult the paper of origin referenced in Table~A1.

The observations and geometrical models of the pulsars are presented in the tables and figures of the Appendix.  Table~A1 describes each pulsar's dispersion (DM) and rotation (RM) measures, the MJDs, lengths and bin numbers of our Arecibo observations, and then gives the sources for the 100-MHz band observations and profile measurements. The PRAO LPA originally operated at 102.5 MHz and was later raised to 111 MHz, and neither KL99 nor MM10 clearly specify which frequency was used.  For the LOFAR observations we indicate the lowest frequency we used.  Table~A2 gives the physical parameters of each pulsar that can be computed from the period and spindown rate \citep[version 1.67]{ATNF}: the energy loss rate, spindown age, surface magnetic field, the acceleration parameter {$B_{12}/P^2$} and the reciprocal of \citet{beskin}'s $Q$ (=$0.5\ 10^{15} \dot P^{0.4} P^{-1.1}$) parameter, which also scales roughly with the spindown energy . The Gaussian fits use Michael Kramer's bfit code \citep{kramer+94,kramer94}.  The geometrical models are given in Table~A3 as will be described below.  Plots then follow showing the behavior of the geometrical model over the frequency interval for which observations are available, as well as Arecibo polarized average profiles where
available.

\section{Core and Conal Beams}
\label{sec:ccbeams}
A full recent discussion of the core/double-cone beam model and its use in computing geometric beam models is given in \citet{paperiv}.

Canonical pulsar average profiles are observed to have up to five components \citep{rankin1983a}, leading to the conception of the core/double-cone beam model \citep{backer}. Pulsar profiles then divide into two families depending on whether core or conal emission is dominant at about 1 GHz.  Core single ({\textbf S$_{t}$}) profiles consist of an isolated core component, often flanked by a pair of outriding conal components at high frequency, triple ({\textbf T}) profiles show a core and conal component pair over a wide band, and five-component ({\textbf M}) profiles have a central core component flanked by both an inner and outer pair of conal components. 

By contrast, conal profiles can be single ({\textbf S$_{d}$}) or double ({\textbf D}) when a single cone is involved, or triple (c{\textbf T}) or quadruple (c{\textbf Q}) when the sightline encounters both conal beams. Outer cones tend to have an increasing radius with wavelength, while inner cones tend to show little spectral variation.  Periodic modulation often associated with subpulse ``drift'' is a common property of conal emission and assists in defining a pulsar's beam configuration \citep[\eg][]{et3}. 

Profile classes tend to evolve with frequency in characteristic ways:  ({\textbf S$_{t}$}) profiles often show conal outriders at high frequency, whereas ({\textbf S$_{d}$}) profiles often broaden and bifurcate at low frequency.  ({\textbf T}) profiles tend to show their three components over a broad band, but relative intensities can change greatly.  ({\textbf M}) profiles usually show their five components most clearly at meter wavelengths, while at high frequency they become conflated into a ``boxy'' form, and at low frequency they become triple because the inner cone often weakens relative to the outer one.

Application of spherical geometry to the measured profile dimensions provides a means of computing the angular beam dimensions---resulting in a quantitative emission-beam model for a given pulsar.  Two key angles describing the geometry are the magnetic colatitude (angle between the rotation and magnetic axes) $\alpha$ and the sightline-circle radius (the angle between the rotation axis and the observer’s sightline) $\zeta$, where the sightline impact angle $\beta$ = $\zeta-\alpha$.  The three beams are found to have regular angular dimensions at 1 GHz in terms of a pulsar's polar cap angular diameter, {$\Delta_{PC}$} = $2.45\degr P^{-1/2}$ \citep{rankin1990}.  The outside half-power radii of the inner and outer cones, {$\rho_{i}$} and {$\rho_{o}$} are  well described by $4.33\degr P^{-1/2}$ and $5.75\degr P^{-1/2}$ \citep{rankin1993a}.  

$\alpha$ can be estimated from the core-component width when present, as its half-power width at 1 GHz, $W_{\rm core}$ has been shown to scale as {$\Delta_{\rm PC}/\sin\alpha$} (ET IV).  The sightline impact angle $\beta$ can then be estimated from the polarization position angle (PPA) sweep rate). $R$=$|d\chi/d\varphi|$ measures the ratio $\sin\alpha/\sin\beta$.  Conal beam radii can similarly be estimated from the outside half-power width of a conal component or conal component pair at 1 GHz $W_{\rm cone}$ together with $\alpha$ and $\beta$ using eq.(4) in ET VIa:  
\begin{equation} \label{eq1}
    \rho = \text{cos} ^{-1} [ \text{cos }\beta  - 2\text{sin }\alpha \text{ sin }\zeta \text{ sin }^{2} (\Delta\psi/4)]
\end{equation}
where $\Delta\psi$ is the total half-power width of the conal components measured in degrees longitude.  The characteristic height of the emission can then be computed assuming dipolarity using eq.(6). 

The outflowing plasma responsible for a pulsar's emission is partly and or fully generated by a polar ``gap" \citep{ruderman}, just above the stellar surface.  \citet{Timokhin} find that this plasma is generated in one of two pair-formation-front (PFF) configurations:  for the younger, energetic part of the pulsar population, pairs are created at some 100 m above the polar cap in a central, uniform (1-D) gap potential---thus a 2-D PFF, but for older pulsars the PFF has a lower, annular shape extending up along the polar fluxtube, thus having a 3-D cup shape. 

An approximate boundary between the two PFF geometries is plotted on the $P$-$\dot P$ diagram of Fig~\ref{fig1}, so that the more energetic pulsars are to the top left and those less so at the bottom right.  Its dependence is $\dot P=$3.95$\times$$10^{-15}P^{11/4}$.  Pulsars with dominant core emission tend to lie to the upper left of the boundary, while the conal population falls to the lower right.  In the parlance of ET VI, the division corresponds to an acceleration potential parameter $B_{12}/P^2$ of about 2.5, which in turn represents an energy loss $\dot E$ of 10$^{32.5}$ ergs/s.  This delineation also squares well with \citet{Weltevrede2008}'s observation that high energy pulsars have distinct properties and \citet{basu2016}'s demonstration that conal drifting occurs only for pulsars with $\dot E$ less than about $10^{32}$ ergs/s.  Table~A2 gives the physical parameters that can be computed from the period $P$ and spindown $\dot P$, including the $\dot E$ and $B_{12}/P^2$ (see the sample below in Table~\ref{tab2}).

\section{Computation and Presentation of Geometric Models} 
\label{sec:models}
 Two key observational values underlie the computation of conal radii at each frequency and thus the model overall:  the conal component width(s) and the polarization position angle (PPA) sweep rate; the former gives the angular scale of the conal beam(s) while the latter gives the impact angle $\beta$ showing how the sightline crosses the beam(s).  Figures~A7--A30 show our Arecibo (or other) profiles and Table~A1 describes them as well as referencing any 100-MHz band published profiles (see the sample below in Table~\ref{tab1}).  Following the analysis procedures of ET VI, we have measured outside conal half-power (3 db or FWHM) widths and half-power core widths wherever possible. The measurements are given in Table~A3 for the 1.4-GHz and 327-MHz bands and for the 100-200 MHz regime (see the sample below in Table~\ref{tab3}).  

These provide the bases for computing geometrical beaming models for each pulsar, which are also shown in the above figures and Table~A3.  However, we do not plot these directly.  Rather we use the widths to model the core and conal beam geometry as above, but here emphasizing as low a frequency range as possible. The model results are given in Table~A3 for the 1-GHz and  100-200-MHz band regimes.  $W_{c}$,  $\alpha$, $R$ and $\beta$ are the 1-GHz core width, the magnetic colatitude, the PPA sweep rate and the sightline impact angle;  $W_i$/$W_o$ and $R=$ $\rho_i$/$\rho_o$ are the respective inner and outer conal component widths and the respective beam radii, at 1 GHz, 327 MHz, and the lowest frequency values in the 100-MHz bands.  The $\alpha$ values are bolded in Table~A3 when they could be determined as above by comparing the core width at around 1 GHz with the 2.45\degr/$P^{1/2}$ intrinsic angular diameter of the polar polar cap.

We depart from past practices by presenting our results in terms of core and conal beam dimensions as a function of frequency.  The results of the model for each pulsar are then plotted in Figures~A7 to A30.  The plots are logarithmic on both axes, and labels are given only for values in orders of 1, 2 and 5. For each pulsar the plotted values represent the {\bf scaled} inner and outer conal beam radii and the core angular width, respectively.  The scaling plots each pulsar's beam dimensions as if it were an orthogonal rotator with a 1-sec rotation period---thus the conal beam radii are scaled by a factor of $\sqrt{P}$ and the core width (diameter) by $\sqrt{P}\sin{\alpha}$ (\eg for B0045+33's cones and cores the factors would be 1.103 and 0.794, respectively). This scaling then gives each pulsar the same expected model beam dimensions, so that similarities and differences can more readily be identified.  The scaled outer and inner conal radii are plotted with blue and cyan lines and the core diameter in red.  The nominal values of the three beam dimensions at 1 GHz are shown in each plot by a small triangle.  Please see the text and figures of Rankin (2022) for a full explanation. 

Estimating and propagating the observational errors in the width values is very difficult.  Instead of quoting the individual measurement errors, we provide error bars reflecting the beam radii errors for a 10\% uncertainties in the conal width values, the PPA sweep rate, and the error in the scaled core width.  The conal error bars shown reflect the {\it rms} of the first two sources with the former indicated in the lower bar and the latter in the upper one.  

\begin{table*}
\caption{Sample: Observation Information} 
\begin{tabular}{lcc|ccc|ccc|l}
    \hline
     & & & & \mbox{\textbf{L-band}} & & \multicolumn{3}{c|}{\textbf{(P-band or Refs.)}} & \mbox{\textbf{100 MHz}} \\
\hline
 Pulsar  & DM & RM  & MJD & $N_{pulses}$ & Bins & MJD & $N_{pulses}$ & Bins & References  \\
         & pc/$cm^{3}$ & rad/$m^{2}$ & & & & \multicolumn{3}{c|}{\textbf{GL98 and other refs.)}}  & (MHz) \\
\hline
\hline
B0045+33     & 39.9 &--82.3 & 52837   &  1085  & 1188   &   53377   &   1085  &  1188   & BKK+, 129; KL99\\
B0820+02    & 23.7 & 13  &  \multicolumn{3}{c|}{\textbf{Hankins \& Rankin (2010)}}  &   54781   &   1388  &  1017   & MM10; XBT+; PHS+: KTSD, 50 \\
B0940+16     & 20.3 & 53  & 52854   &  1048  & 1024   &   53490   &    827  &  1175   & BKK+, 149 \\
B1534+12    & 11.62 & 10.6 & 55637 & 15835 & 256 & \multicolumn{3}{c|}{\textbf{\citet{GL98}}} & KL99, MM10     \\
B1726--00   & 41.1 & 20  & 56415   &  1554  & 1024   &   52930   &   6217  &   752   & MM10 \\
\\[-2pt]
B1802+03    & 80.9 & 38.9 & 56769   &  2743  & 1096   & \multicolumn{3}{c|}{\textbf{\citet{mcewen}}} & MM10 \\
B1810+02   & 104.1& --25 & 56406   &  1032  & 1024   &   53378   &   1032  &   775   & MM10 \\
B1822+00    & 62.2 & 158 & 56406   &  1052  & 1014   &   53378   &   1052  &   762   & MM10 \\
B1831-00    & 88.65 & --- & 52735  &  1151  &  256   &   53377   &   1151  &   256   &  \\                  
B1848+04     & 115.5&  86  & 56768   &  2102  &  512   & \multicolumn{3}{c|}{\textbf{\citet{boriakoff}}}  & MM10 \\
\hline
\end{tabular}
\label{tab1} 
\end{table*}

\begin{table}
\setlength{\tabcolsep}{2pt}
\caption{Sample: Pulsar Parameters}
\begin{center}
\begin{tabular}{lccccccc}
\hline
 Pulsar &  P & $\dot{P}$ & $\dot{E}$ & $\tau$ & $B_{surf}$ & $B_{12}/P^2$ & 1/Q  \\
 (B1950) & (s) & ($10^{-15}$ & ($10^{32}$  & (Myr) & ($10^{12}$ &   &   \\
 & & s/s) & ergs/s) & & G) &   &    \\
\hline
\hline
B0045+33 & 1.2171 & 2.35 & 0.52 & 8.19 & 1.71 & 1.2 & 0.6 \\
B0820+02 & 0.8649 & 0.10 & 0.06 & 131 & 0.30 & 0.4 & 0.2 \\
B0940+16 & 1.0874 & 0.09 & 0.03 & 189 & 0.32 & 0.3 & 0.2 \\
B1534+12 & 0.0379 & 0.00 & 18.0 & 248 & 0.01 & 6.8 & 1.6 \\
B1726--00 & 0.3860 & 1.12 & 7.70 & 5.45 & 0.67 & 4.5 & 1.5 \\
\\[-2pt]
B1802+03 & 0.2187 & 1.00 & 38.0 & 3.47 & 0.47 & 9.9 & 2.7 \\
B1810+02 & 0.7939 & 3.60 & 2.80 & 3.49 & 1.71 & 2.7 & 1.1 \\
B1822+00 & 1.3628 & 1.75 & 0.27 & 12.40 & 1.56 & 0.8 & 0.4 \\
B1831--00 & 0.5210 & 0.01 & 0.03 & 784 & 0.07 & 0.3 & 0.2	\\
B1848+04 & 0.2847 & 1.09 & 19.0 & 4.14 & 0.56 & 6.9 & 2.1 \\
\hline
\end{tabular}
\label{tab2} 
\end{center}
Notes: Values from the ATNF Pulsar Catalog \citep{ATNF}, \textbf{Version 1.67} .
\end{table}

\begin{table*}
\caption{Sample: Emission-Beam Model Geometry}
\setlength{\tabcolsep}{3pt} 
\begin{tabular}{lc|ccc|ccccc|ccccc|ccccc}
    \toprule
    Pulsar &  Class & $\alpha$ & $R$ & $\beta$ & $W_c $ & $W_i$ & $\rho_i$ & $W_o$  & $\rho_o$ & $W_c $ & $W_i$ & $\rho_i$ & $W_o$  & $\rho_o$ & $W_c $ & $W_i$ & $\rho_i$ & $W_o$  & $\rho_o$  \\
          &   & (\degr) & (\degr/\degr) & (\degr) & (\degr) & (\degr) & (\degr) & (\degr) & (\degr) & (\degr) & (\degr) & (\degr) & (\degr) & (\degr) & (\degr) & (\degr) & (\degr) & (\degr) & (\degr) \\
    \midrule
    & & \multicolumn{3}{c|}{(1-GHz  Geometry)} & \multicolumn{5}{c|}{(1.4-GHz Beam Sizes)} & \multicolumn{5}{c|}{(327-MHz Beam Sizes)} & \multicolumn{5}{c}{(100-MHz Band Beam Sizes)} \\
    \midrule
    \midrule
B0045+33 & D & 46 & -15 & +2.7 &  --- & 7.8 & 4.0 &  --- &  --- &  --- & 6.0 & 3.5 &  --- &  --- &  --- & 5.9 & 3.5 &  --- &  --- \\
B0820+02 & Sd & 71 & +12 & +4.5 &  --- &  --- &  --- & 9.0 & 6.2 &  --- & 0 &  --- & 9.5 & 6.4 &  --- &  --- &  --- & 15.4 & 8.6 \\
B0940+16 & Sd/PC & 25 & +6 & +4.0 &  --- &  --- &  --- & 16.0 & 5.4 &  --- &  --- &  --- & 24.9 & 6.9 &  --- &  --- &  --- & 28.5 & 7.6 \\
B1534+12 & ?? & 60 & -8 & +6.2 & 6.3 & $\sim$48 & 22.2 &  --- &  --- &  --- &  --- &  --- &  --- &  --- & $\sim$8 & $\sim$55 & 25.2 &  --- &  --- \\
B1726--00 & T? & {\bf 26} & $\sim$5 & +5.0 & $\sim$9 & 20.3 & 7.0 &  --- &  --- &  --- & 19.8 & 6.9 &  --- &  --- &  --- & $\sim$28 & 8.3 &  --- &  --- \\
\\
B1802+03 & St & {\bf 44} & +4.2 & +9.6 & 7.5 &  --- &  --- & 20.0 & 12.2 & $\sim$7 &  --- &  --- &  --- &  --- & $\sim$14 &  --- &  --- &  --- &  --- \\
B1810+02 & St & {\bf 25} & +36 & +0.7 & 6.7 &  --- &  --- &  --- &  --- & 6.5 &  --- &  --- &  --- &  --- & $\sim$13 &  --- &  --- &  --- &  --- \\
B1822+00 & cT? & 27 & -7.5 & +3.5 &  --- & $\sim$6 & 3.8 & 13.7 & 5.0 &  --- & $\sim$7 & 3.9 & $\sim$15 & 5.0 &  --- &  --- &  --- & $\sim$27 & 7.3 \\
B1831-00 & Sd & 12 & +2.4 & +5.0 &  --- & $\sim$25 & 5.8 &  --- &  --- &  --- & $\sim$20 & 5.5 &  --- &  --- &  --- &  --- &  --- &  --- &  --- \\
B1848+04 & T & {\bf 8} & +2.4 & +3.4 & $\sim$32 & 88 & 8.1 &  --- &  --- & $\sim$32 & $\sim$99 & 8.9 &  --- &  --- &  --- & $\sim$101 & 9.0 &  --- &  --- \\
\hline
\end{tabular}
\label{tab3} 
\end{table*}

\section{Low-Frequency Scattering Effects} 
\label{sec:scattering}
No competent interpretation of pulsar profiles at low frequency can be made without also considering the level of distortion on particular observations at particular frequencies.  Here, we use the \citet{kmn+15} compendium of scattering times which draws on \citet{kuzmin_LL2007}'s measurements of 100-MHz scattering times as well as other studies.  These are shown on the model plots as double-hatched orange regions where the boundary reflects the scattering timescale at that frequency in rotational degrees. For pulsars having no scattering study, we use the mean scattering level determined from the dispersion measure following \citet{Kuzmin2001} and \citet[KLL07]{kuzmin_LL2007}, though some pulsars have scattering levels up to about ten times greater or smaller than the average level.  Our model plots show the average scattering level (where applicable) as yellow single hatching and with an orange line indicating ten times this value as a rough upper limit.

\section{Analysis and Discussion} 
\label{sec:discussion}
\noindent\textit{\textbf{Core/double-cone Modeling Results}}:  The 76 pulsars show beam configurations across all of the core/double-cone model classes.  About half of this group have $\dot E$ values $\ge$ $10^{32.5}$ ergs/s and either core-cone triple \textbf{T} or core-single \textbf{S$_t$} profiles.  The remainder tend to have profiles dominated by conal emission---that is, conal single \textbf{S$_d$}, double \textbf{D}, triple c\textbf{T}, or quadruple c\textbf{Q} geometries. This small population again displays the $\dot E$ boundary between core and conal dominated profiles and the emission beams that produce them. 

Conal profiles tend to show periodic modulation, so fluctuation spectral features identifying such effects are often be useful in identifying conal emission.  Here, we made use of the pulse-modulation studies of \citet{WES06,WSE07} and in some cases carried out analyses of our own---\eg see the discussion of pulsar B1919+14 below and Fig.~A35.

We were able to construct quantitative beam geometry models for all but two of the pulsars; however, some are better established than others on the basis of the available information.  Lack of reliable PPA rate estimates was a limiting factor in a number of cases, either due to low fractional linear polarization or difficulty interpreting it.  Generally, it was possible to trace the sightline's traverse through the emission beam(s) across the three bands, sometimes despite very different spectral behavior, but in many cases scattering obliterated profile structure at the lowest frequencies making it impossible to discern the profile structure.  In rare cases the profiles may be dominated by different profile modes---and this may well account for the one pulsar (B1915+22) for which we were unable to identify its beam configuration.

Common structures were usually recognizable between profiles at different frequencies.  Core components and beams usually showed the expected geometric 2.5\degr\ width of the pulsar polar cap at 1 GHz and little escalation with wavelength.  In one case (B1918+26) the core width narrowed at meter wavelengths, an unusual behavior seen in a few other objects.  In cases where $\alpha$ could be estimated from the intrinsic polar cap diameter (see above), the inner and outer conal radii tended to assume values very close to their expected scaled dimensions of 4.33\degr\ and 5.75\degr.  Inner cones tended to change little with wavelength whereas the outer ones often showed some intrinsic increases before scattering overtook them.  When no core feature was discernible, $\alpha$ was adjusted such that the conal radius had the inner or outer value.  For both conal triple (c\textbf{T}) or quadruple (c\textbf{Q}) profiles, $\alpha$ was determined by the presence of both cones.  For \textbf{S$_d$} and (\textbf{D}) profiles, it was often difficult to determine whether an inner or outer conal beam was involved: sometimes width increases with frequency suggests and outer cone and rarely one is excluded because $\alpha$ cannot exceed 90\degr. 
\vskip 0.1in
\begin{figure}
\begin{center}
\includegraphics[width=75mm,angle=0.]{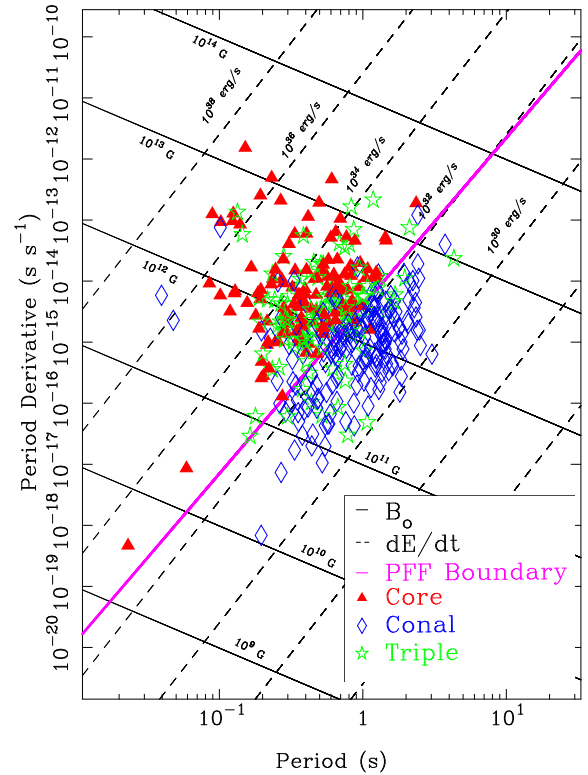}
\caption{P-$\dot P$ Diagram showing the distribution of the ``B'' pulsar population in relation to the PFF boundary. Core-emission-dominated pulsars tend to lie to the upper left of the boundary line, whereas those with mainly conal emission fall to the lower right (see text). Pulsars with core-cone triple {\textbf T} are distributed across both regions.  Of particular interest are the three energetic pulsars that seem to be conal emitters, B1259--63, B1823--13 and B1951+32.}
\label{fig1}
\end{center}
\end{figure}
\noindent\textit{\textbf{``B'' Populations}}:  The ATNF Catalog\footnote{https://www.atnf.csiro.au/research/pulsar/psrcat/} lists some 487 normal (rotation-powered) pulsars with ``B'' discovery names---that is, sources that were discovered before the mid-1990s or so.  These are an interesting population because many were discovered with---and all are accessible to---either the 70-80-meter-class Jodrell Bank Lovell or the Parkes Telescopes.  Of these, 130 fall within the Arecibo sky---that is between declinations of --1.5\degr\ and +38.5\degr---and 100 were included in the GL98 survey.  Here we consider a large proportion, 76, of this population\footnote{We also account for the 12 objects we were not able to include; see the Note in Table~A1}---in general, a weaker less prominent and studied group.  We have treated the best known, generally brighter group of 42 similarly in \citet{olszankski+22}, so overall some 90\% of the Arecibo ``B'' population has been observed and studied.  The population of `B'' pulsars outside the Arecibo sky has also been similarly studied (Rankin 2022).  Combining these studies, Fig.~\ref{fig1} shows the P-{$\dot P$} distribution of the overall ``B'' population.  Nearly 2/3 of this population is core-dominated---that is, mostly \textbf{S$_t$}.  

Further, 15 of the pulsars are included in the LOFAR High Band Survey \citep{bilous2016} along with a number of more recently discovered objects, and a full study of this population is in preparation \citet{Wahl+22}.
\vskip 0.1in
 
\noindent\textit{\textbf{Galactic Distribution of Arecibo ``B'' Pulsars}}:  The partial sky coverage of the Arecibo telescope is well known, but it is useful to remind ourselves of the specifics.  The declination limits of --1.5\degr\ to +38.5\degr\ has the effect of giving access to only two narrow portions of the Galactic plain, one near 5$^h$ RA and another around 19$^h$.  At other right ascensions pulsars have large Galactic latitudes reflecting relatively local objects.  The Galactic anticenter region is largely accessible within the Arecibo sky, whereas the center region can only be accessed down to about 30\degr\ Galactic longitude.  
\vskip 0.1in
\begin{figure}
\begin{center}
\includegraphics[width=75mm,angle=0.]{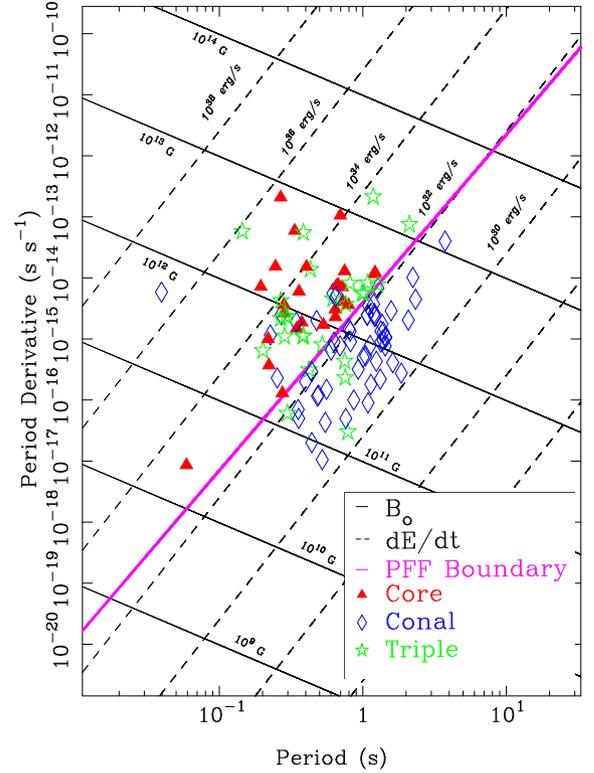}
\caption{P-$\dot P$ Diagram showing the distribution of the Arecibo pulsar population in relation to the PFF boundary as in Fig.~\ref{fig1}. Here we see a somewhat smaller fraction of core-dominated pulsars in the Arecibo sky.}
\label{fig2}
\end{center}
\end{figure}

This largely local and Galactic anticenter population of pulsars within the Arecibo sky seems to give them particular characteristics:  many are older perhaps having taken some time to move away from the plane where they can be detected at largish Galactic latitudes, and some may be intrinsically less luminous given that although relative close many remain quite bright.  Or put differently, the Arecibo population has a somewhat larger fraction of conal dominated pulsars---around 50\%---as shown in the P-{$\dot P$} distribution of Fig.~\ref{fig2}---as opposed to about 1/3 of the ``B'' pulsars in Fig.~\ref{fig1}.  There are only a few bright core dominated objects within this sky, whereas many are found near the plane in the inner Galaxy.  It is also useful to keep in mind that the Arecibo sky is 75\% of the northern sky, as only an interval of the plane above about 38\degr\ declination is missing---that between about 20$^h$ and 4$^h$---however, some 50 ``B'' pulsars reside in this region.  

Finally, while Arecibo lacked access to the Galactic center region, so do most northern instruments to one extent or another.  Arrays at high northern latitudes lose sensitivity toward the equator and even such a fully steerable telescope as the Lovell has access only down to --35\degr\ declination.  
\vskip 0.1in

\noindent\textit{\textbf{Pulsars With Interesting Characteristics}}

\noindent{\textbf{B0045+33}} seems to modulate its single pulses in two different modes as shown in Fig~A1, one with a 2.18-$P$ periodicity and another with three times this. This modulation is indicative of conal radiation \citep[\eg][and the cited references]{DR01}. 

\noindent{\textbf{B0820+02}} has a stable 4-5-$P$ phase modulation across it conal single profile; see Fig~A2.

\noindent{\textbf{B1822--00}} shows a 5.5-$P$ amplitude periodicity that modulates parts of its profile at different phases as shown in Fig~A3.  

\noindent{\textbf{B1854+00}} shows clear driftbands in Fig.~A4 but no fluctuation-spectral feature, perhaps due to irregularity or its weakness. 

\noindent{\textbf{B1901+10}} exhibits a 23-$P$ amplitude modulation as depicted in the folded pulse sequence in Fig.~A6.  

\noindent{\textbf{B1907+03}} is modulated at a 2.4-$P$ cycle---or perhaps one four times as long---such that different profile regions are bright at different phases of its cycle supporting the conal triple or quadruples identification; see Fig~A32.

\noindent{\textbf{B1913+167}} exhibits a strong 65.4-$P$ fluctuation feature that interestingly is produced by a regular alternating pattern of emission in the leading and trailing conal components with core emission at the beginning of both intervals, as shown in Fig~A33. 

\noindent{\textbf{B1919+14}}, remarkably, shows a 10.24-$P$ coherent phase and amplitude modulation, and sidebands spaced at 1/4 this frequency are also clearly evident.  This seems to indicate a stable pattern of 4 ``beamlets'' as shown in Fig~A35, perhaps in the manner of a carousel configuration as in B0943+10 \citep{DR99,DR01}.

\noindent{\textbf{B1919+20}}'s single pulses show an 8.3-$P$ amplitude modulation in Fig~A36.

\noindent{\textbf{B1930+22}} shows a persistent 42.7-$P$ modulation, such that several distinct regions are illuminated during the cycle; see Fig.~A37. 

\noindent{\textbf{B1930+13}} has a conal double profile with a 4.92-$P$ part phase and part amplitude modulation common to both components as shown in Fig.~A38.

\noindent{\textbf{B1942--00}} exhibits an 8.26-$P$ phase-modulated periodicity that is common to both components with a different phase in each component, as shown in Fig.~A39. 

\noindent{\textbf{B2000+32}} emits very intense narrow single pulses against a background of much weaker ones as shown in Fig.~A40;  many of these are probably ``giant'' pulses.

\noindent{\textbf{B2053+21}} has an interesting and unusual subpulse modulation pattern, where subpulses in the first component have a fixed longitude; whereas, in the second component they often show ``drift''; see Fig.~A41.

\noindent{\textbf{B2210+29}} has a classic five-component profile with a 5.45-$P$ periodicity modulating its conal components at different phases (Fig.~A42) and a 6-4-$P$ cycle in the central core region. 
\vskip 0.1in

\noindent\textit{\textbf{Scattering Levels of Arecibo Population Pulsars}}: The above analyses have provided opportunity to study the levels of interstellar scattering for the entire population of ``B'' pulsars.  A large proportion of the objects in this population have measured scattering times, many following from the PRAO work of \citet{kuzmin_LL2007} as updated by \citet{kmn+15}.  The analysis in \citet{paperiv} showed that scattering levels are higher or much higher in the Galactic Center region, whereas in the anticenter direction or at higher Galactic latitudes they are generally less severe.  Only a small portion of the Arecibo sky approached the Galactic Center direction within 20-30\degr, and it is only here where severe scattering is encountered.  Generally, then the average scattering level of \citet{Kuzmin2001} is appropriate for this work---the more so in being based on 100-MHz PRAO measurements.  This all said, we can be reminded that scattering is very ``patchy'' over the entire sky, and some particular directions encounter very little scattering well into the decameter band as seen in the recent work of \citet{Zakharenko2013}.

\vskip 0.1in

\noindent\textit{\textbf{Particular Significance of Decametric Pulsar Observations}}:  Our analyses in this and previous works are framed by efforts to interpret pulsar beamforms at the lowest possible frequencies---and this line of investigation gives pride of place to the 25 and 20-MHz pulsar surveys from the UTR-2 instrument in Kharkiv, Ukraine, along with highly significant results from use of LOFAR's Low Band.  We have therefore pointed to one or another of the 40 profiles in the \citet{Zakharenko2013} survey or others from \citet{bilous2019,pilia2016} wherever possible, and we are cogisant of the recent second survey of \citet{Kravtsov22} as well.  
This is very difficult work technically for any number of reasons, so only a minority of the actual detections provide relatively intrinsic profiles that can be interpreted reliably.  Many suffer from scattering distortion and/or poor definition due to spectral turnovers and escalating Galactic noise---to say nothing of interference.  The largest overlap is in Paper I and \citet{paperiv} but we note several here and in a subsequent paper \citep{paperiii} also. Most of the detections so far have been of ``B'' pulsars, but fully half of the recent Kharkiv survey are of pulsars discovered with LOFAR \citet{scb+19,tbc+20} and not yet well studied at higher frequencies.

The few pulsars with good quality, relatively undistorted profiles in the decameter band provide unique information on pulsar emission physics and beaming configurations.  Between the two above Kharkiv surveys more than 50 pulsars have been detected, most lie out of the Galactic plane and many in the anticenter direction.  These detections represent a further population of pulsars that can be discovered and observed at low frequency, some with high quality profiles.  Moreover, scattering distortion can be alleviated in some cases using deconvolution methods as did \citet{kuzmin1999}.  Further development of these techniques promises to reveal more about pulsar beaming as well as the characteristics of the scattering itself.

\section{Summary} 
\label{sec:summary}
We have provided analyses of the beam structure of 76 ``B'' pulsars that were included in the GL98 survey as well as a number in the LOFAR High Band survey.  These compliment a group of the most-studied pulsars within the Arecibo sky that were similarly treated in Olszanski  \etal\ (2022).  This group also includes almost all of the ``B'' pulsars that have been detected at frequencies below 100 MHz.  It also compliments a large group of objects lying outside the Arecibo sky in Rankin (2022), and a number are included in the LOFAR High Band Survey \citep{bilous2016}.  Our analysis framework is the core/double-cone beam model, and we took the opportunity not only to review the models for these mostly venerable pulsars but to point out situations where the modeling is difficult or impossible.  As an Arecibo population, many or most of the objects tend to fall in the Galactic anticenter region or at high Galactic latitudes, so overall it includes a number of nearer, older pulsars.  We found a number of interesting or unusual characteristics in some of the pulsars that would benefit from additional study.   Overall, the scattering levels encountered for this group are low to moderate, apart from a few pulsars lying in directions toward the inner Galaxy.

\section*{Acknowledgements}
HMW gratefully acknowledges a Sikora Summer Research Fellowship. Much of the work was made possible by support from the US National Science Foundation grants AST 99-87654 and 18-14397. We especially thank our colleagues who maintain the ATNF Pulsar Catalog and the European Pulsar Network Database as this work drew heavily on them both.  The Arecibo Observatory is operated by the University of Central Florida under a cooperative agreement with the US National Science Foundation, and in alliance with Yang Enterprises and the Ana G. M\'endez-Universidad Metropolitana. This work made use of the NASA ADS astronomical data system.

\section{Observational Data availability}
The profiles will be available on the European Pulsar Network download site, and the pulses sequences can be obtained by corresponding with the lead author.




%
%

\bibliography{biblio.bib}

\newpage

\twocolumn

\appendix
\setcounter{figure}{0}
\renewcommand{\thefigure}{A\arabic{figure}}
\renewcommand{\thetable}{A\arabic{table}}
\setcounter{table}{0}
\renewcommand{\thefootnote}{A\arabic{footnote}}
\setcounter{footnote}{0}

\section{Pulsar Tables, Models and Notes}

\begin{figure}
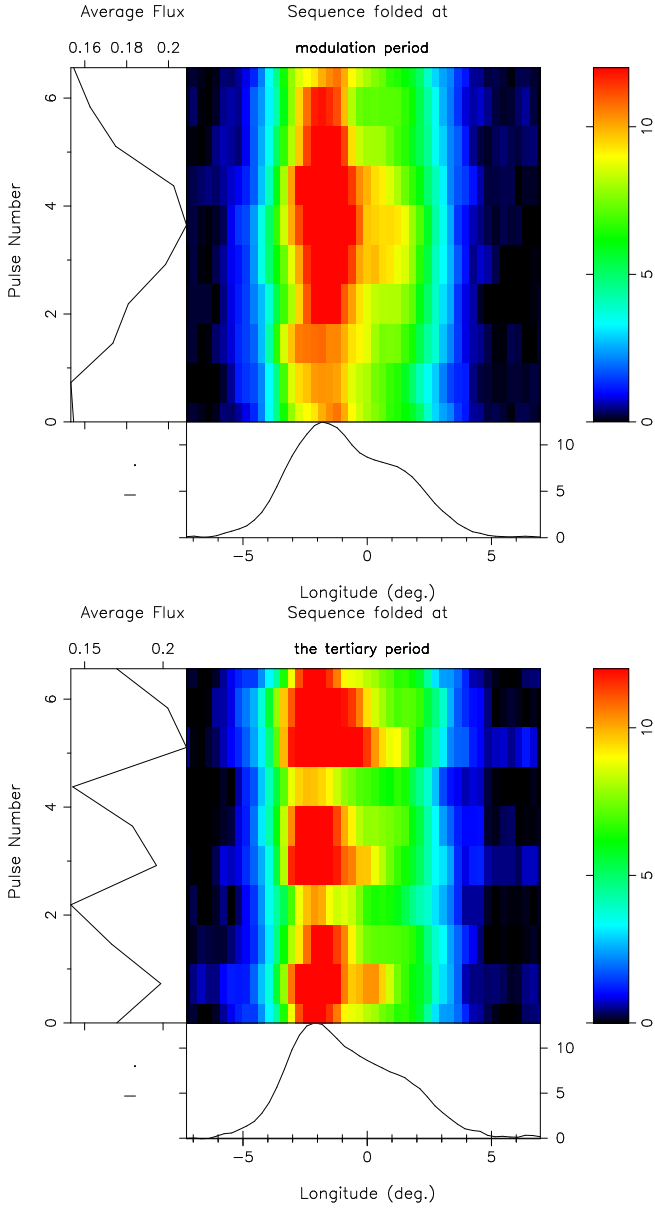

\begin{center}
\includegraphics[height=85mm,angle=-90.]{plots/PQB0045+33.53377p_modfold_6.56P_512p.ps} \\
\includegraphics[height=85mm,angle=-90.]{plots/PQB0045+33.53377p_modfold_2.18P_256p.ps}
\caption{Diagram showing what seem to be two amplitude-modulation modes in pulsar B0045+33 at 327 MHz, one at 6.56 $P$ in one 512-pulses section and a second at three times this rate at 2.18 $P$ in a different section.}
\label{figA100}
\end{center}
\end{figure}

\noindent\textit{\textbf{B0045+33}}: At LOFAR frequencies (see BKK+), this pulsar has one sharp, narrow component; as the frequency increases, it develops a second component that never fully separates from the first. Unusually, the 1.4-GHz profile is wider than those at lower frequencies.  Its geometry is modeled as a narrow \textbf{D} inner cone.  Fluctuation spectra  (see \citet{DR01} and it's references for explanation) for two sections of the 327-MHz observation show different modulations, one at 2.18-$P$ (lower panel) and a second at three times this rate (6.56 $P$, upper panel), both primarily amplitude modulations as shown in Fig~\ref{figA100}. 
\vskip 0.1in

\noindent\textit{\textbf{B0531+21}}: The famous Crab Pulsar was the first MSP, and the beamforms corresponding to its several components remain a complex mystery.  Moreover, apart from the low frequency precursor component, the others show little polarization, so we see no systematic PPA traverse.  The main pulse and so-called interpulse are both narrow and of comparable intensity at meter wavelengths. but their spacing is just under 150\degr, far less than the half period suggestive of a two-pole interpulsar.  The precursor has about the right width (and softer spectrum) to be a core component if $\alpha$ is about 90\degr\ \citep{rankin1990}; however if so, it is surprisingly weak relative to the putatively conal MP and IP, perhaps again suggesting that our sightline has a large impact angle.  Aberration/retardation might also be expected to be a strong factor in the profile structure, but it can only shift components relative to each other by 57\degr\ or less, and no such shift provides and recognizable core-cone structure.  And this fails to consider the structures and phenomena observed at high frequencies by \citet{hje15}.  All this said, the Crab is an MSP, and many MSPs show profiles with no obvious such structure. 
\vskip 0.1in

\begin{figure}
\begin{center}
\includegraphics[width=75mm,angle=-90.]{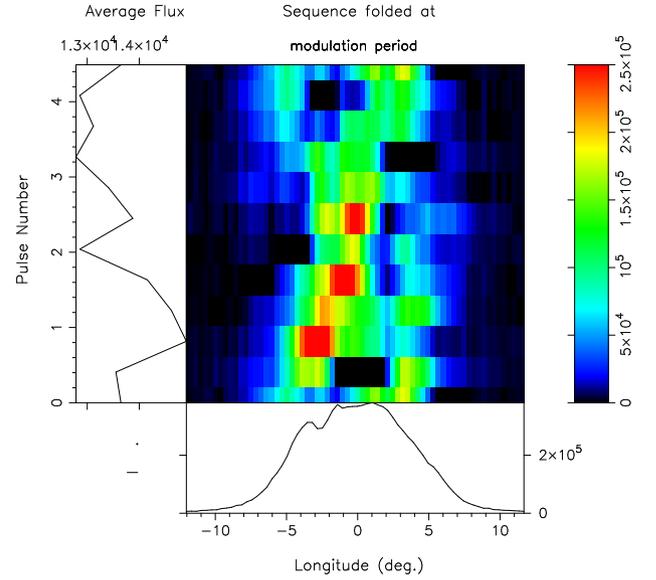}
\caption{B0820+02: The pulsar shows a phase modulation associated with drifting subpulses on a 4.5-$P$ cycle.  The plots shows a 512-pulse interval folded at this period with the unvarying ``base'' removed.}
\label{figA101}
\end{center}
\end{figure}
\noindent\textit{\textbf{B0820+02}}: PSR B0820+02 exhibits a conal single profile that bifurcates at very low frequency, and it shows an accurate drift modulation as shown in Fig~\ref{figA101}.  It could be either and inner or outer; the substantial width increase at low frequency suggests it has an outer cone.  \citet{IMS89} provide a 102-MHz observation, and \citet{KTSD23} down to 50 MHz; their profiles cannot be measured accurately but suggest that the former is too large, perhaps due to poor resolution.  (The large putative width of the MM10 profile suggests some error.)  
\vskip 0.1in

\noindent\textit{\textbf{B0940+16}}: This pulsar's main pulse has been difficult to classify with various attempts as \textbf{M} \citep{et3} and a \textbf{D} \citep{etx}, Its possible post-cursor and the bridge of connecting emission are also perplexing. In addition, we find flat PPA traverses, whereas GL98 suggests a value of perhaps +6\degr/\degr.  Two main pulse (hereafter, MP) components are present at all frequencies; at LOFAR frequencies, the leading component is much stronger than the trailing and remains so at higher frequencies with less disparity.  Here we model the MP geometry as \textbf{$S_d$} in part because \cite{Deich} identified drifting subpulses and use GL98's $R$ value (see \S\ref{sec:ccbeams}) of +6\degr/\degr.  \citet{Zakharenko2013} detected the pulsar at both 25 and 20 MHz with widths near 62 and 85\degr\ that may be compatible with an outer conal evolution. 
\vskip 0.1in
\noindent\textit{\textbf{B1534+12}} is a 38-ms binary interpulsar, and it exemplifies the issues in modeling most such pulsars with a core cone model.  Its main-pulse profile appears to have what is a central core component flanked by two conal outriders.  However, a first issue is that its putative core has a width (some 6.3\degr) far less than the angular diameter of its polar cap (12.6\degr) assuming a magnetic dipole field centered in the star.  This strongly argues that it is not centered, more like the B field of a sunspot.  One can nonetheless model the conal geometry (as we have done), but this is meaningless.  \citet{Arzoumanian96} made an attempt to model the geometry based on least-squares fits to the PPA traverse---which remarkably can be traced over most of the pulsar's rotation cycle.  They, however, run into some of the same problems we do. An interesting sidelight is that both KL99 and MM10 detect the pulsar at 103/111 MHz, and the former both resolve it better and fit three Gaussians to the profile---and these dimensions square well with 
those at higher frequencies, suggesting that (as for many MSPs) the profile changes little over a broad frequency range.
\vskip 0.1in

\noindent\textit{\textbf {B1726--00}}: The 1.4-GHz profile has a filled double form, whereas the 327-MHz one shows a probably three components in a manner consistent with the usual steeper spectrum of the core.  We thus model the geometry using a core-cone triple \textbf{$S_t$} configuration.  Estimating the core width at about 9\degr\ indicates an inner cone.  The PPA traverse is difficult to interpret with both positive and negative intervals, but use of either gives roughly comparable results.  No scattering time measurement has been reported. 
\vskip 0.1in

\noindent\textit{\textbf {B1802+03/J1805+0306}}: This pulsar has two bright components and seemingly a weak trailing one at 1.4 GHz.  Only the central component is seen at lower frequencies in GL98 and \citet{mcewen}.  We model it as an outer cone \textbf{$S_t$} but the PPA rate is poorly determined. A rough quadrature correction has been applied to \citet{malov2010}'s profile width.
\vskip 0.1in

\noindent\textit{\textbf{B1810+02}}: The profile of PSR B1810+02 appears to have a single core component in all three bands.  Our 1.4-GHz profile (as well as some of the GL98's) show bifurcation, but the dimensions are too small for a conal interpretation. It is possible that this is another rare example of a bifurcated core component.  \citet{malov2010}'s profile is difficult to interpret, but we take the full width and roughly correct it as for the previous pulsar.  

\begin{figure}
\begin{center}
\includegraphics[height=85mm,angle=-90.]{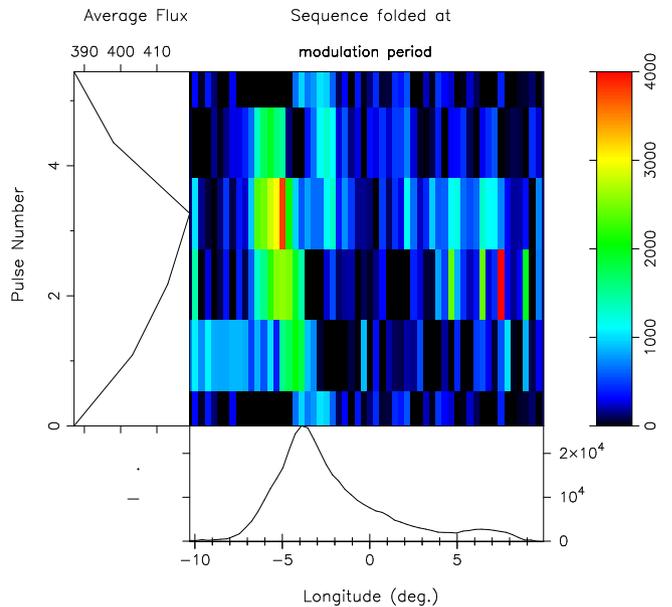}
\caption{Pulsar B1822+00 has a 5.45-$P$ modulation cycle that moves through its components as we see from the different phases at different longitudes.
The display shows the dynamic variations above a static ``base'', some 10\% of the overall amplitude.}
\label{figA102}
\end{center}
\end{figure}
\vskip 0.1in

\noindent\textit{\textbf{B1822+00}}: The profile of this pulsar exhibits a 5.5-$P$ amplitude periodicity that modulates parts of its profile at different phases as shown in Fig~\ref{figA102}.  Given this together with the several ``breaks'' in its profiles, we have modeled its profiles as having a conal quadruple c\textbf{Q} geometry.  
\vskip 0.1in

\noindent\textit{\textbf{B1831--00}}: A strong 2.1-$P$ fluctuation feature points to the emission being conal, and the lack of width growth from 1.4-GHz to 350 MHz \citep{mcewen} (at 408 MHz, GL98) suggests an inner cone---or maybe a narrowing cone as seen in a few conal profiles such as B0809+74.  The PPA traverse suggests a 90\degr\ flip, and when this is repaired we see a small positive slope.  This is the basis for the \textbf{S$_d$} model. 
\vskip 0.1in

\noindent\textit{\textbf{B1848+04}}: A very broad triple \textbf{T} profile is visible throughout the available observations of PSR B1848+04.  We use Boriakoff's 318-MHz profile \citet{boriakoff} for the needed 327 MHz width, and it clearly shows an interpulse that we were not able to see clearly in our observation.  The core width cannot be estimated accurately from any available profile, but a value around 30\degr\ is plausible.  Our small model $\alpha$ value and the apparent varying spacing of the interpulse suggest that this is a single-pole interpulsar.  Scattering at 102-MHz has little effect because of the very broad profile.    
\vskip 0.1in

\noindent\textit{\textbf{B1849+00}}: The 1.4-GHz profiles (JK18,W+04,GL98) show it to be highly scattered, and one can doubt whether the latter's 606-MHz is actually a detection.  The \citet{kkwj98} 4.9-GHz profile suggests a triple structure, but there is no polarimetry; however, JK18's leading PPA ramp suggests a flattened steep rate.  We thus model this latter profile as if it were at 1.4 GHz, and an inner cone/core model has the right dimensions quantitatively if the core width is about 7\degr\ as is quite plausible.  Gl98's 1.6-GHz profile may also show the triple structure, and correcting it for scattering, similar core and conal dimensions obtain.
\vskip 0.1in

\noindent\textit{\textbf{B1853+01}}: This pulsar seems to be primarily core with the possibility of some conal emission emerging at 1.4 GHz and above.  The 1.4-GHz profiles suggest a steep PPA traverse, and we model it as a core-cone triple beam structure. The 4.9-GHz profiles \citep{sgg+95,kkwj98} are broader, one with a hint of triple structure, but cannot be usefully measured.  \citet{malov2010} do not provide a profile, but their index implies that it would be scattered out at 111 MHz.  There is a suggestion of scattering even at 327 MHz.  
\vskip 0.1in

\begin{figure}
\begin{center}
\includegraphics[height=85mm,angle=-90.]{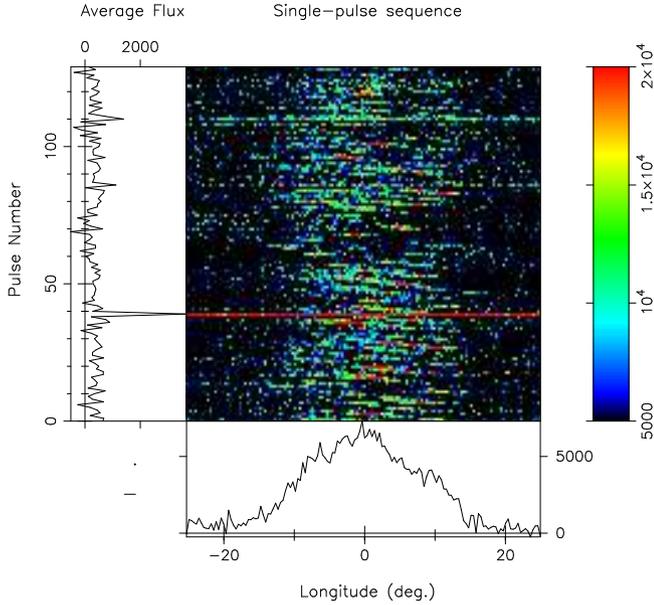}
\caption{Pulsar B1854+00 seems to be a irregular drifter---at least in this 
interval---but some subbands are clear. Strong RFI is seen at about pulse 48, and indeed no meaningful fluctuation spectra can be computed because of this occasional strong RFI in the observation.}
\label{figA103}
\end{center}
\end{figure}

\noindent\textit{\textbf{B1854+00/J1857+0057}}: Superficially, this pulsar seems to have a conal single \textbf{S$_d$} profile, and a fluctuation spectrum of our 1.4-GHz observations (not shown) seems to support this, as does the apparent drift modulation of its individual pulses (Fig~\ref{figA103}). \citet{Weisberg2004}'s 430-MHz profile provides a meter-wavelength width (as does \citet{mcewen}'s 350-MHz profile), and the 111-MHz profile of MM10 shows a double form. Thus, the pulsar seems to show the usual conal single evolution with frequency.  However, the structure of the \citet{Weisberg2004} profile (as well as the poor GL98 profiles---mislabeled B1953+00) is peculiar, but its width and PPA slope are compatible with the other profiles.  The meter-wavelength filled structure suggests a more complex beam encounter, and further study may show this to involve both cones, perhaps in a c\textbf{T} configuration.
\vskip 0.1in

\noindent\textit{\textbf{B1855+02}}. The \citet{GL98} 21-cm profile may have a trailing component that might suggest a triple configuration; however, their lower frequency profiles show progressive scattering tails.  Nor does our profile (see Fig.~\ref{figA30}) show a clear trailing feature, although there seems to be an inflection possibly associated with the $L$ peak.  Given the strong probability that the emission is core dominated, we tilt towards an aspirational triple beam model.  The \citet{kkwj98} 4.9-GHz profile shows two components, but the low S/N would obscure a weak trailing feature.    
\vskip 0.1in

\noindent\textit{\textbf{B1859+01}}: Both our 1.4-GHz and GL's 606-MHz profiles show a clear triple form.  We therefore model them with a core-cone triple \textbf{T} beam geometry.  The usual inner cone geometry seems to imply an unresolved very steep PPA traverse that we take here as infinite, although the W+04 1.4-GHz PPA rates seems more like --13\degr/\degr.  The lower frequency profile has very similar dimensions, supporting the inner cone model.  
\vskip 0.1in

\begin{figure}
\begin{center}
{\mbox{\includegraphics[width=75mm]{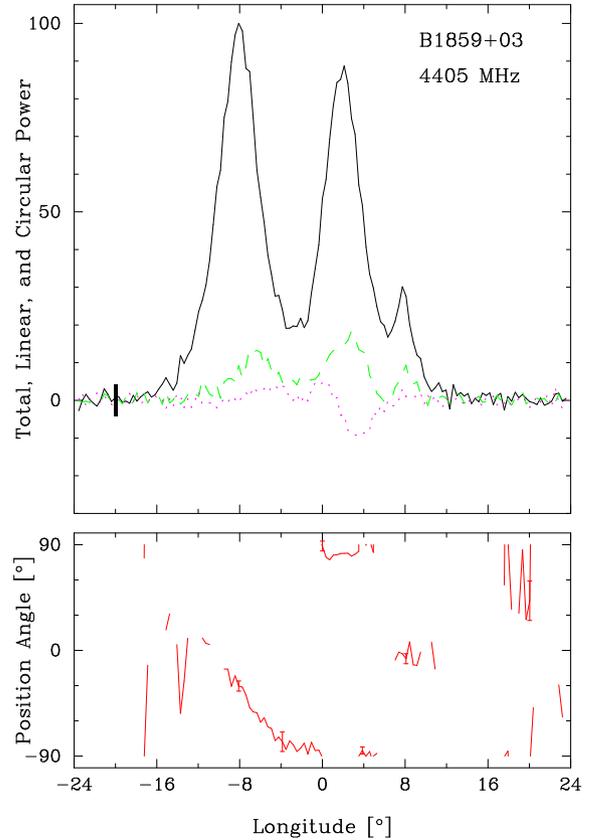}}}
\caption{Pulsar B1859+03 exhibits a clear triple profile at 4.4 GHz, unlike the situation at 1.4 GHz where the weak trailing conal outrider is conflated with the trailing part of the core component.}
\hspace{-0.55 cm}
\label{figA104a}
\end{center}
\end{figure}

\noindent\textit{\textbf{B1859+03}}:  We follow ET VI in modeling this pulsar as having a core-single \textbf{S$_t$} beam configuration.  At 1.4 GHz the trailing outrider can hardly be discerned, but at 4.4 GHz, the three components can be distinguished very clearly (see Fig.~\ref{figA104a}).  At lower frequencies scattering sets in and the profile structures become inscrutable.  
\vskip 0.1in

\noindent\textit{\textbf{B1859+07}}:  This pulsar is the most prolific ``swisher'', and these events must be accommodated before any beam modeling is appropriate \citep{bistable}.  We model it using the reconstructed profiles in the foregoing paper, and prefer a core single or core-cone triple model; however, a conal single model can be modeled with almost identical geometry.  We know little about the meter-wavelength profile as single-pulse sequences are needed to assess the effects of the ``swooshes''.  However, the GL98 profiles seems to have little width growth---thus we surmise a similar behavior at 327-MHz.
\vskip 0.1in

\onecolumn
\begin{table*}
\caption{Observation Information} 
\begin{tabular}{lcc|ccc|ccc|l}
    \hline
     & & & & \mbox{\textbf{L-band}} & & \multicolumn{3}{c|}{\textbf{(P-band or Refs.)}} & \mbox{\textbf{100 MHz}} \\
\hline
 Pulsar  & DM & RM  & MJD & $N_{pulses}$ & Bins & MJD & $N_{pulses}$ & Bins & References  \\
         & pc/$cm^{3}$ & rad/$m^{2}$ & & & & \multicolumn{3}{c|}{\textbf{GL98 and other refs.)}}  & (MHz) \\
\hline
\hline
B0045+33     & 39.9 &--82.3 & 52837   &  1085  & 1188   &   53377   &   1085  &  1188   & BKK+, 129; KL99\\
B0820+02    & 23.7 & 13  &  \multicolumn{3}{c|}{\textbf{Hankins \& Rankin (2010)}}  &   54781   &   1388  &  1017   & IMS89;\ XBT+;\ PHS+;\ KTSD,\ 50	\\
B0940+16     & 20.3 & 53  & 52854   &  1048  & 1024   &   53490   &    827  &  1175   & BKK+, 149 \\
B1534+12    & 11.62 & 10.6 & 55637 & 15835 & 256 & \multicolumn{3}{c|}{\textbf{\citet{GL98}}} & KL99, MM10     \\
B1726--00   & 41.1 & 20  & 56415   &  1554  & 1024   &   52930   &   6217  &   752   & MM10 \\
\\[-2pt]
B1802+03    & 80.9 & 38.9 & 56769   &  2743  & 1096   & \multicolumn{3}{c|}{\textbf{\citet{mcewen}}} & MM10 \\
B1810+02   & 104.1& --25 & 56406   &  1032  & 1024   &   53378   &   1032  &   775   & MM10 \\
B1822+00    & 62.2 & 158 & 56406   &  1052  & 1014   &   53378   &   1052  &   762   & MM10 \\
B1831-00    & 88.65 & --- & 52735  &  1151  &  256   &   53377   &   1151  &   256   &  \\                  
B1848+04     & 115.5&  86  & 56768   &  2102  &  512   & \multicolumn{3}{c|}{\textbf{\citet{boriakoff}}}  & MM10 \\
\\[-2pt]
B1849+00    & 787 & 341 & --- & --- & --- & \multicolumn{3}{c|}{\textbf{KKWJ, JK18}}    &        \\
B1853+01    & 96.7 &--140 & 57941   &  1041  & 1024   &   57982   &   1666  &   864   &  \\
B1854+00     & 82.4 & 104 & 52735   &  1679  & 1024   & \multicolumn{3}{c|}{\textbf{\citet{Weisberg2004}}} & MM10 \\
B1855+02    & 506.8 & 423 & 52739 & 1080 & 812 &   \multicolumn{3}{c|}{\textbf{KKWJ}}   &             \\
B1859+01    & 105.4 & --122 & 54842 & 2080 & 1125 & 53377 & 1068 & 522 &   \\
\\[-2pt]
B1859+03    & 402.1 & --237.4 & 56768 & 1021 & 1643 & \multicolumn{3}{c|}{\textbf{\citet{GL98}}}   &         \\
B1859+07    & 252.8 & 282 & 57121 & 12190 & 1257 &    \multicolumn{3}{c|}{\textbf{KKWJ}}   &             \\
B1900+05    & 177.5 & --113 & 54842 & 1045 & 933 &    \multicolumn{3}{c|}{\textbf{SGG+95}}   &     \\
B1900+06    & 502.9 & 552.6 & 55633 & 1037 & 1024 &    \multicolumn{3}{c|}{\textbf{SGG+95,KKWJ}}   &     \\
B1900+01    & 245.7&  72.3 & 57940 &  715 & 1024 & 57982 &  809 & 1024 & IMS89   \\
\\[-2pt]
B1901+10    & 135 & --98.1 & 56563 & 1029 & 1075 & 53777 & 538 &  1024 &   \\
B1902--01   & 229.1& 142 & 52735   &  1026  & 1024   &   53778   &   1555  &   628   & MM10 \\
B1903+07    & 245.3 & 272.7 & 57115 & 721 & 1029 \\
B1904+06    & 472.8 & 372 & 56768 & 2239 & 1044 \\
B1907+00     & 112.8& --40 & 54540   &  1033  &  993   &   53377   &   1057  &   786   & PHS+, 135 \\
\\[-2pt]
B1907+02     & 171.7& 254 & 57134   &  1025  & 1050   &   53377   &   1031  &   966   & MM10; PHS+, 135 \\
B1907+10    & 145.1& 150 & 54537   &  3987  & 1024   &   54631   &   2108  &   864   & MM10; PHS+ 149 \\
B1907+12    & 258.6& 978 & 52739    &  297  &    1407 & 52942 & 1619   &  1024 & MM10 \\
B1907+03    & 82.9 &--127 & 55637   &  1029  & 1024   & \multicolumn{3}{c|}{\textbf{\citet{hankins2010}}} & KL99 \\ 
B1911+09    & 157 & --- & 56769 & 1025 & 1212 & \multicolumn{3}{c|}{\textbf{\citet{w80}}} & \\
\\[-2pt]
B1911+13     & 145.1& 435 & 52735   &  1150  & 1020   &   53966   &   1147  &   613   & MM10 \\
B1911+11    & 100 & 361 & 52738 & 1015 & 1173 &   \multicolumn{3}{c|}{\textbf{\citet{GR78}}}   &   \\
B1913+10    & 242 & 430 & 54538 & 2077 & 1024  &   \multicolumn{3}{c|}{\textbf{\citet{GR78}}}    &     \\
B1913+16    & 168.8 & 357 & 56171 & 15241 & 661 &   \multicolumn{3}{c|}{\textbf{\citet{Blaskiewic}}} & \\
B1913+167   & 62.6 & 172 & 52947 & 1112 & 1052 & 55637 & 1028 & 1024 &     \\
\\[-2pt]
B1914+13   & 237. & 280 & 54842   &  2128  & 1100   &   55433   &   1662  &   550   & MM10 \\
B1915+22     & 134.9& 192 & 58288   &  2107  & 1035   &   58274   &   1391  &   1100  & BKK+, 149  \\
B1916+14    & 27.2 & -41.7 & 57942 & 1068 & 1024 & 56415 & 1031 & 1024 &     \\
B1917+00	 & 90.3	& 120 & 58654   &  2166  & 1156   &   58679   &   1642  &   1024  & PHS+; KTSD, 50  \\
B1918+26    & 27.6 & 20.8 & 56419   &  1031  & 1024   &   53967   &   1044  &   1208  & BKK+, 129; PHS+; MM10\\
\\[-3pt]
B1918+19    & 153.9	& 160 & 53778 & 3946 & 922 & 54541 & 731 & 1001 & IMS89 \\
B1919+14     & 91.6 & 165 & 56768   &  1051  & 1212   & \multicolumn{3}{c|}{\textbf{\citet{hankins2010}}} & MM10 \\
B1919+20    & 101 & 128 & 56419 & 1027 & 1024 & \multicolumn{3}{c|}{\textbf{\citet{Weisberg2004}}} & \\
B1920+20    & 203.3	& 301 & \multicolumn{3}{c|}{\textbf{GL98}}  &\multicolumn{3}{c|}{\textbf{\citet{Weisberg2004}}} & \\
B1920+21	 & 217.1& 282 & 52737   &  505   & 1052   &   52941   &   2220 &  1024  & PHS+, 143; KL99 \\
\\[-2pt]
B1921+17    & 142.5 & 380 & 56768 & 1091 & 1068 & \multicolumn{3}{c|}{\textbf{\citet{Weisberg2004}}} & \\
B1923+04	 & 102.2 & -39.5 & \multicolumn{3}{c|}{\textbf{\citet{W99}}} & 53377 & 837 & 1024 & PHS+, 135\\
B1924+14    & 211.4& 249 & 57004   &   508  & 1024   &   52949   &   845   &    646  & MM10 \\
B1924+16    & 176.9 & 320 & 55982   & 1029  &  1004  &   57942   &  1115 &   1024  &           \\
B1925+18    & 254 & 417 & 56769 & 1237 & 1024 & \multicolumn{3}{c|}{\textbf{\citet{Weisberg2004}}} & \\
\\[-3pt]
B1925+188   & 99 & 74.4 & 56769 & 2006 & 256 \\
B1925+22    & 180 &	215.7 & 52949 & 880 & 698 & 57126 & 1022 & 1029 &       \\
B1926+18    &  112 & 174  & \multicolumn{3}{c|}{\textbf{\citet{W99}}} & 53967 & 1023 & 1024 & \\
B1927+13	 & 207.3 & -2.3 & \multicolumn{3}{c|}{\textbf{\citet{W99}}} & \multicolumn{3}{c|}{\textbf{\citet{Weisberg2004}}} \\
B1929+20	 & 211.1 & 10  & \multicolumn{3}{c|}{\textbf{\citet{JK18}}} & \multicolumn{3}{c|}{\textbf{\citet{WSE07}}} &  \\
\end{tabular}
\label{tabA1}
\end{table*}
\begin{table*}
\noindent\textbf{Table~\ref{tabA1}.} Observation Information (cont'd) 
\begin{center}
\begin{tabular}{lcc|ccc|ccc|l}
    \hline
     & & & & \mbox{\textbf{L-band}} & & \multicolumn{3}{c|}{\textbf{(P-band and/or Refs.)}} & \mbox{\textbf{100 MHz}} \\
\hline
 Pulsar  & DM & RM  & MJD & $N_{pulses}$ & Bins & MJD & $N_{pulses}$ & Bins & References  \\
         & pc/$cm^{3}$ & rad/$m^{2}$ & & & & \multicolumn{3}{c|}{\textbf{GL98 and other refs.)}}  & (MHz) \\
\hline
\hline
B1930+22     & 219.2& 173 & 54540   &  4151  &  564   & \multicolumn{3}{c|}{\textbf{\citet{ETIX}}} & MM10 \\
B1930+13    & 177.9& --120 & 58278   &  1071  &  1031  &   58274   &  1068   &  1042  &  BKK+, 149 \\
B1931+24    & 106.03 & 114.8 & 58679 & 2449 & 1051 & 55982 & 1027 & 1005 &      \\ 
B1942--00    & 59.7 & --45  & 55982   &   815  &  1021  &   52930   &  1146   &  1024  & MM10 \\
B1944+22    & 140 & 2 & 55276 & 932 & 1026 & 57112 & 1029 & 512 &      \\
\\
B1949+14     & 31.5 & --21  & 57639   &  3634  &  1024  &   57878   &  4308   &  1015  & BKK+, 149 \\
B1951+32    & 45.0 & --182 & 55632   & 15171  &   256  & \multicolumn{3}{c|}{\textbf{\citet{Weisberg2004}}} & MM10 \\
B1953+29    & 104.5 & 22.4 & 56585 & 391270 & 162 \\  
B2000+32    & 122.2&--90.2 & 57110   &  1028  &  1024  &   54632   &  2067   &  1024  & MM10 \\
B2002+31    & 234.8 & 31.5 & 56444 & 1297 & 2020 & 56445 & 1245 & 1024 &     \\
\\
B2027+37   & 190.7&  --6  & 57115   &  1030  &  1194  &   57690   &   714   &  1024  & MM10 \\
B2025+21    & 96.8 &--212 & 56585   &  2505  &   512  &   55433   &  1061   &   777  &  BKK+, 149; MM10 \\
B2028+22	 & 71.8	& --192  & 54540   &  1027  &  1231  &   52947   &  1581   &  1024  & BKK+, 149; MM10 \\
B2034+19     &  36. & --97 & 52735   &  1341  &  1187  &   56353   &  2048   &  1014  & BKK+, 149\\
B2035+36     & 93.6 &  252  & 57109   &  1029  &  1208  &   52931   &  2808   &  1208  & MM10 \\
\\
B2044+15     & 39.8 &--100  & 55633   &  1576  &  1024  &   55639   &  2058   &  1024  & BKK+, 129: PHS+, 135; KL99  \\
B2053+21    & 36.4 & --100 & 56586   &  1976  &  1024  &   53378   &  2944   &  1000  & BKK+; MM10; KTSD, 50  \\
B2053+36    & 97.3 & --68  & 57110   &  2703  &  1105  &   57981   &  3433   &   715  & BKK+ 129; MM10  \\
B2113+14    & 56.2 &  --25 & 57115   &  1358  &  1023  &   52920   &  2996   &   429  & KL99; BKK+, 129; PHS+, 143 \\
B2122+13     & 30.1 &  --48 & 57115   &  1032  &  1035  &   52931   &  1583   &  1024  & PHS+, 135; BKK+, 149; MM10 \\
\\
B2210+29     & 74.5 & --168 & 52837   &  1111  &   980  &   52931   &   981   &  1024  & KL99/MM10; BKK+, 129 \\
\hline
\end{tabular}
\end{center}
\vskip 0.05in
\textbf{Notes}: BKK+: \cite{bilous2016}; GL98: \cite{GL98}; JK18: \cite{JK18}; KKWJ: \cite{kkwj98}; KTSD: \citet{KTSD23}; KL99: \cite{kuzmin1999}; MM10: \cite{malov2010}; PHS+: \cite{pilia2016}; SGG+95: \cite{sgg+95}; XBT+: \citet{Xue+17}.  Values from the ATNF Pulsar Catalog \citep{ATNF}.

\textbf{``B'' Pulsars with Inadequate Observational Information}: Pulsars B1922+20, B1929+15, B1933+15, B1933+17, B1937+24 and B1939+17 were observed polarimetrically by \citet{GL98}, \citet{W99} and/or \citet{Weisberg2004} but without any usable PPA traverse information.  Published profiles do not seem to be available for pulsars B1852+10, B1904+12, B1913+105, B1924+19, B1943+18 or B2127+11A. 
\end{table*}

\twocolumn

\begin{table}
\setlength{\tabcolsep}{2pt}
\caption{Pulsar Parameters}
\begin{center}
\begin{tabular}{lccccccc}
\hline
 Pulsar &  P & $\dot{P}$ & $\dot{E}$ & $\tau$ & $B_{surf}$ & $B_{12}/P^2$ & 1/Q  \\
 (B1950) & (s) & ($10^{-15}$ & ($10^{32}$  & (Myr) & ($10^{12}$ &   &   \\
 & & s/s) & ergs/s) & & G) &   &    \\
\hline
\hline
B0045+33 & 1.2171 & 2.35 & 0.52 & 8.19 & 1.71 & 1.2 & 0.6 \\
B0820+02 & 0.8649 & 0.10 & 0.06 & 131 & 0.30 & 0.4 & 0.2 \\
B0940+16 & 1.0874 & 0.09 & 0.03 & 189 & 0.32 & 0.3 & 0.2 \\
B1534+12 & 0.0379 & 0.00 & 18.0 & 248 & 0.01 & 6.8 & 1.6 \\
B1726--00 & 0.3860 & 1.12 & 7.70 & 5.45 & 0.67 & 4.5 & 1.5 \\
\\[-3pt]
B1802+03 & 0.2187 & 1.00 & 38.0 & 3.47 & 0.47 & 9.9 & 2.7 \\
B1810+02 & 0.7939 & 3.60 & 2.80 & 3.49 & 1.71 & 2.7 & 1.1 \\
B1822+00 & 1.3628 & 1.75 & 0.27 & 12.40 & 1.56 & 0.8 & 0.4 \\
B1831--00 & 0.5210 & 0.01 & 0.03 & 784 & 0.07 & 0.3 & 0.2	\\
B1848+04 & 0.2847 & 1.09 & 19.0 & 4.14 & 0.56 & 6.9 & 2.1 \\
\\[-3pt]
B1849+00 & 2.1802 & 96.95 & 3.70 & 0.4 & 14.70 & 3.1 & 1.3	\\
B1853+01 & 0.2674 & 208.4 & 4300 & 0.02 & 7.55 & 106 & 18.1 \\
B1854+00 & 0.3569 & 0.05 & 0.47 & 104 & 0.14 & 1.1 & 0.5 \\
B1855+02 & 0.4158 & 40.27 & 220 & 0.2 & 4.14 & 23.9 & 5.8	\\
B1859+01 & 0.2882 & 2.36 & 39.0 & 1.9 & 0.83 & 10.0 & 2.8	\\
\\[-3pt]
B1859+03 & 0.6555 & 7.46 & 10.0 & 1.4 & 2.24 & 5.2 & 1.8	\\
B1859+07 & 0.6440 & 2.29 & 3.40 & 4.5 & 1.23 & 3.0 & 1.1	\\
B1900+05 & 0.7466 & 12.88 & 12.0 & 0.9 & 3.14 & 5.6 & 1.9	\\
B1900+06 & 0.6735 & 7.71 & 10.0 & 1.4 & 2.31 & 5.1 & 1.7	\\
B1900+01 & 0.7293 & 4.03 & 4.10 & 2.9 & 1.73 & 3.3 & 1.2 \\
\\[-3pt]
B1901+10 & 1.8566 & 0.28 & 0.02 & 107 & 0.72 & 0.2 & 0.2	\\
B1902--01 & 0.6432 & 3.06 & 4.50 & 3.3 & 1.42 & 3.4 & 1.3 \\
B1903+07 & 0.6480 & 4.94 & 7.20 & 2.1 & 1.81 & 4.3 & 1.5	\\
B1904+06 & 0.2673 & 2.14 & 44.0 & 2.0 & 0.77 & 10.7 & 2.9	\\
B1906+09 & 0.8303 & 0.01 & 0.02 & 134 & 0.29 & 0.4 & 0.2 \\
\\[-3pt]
B1907+00 & 1.0169 & 5.52 & 2.10 & 2.9 & 2.40 & 2.3 & 1.0 \\
B1907+02 & 0.9898 & 5.53 & 2.20 & 2.8 & 2.37 & 2.4 & 1.0 \\
B1907+10 & 0.2836 & 2.64 & 46.0 & 1.7 & 0.88 & 10.9 & 2.9 \\
B1907+03 & 2.3303 & 4.47 & 0.14 & 8.3 & 3.27 & 0.6 & 0.4 \\
B1907+12 & 1.4417 & 8.23 & 1.10 & 2.8 & 3.49 & 1.7 & 0.8 \\
\\[-3pt]
B1911+09 & 1.2420 & 0.43 & 0.09 & 45.6 & 0.74 & 0.5 & 0.3	\\
B1911+13 & 0.5215 & 0.80 & 2.20 & 10.3 & 0.66 & 2.4 & 0.9 \\
B1911+11 & 0.6010 & 0.66 & 1.20 & 14.5 & 0.64 & 1.8 & 0.7	\\
B1913+10 & 0.4046 & 15.26 & 91.0 & 0.4 & .51 & 15.3 & 4.0	\\
B1913+16 & 0.0590 & 0.01 & 17.0 & 109 & 0.02 & 6.5 & 1.7	\\
\\[-3pt]
B1913+167 & 1.6162 & 0.41 & 0.04 & 63.2 & 0.82 & 0.3 & 0.2 \\
B1914+13 & 0.2818 & 3.65 & 64.0 & 1.2 & 1.03 & 13.0 & 3.4 \\
B1915+22 & 0.4259 & 2.86 & 14.64 & 2.4 & 1.12 & 6.2 & 1.9 \\
B1916+14 & 1.1810 & 212.36 & 51.0 & 0.1 & 16.00 & 11.5 & 3.6 \\
B1917+00 & 1.2723 & 7.67 & 1.50 & 2.6 & 3.16 & 2.0 & 0.9 \\
\\[-3pt]
B1918+26 & 0.7855 & 0.03 & 0.03 & 362 & 0.17 & 0.3 & 0.2 \\
B1918+19 & 0.8210 & 0.90 & 0.64 & 14.5 & 0.87 & 1.3 & 0.6 \\
B1919+14 & 0.6182 & 5.60 & 9.40 & 1.8 & 1.88 & 4.9 & 1.7 \\
B1919+20 & 0.7607 & 0.05 & 0.04 & 241 & 0.20 & 0.34 & 0.20 \\
B1920+20 & 1.1728 &	0.65 & 0.06 & 28.6 & 0.88 & 0.6 & 0.4 \\ 
\\[-3pt]
B1920+21 & 1.0779 & 8.18 & 2.60 & 2.1 & 3.00 & 2.6 & 1.1 \\
B1921+17 & 0.5472 & 0.04 & 0.10 & 202 & 0.16 & 0.5 & 0.3	\\
B1924+14 & 1.3249 & 0.22 & 0.04 & 95.6 & 0.55 & 0.3 & 0.2 \\
B1924+16 & 0.5798 & 0.18 & 36.4 & 0.51 & 3.27 & 9.7 & 2.9 \\
B1925+18 & 0.4828 & 0.12 & 0.41 & 65.9 & 0.24 & 1.0 & 0.5	\\
\\[-3pt]
B1925+188 & 0.2983 &2.24 & 33.0 & 2.1 & 0.83 & 9.3 & 2.6	\\
B1925+22 & 1.4311 & 0.77 & 0.10 & 29.4 & 1.06 & 0.5 & 0.3 \\
B1926+18 & 1.2205 & 2.36 & 0.51 & 8.2 & 1.72 & 1.15 & 0.57 \\
B1927+13 & 0.7600 & 3.66 & 3.30 & 3.3 & 1.69 & 2.9 & 1.1 \\
B1929+20 & 0.2682 & 4.22 & 86.0 & 1.0 & 1.08 & 15.0 & 3.8 \\
\end{tabular}
\end{center}
\label{tabA2}
\end{table}
\begin{table}
\setlength{\tabcolsep}{2pt}
\textbf{Table~\ref{tabA2}.} Pulsar Parameters (cont'd)
\begin{center}
\begin{tabular}{lccccccc}
\hline
 Pulsar &  P & $\dot{P}$ & $\dot{E}$ & $\tau$ & $B_{surf}$ & $B_{12}/P^2$ & 1/Q  \\
 (B1950) & (s) & ($10^{-15}$ & ($10^{32}$  & (Myr) & ($10^{12}$ &   &   \\
 & & s/s) & ergs/s) & & G) &   &    \\
\hline
\hline
B1930+22 & 0.1445 & 57.57 & 7500 & 0.0 & 2.92 & 140 & 21.2 \\
B1930+13 & 0.9283 & 0.32 & 0.16 & 46.2 & 0.55 & 0.6 & 0.3 \\
B1931+24 & 0.8137 & 8.11 & 5.90 & 1.6 & 2.60 & 3.9 & 1.4 \\
B1942--00 & 1.0456 & 0.53 & 0.18 & 31.0 & 0.76 & 0.7 & 0.4 \\
B1944+22 & 1.3344 & 0.89 & 0.15 & 23.8 & 1.10 & 0.6 & 0.3 \\
\\
B1949+14 & 0.2750 & 0.13 & 2.43 & 34.0 & 0.19 & 2.5 & 0.9 \\
B1951+32 & 0.0395 & 5.84 & 37000 & 0.1 & 0.49 & 311 & 35.4 \\
B1953+29 & 0.0061 & 0.00 & 51.0 & 3270 & 0.00 & 11.5 & 2.1	\\
B2000+32 & 0.6968 & 105.14 & 120.0 & 0.1 & 8.66 & 17.8 & 4.8 \\
B2002+31 & 2.1113 & 74.55 & 3.10 & 0.4 & 12.70 & 2.8 & 1.2 \\
\\
B2025+21 & 0.3982 & 0.20 & 1.27 & 31.1 & 0.29 & 1.8 & 0.7 \\
B2027+37 & 1.2168 & 12.32 & 2.70 & 1.6 & 3.92 & 2.6 & 1.1 \\
B2028+22 & 0.6305 & 0.89 & 1.39 & 11.3 & 0.76 & 1.9 & 0.8 \\
B2034+19 & 2.0744 & 2.04 & 0.09 & 16.1 & 2.08 & 0.5 & 0.3 \\
B2035+36 & 0.6187 & 4.50 & 7.50 & 2.2 & 1.69 & 4.4 & 1.5 \\
\\
B2044+15 & 1.1383 & 0.18 & 0.05 & 98.9 & 0.46 & 0.4 & 0.2 \\
B2053+21 & 0.8152 & 1.34 & 0.98 & 9.6 & 1.06 & 1.6 & 0.7 \\
B2053+36 & 0.2215 & 0.37 & 13.0 & 9.5 & 0.29 & 5.9 & 1.8 \\
B2113+14 & 0.4402 & 0.29 & 1.34 & 24.1 & 0.36 & 1.9 & 0.8 \\
B2122+13 & 0.6941 & 0.77 & 0.91 & 14.3 & 0.74 & 1.5 & 0.7 \\
\\
B2210+29 & 1.0046 & 0.50 & 0.19 & 32.1 & 0.71 & 0.7 & 0.4 \\
\hline  
\end{tabular}
\end{center}
Notes: Values from the ATNF Pulsar Catalog \citep{ATNF}, \textbf{Version 1.67}.
\end{table}

\onecolumn
\begin{table*}
\caption{Emission-Beam Model Geometry}
\setlength{\tabcolsep}{3pt} 
\begin{tabular}{lc|ccc|ccccc|ccccc|ccccc}
    \toprule
    Pulsar &  Class & $\alpha$ & $R$ & $\beta$ & $W_c $ & $W_i$ & $\rho_i$ & $W_o$  & $\rho_o$ & $W_c $ & $W_i$ & $\rho_i$ & $W_o$  & $\rho_o$ & $W_c $ & $W_i$ & $\rho_i$ & $W_o$  & $\rho_o$  \\
          &   & (\degr) & (\degr/\degr) & (\degr) & (\degr) & (\degr) & (\degr) & (\degr) & (\degr) & (\degr) & (\degr) & (\degr) & (\degr) & (\degr) & (\degr) & (\degr) & (\degr) & (\degr) & (\degr) \\
    \midrule
    & & \multicolumn{3}{c|}{(1-GHz  Geometry)} & \multicolumn{5}{c|}{(1.4-GHz Beam Sizes)} & \multicolumn{5}{c|}{(327-MHz Beam Sizes)} & \multicolumn{5}{c}{(100-MHz Band Beam Sizes)} \\
    \midrule
    \midrule
B0045+33 & D & 46 & -15 & +2.7 &  --- & 7.8 & 4.0 &  --- &  --- &  --- & 6.0 & 3.5 &  --- &  --- &  --- & 5.9 & 3.5 &  --- &  --- \\
B0820+02 & Sd & 71 & +12 & +4.5 &  --- &  --- &  --- & 9.0 & 6.2 &  --- &  --- &  --- & 9.5 & 6.4 &  --- &  --- &  --- & $\sim$19 & 10.2 \\
B0940+16 & Sd/PC & 25 & +6 & +4.0 &  --- &  --- &  --- & 16.0 & 5.4 &  --- &  --- &  --- & 24.9 & 6.9 &  --- &  --- &  --- & 28.5 & 7.6 \\
B1534+12 & ?? & 60 & -8 & +6.2 & 6.3 & $\sim$48 & 22.2 &  --- &  --- &  --- &  --- &  --- &  --- &  --- & 8 & 55 & $\sim$25.2 &  --- &  --- \\
B1726-00 & T? & {\bf 26} & $\sim$5 & +5.0 & $\sim$9 & 20.3 & 7.0 &  --- &  --- &  --- & 19.8 & 6.9 &  --- &  --- &  --- & $\sim$28 & 8.3 &  --- &  --- \\
\\[-4pt]
B1802+03 & St & {\bf 44} & +4.2 & +9.6 & 7.5 &  --- &  --- & 20.0 & 12.2 & $\sim$7 &  --- &  --- &  --- &  --- & $\sim$14 &  --- &  --- &  --- &  --- \\
B1810+02 & St & {\bf 25} & +36 & +0.7 & 6.7 &  --- &  --- &  --- &  --- & 6.5 &  --- &  --- &  --- &  --- & $\sim$13 &  --- &  --- &  --- &  --- \\
B1822+00 & cT? & 27 & -7.5 & +3.5 &  --- & $\sim$6 & 3.8 & 13.7 & 5.0 &  --- & $\sim$7 & 3.9 & $\sim$15 & 5.0 &  --- &  --- &  --- & $\sim$27 & 7.3 \\
B1831-00 & Sd & 12 & +2.4 & +5.0 &  --- & $\sim$25 & 5.8 &  --- &  --- &  --- & $\sim$20 & 5.5 &  --- &  --- &  --- &  --- &  --- &  --- &  --- \\
B1848+04 & T & {\bf 8} & +2.4 & +3.4 & $\sim$32 & 88 & 8.1 &  --- &  --- & $\sim$32 & $\sim$99 & 8.9 &  --- &  --- &  --- & $\sim$101 & 9.0 &  --- &  --- \\
\\[-4pt]
B1849+00 & T & 14.1 & $\infty$ & 0.0 & 6.8 & 24.0 & 2.9 &  --- &  --- &  --- &  --- &  --- &  --- &  --- &  --- &  --- &  --- &  --- &  --- \\
B1853+01 & St & {\bf 75} &  --- &  --- & 6.1 &  --- &  --- &  --- &  --- & 4.9 &  --- &  --- &  --- &  --- & $\sim$36 &  --- &  --- &  --- &  --- \\
B1854+00 & Sd/cT? & 41 & +6 & +6.3 &  --- &  --- &  --- & 21.2 & 9.7 &  --- &  --- &  --- & 26.7 & 11.2 &  --- &  --- &  --- & $\sim$42 & 15.8 \\
B1855+02 & St & {\bf 27} & -5 & -5.1 & 8.5 & 22.0 & 6.8 &  --- &  --- &  --- &  --- &  --- &  --- &  --- &  --- &  --- &  --- &  --- &  --- \\
B1859+01 & T & {\bf 66} & $\infty$ & 0 & $\sim$5 & $\sim$18 & 8.2 &  --- &  --- & 4.8 &  --- &  --- &  --- &  --- &  --- &  --- &  --- &  --- &  --- \\
\\[-4pt]
B1859+03 & St & {\bf 35} & -10 & -3.3 & 5.3 & $\sim$16 & 5.5 &  --- &  --- &  --- &  --- &  --- &  --- &  --- &  --- &  --- &  --- &  --- &  --- \\
B1859+07 & St/T? & {\bf 31} & +6 & +4.9 & $\sim$6 &  --- &  --- & 18.9 & 7.1 & $\sim$6 &  --- &  --- & 18.9 & 7.1 &  --- &  --- &  --- &  --- &  --- \\
B1900+05 & St & {\bf 28} & $\infty$ & 0 & 6 & $\sim$20 & 4.7 &  --- &  --- &  --- &  --- &  --- &  --- &  --- &  --- &  --- &  --- &  --- &  --- \\
B1900+06 & St? & {\bf 84} & +15 & 3.8 & $\sim$3 & $\sim$7 & 5.2 &  --- &  --- &  --- &  --- &  --- &  --- &  --- &  --- &  --- &  --- &  --- &  --- \\
B1900+01 & T/cT & {\bf 60} & +45 & +1.1 & 3.3 & 11.3 & 5.1 &  --- &  --- &  --- &  --- &  --- &  --- &  --- & $\sim$14 &  --- &  --- &  --- &  --- \\
\\[-4pt]
B1901+10 & D & 15 & -8 & -1.9 &  --- &  --- &  --- & 31.3 & 4.2 &  --- &  --- &  --- & 29.0 & 4.0 &  --- &  --- &  --- &  --- &  --- \\
B1902-01 & St & {\bf 80} & $\infty$ & 0.0 & 3.1 & $\sim$11 & 5.4 &  --- &  --- & $\sim$7 &  --- &  --- &  --- &  --- & $\sim$35 &  --- &  --- &  --- &  --- \\
B1903+07 & Sd? & 19 & +4.8 & +3.9 &  --- & 21.1 & 5.4 &  --- &  --- &  --- &  --- &  --- &  --- &  --- &  --- &  --- &  --- &  --- &  --- \\
B1904+06 & T & {\bf 32} & +4 & +7.6 & $\sim$9 & 28.0 &  --- & 28.0 & 11.1 &  --- &  --- &  --- & 29.0 & 11.3 &  --- &  --- &  --- &  --- &  --- \\
B1906+09 & D? &  --- &  --- &  --- &  --- &  --- &  --- &  --- &  --- &  --- &  --- &  --- &  --- &  --- &  --- &  --- &  --- &  --- &  --- \\
\\[-4pt]
B1907+00 & T & {\bf 69} & $\infty$ & 0 & 2.2 &  --- &  --- & 12.0 & 5.6 & 2.6 &  --- &  --- & 13.3 & 6.2 & $\sim$16 &  --- &  --- &  --- &  --- \\
B1907+02 & T & {\bf 48} & $\infty$ & 0 & 3.3 & 11.9 & 4.4 &  --- &  --- & 3.7 & $\sim$14 & 5.2 &  --- &  --- & $\sim$26 &  --- &  --- &  --- &  --- \\
B1907+10 & St/D? & {\bf 62} & +8 & +6.7 & 5.2 & $\sim$11 & 8.3 &  --- &  --- & 5.6 & $\sim$22 & 12.0 &  --- &  --- & $\sim$50 &  --- &  --- &  --- &  --- \\
B1907+03 & cQ/cT & 6 & -4 & +1.4 &  --- &  --- &  --- & 60.8 & 3.7 &  --- &  --- &  --- & 66.8 & 4.0 &  --- &  --- &  --- & $\sim$77 & 4.5 \\
B1907+12 & Sd/St? & {\bf 36} & -15 & +2.2 & 3.5 & $\sim$10 & 3.7 &  --- &  --- &  --- & $\sim$10 & 3.7 &  --- &  --- &  --- & $\sim$50 & 9.1 &  --- &  --- \\
\\[-4pt]
B1911+09 & Sd & 34 & +7.5 & +4.3 & 0 &  --- &  --- & 9.8 & 5.2 & 0 &  --- &  --- & 13.0 &  --- &  --- &  --- &  --- &  --- &  --- \\
B1911+13 & T & {\bf 62} & +11 & +4.6 & 3.7 &  --- &  --- & 14.9 & 8.1 & 4.0 &  --- &  --- & 18.1 & 9.3 &  --- &  --- &  --- & $\sim$42 & 19.4 \\
B1911+11 & D & 50 & -12 & 3.7 &  --- & 11.0 & 5.7 &  --- &  --- &  --- &  --- &  --- &  --- &  --- &  --- &  --- &  --- &  --- &  --- \\
B1913+10 & St? & {\bf 64} &  --- &  --- & 4.3 &  --- &  --- &  --- &  --- &  --- &  --- &  --- &  --- &  --- &  --- &  --- &  --- &  --- &  --- \\
B1913+16 & St & {\bf 46} & +10 & +4.0 & 14 & 47 & 17.9 &  --- &  --- & 17 & 53.6 & 20.3 &  --- &  --- &  --- &  --- &  --- &  --- &  --- \\
\\[-4pt]
B1913+167 & cT & {\bf 46} & -30 & +1.4 & $\sim$3 & 8.5 & 3.4 &  --- &  --- & $\sim$3 & $\sim$11 & 4.2 &  --- &  --- &  --- & 0.0 & 1.4 &  &  \\
B1914+13 & St & {\bf 67} & +8 & +6.6 & $\sim$5 & 9.6 & 8.0 &  --- &  --- & 11.9 &  --- &  --- &  --- &  --- & $\sim$80 &  --- &  --- &  --- &  --- \\
B1915+22 & ?? & 30 & -5 & +5.7 &  --- &  --- &  --- & 25.0 & 8.9 &  --- &  --- &  --- & 25.0 & 8.9 &  --- &  --- &  --- &  --- &  --- \\
B1916+14 & T & {\bf 79} & +36 & +1.6 & 2.3 & 7.7 & 4.1 &  --- &  --- & 2.5 & 8.1 & 4.3 &  --- &  --- & $\sim$3 & 8.1 & 4.3 &  --- &  --- \\
B1917+00 & T & {\bf 81} & -45 & +1.3 & $\approx$2? &  --- &  --- & 10.0 & 5.1 &  --- &  --- &  --- & 10 & 5.1 &  --- &  --- &  --- & 16.0 & 8.0 \\
\\[-4pt]
B1918+26 & T/M & {\bf 54} & -11 & +4.2 & 3.4 &  --- &  --- & 11.7 & 6.5 & 2.70 &  --- &  --- & 13.6 & 7.1 & 2.7 &  --- &  --- & 15.7 & 7.8 \\
B1918+19 & cQ & 13.7 & -3.2 & -4.2 &  --- & $\sim$22 & 4.8 & 49.0 & 6.4 &  --- & $\sim$22 & 4.8 & $\sim$59 & 7.2 &  --- &  --- &  --- & $\sim$12 & 12.0 \\
B1919+14 & Sd & 21 & -4.3 & +4.8 &  --- & 14.9 & 5.6 &  --- &  --- &  --- & 14.9 & 5.6 &  --- &  --- &  --- & $\sim$20 & 6.2 &  --- &  --- \\
B1919+20 & D & 44 & +15 & +2.7 &  --- &  --- &  --- & 17.1 & 6.6 &  --- &  --- &  --- & 19.3 & 7.4 &  --- &  --- &  --- &  --- &  --- \\
B1920+20 & cQ? & 35 & +20 & +1.6 &  --- &  --- &  --- & 17.0 & 5.2 &  --- & $\sim$12 & 3.9 & $\sim$19 & 5.8 &  --- &  --- &  --- &  --- &  --- \\
\\[-4pt]
B1920+21 & T & {\bf 44} & -36 & +1.1 &  --- &  --- &  --- & 16.0 & 5.7 & 3.4 &  --- &  --- & 17.5 & 6.2 & $\sim$19 &  --- &  --- &  --- &  --- \\
B1921+17 & D/T? & {\bf 60} & $\sim$90 & +0.6 & $\sim$4 & 13.6 & 5.9 &  --- &  --- &  --- & 16.4 & 7.1 &  --- &  --- &  --- &  --- &  --- &  --- &  --- \\
B1924+14 & D & 15.5 & +20 & +0.8 &  --- &  --- &  --- & 36.3 & 5.0 &  --- &  --- &  --- & 49.1 & 6.7 &  --- &  --- &  --- &  --- &  --- \\
B1924+16 & St & {\bf 34} & +5.2 & +5.0 & 5.7 &  --- &  --- &  --- &  --- &  --- &  --- &  --- &  --- &  --- &  --- &  --- &  --- &  --- &  --- \\
B1925+18 & D & 32 & -4.5 & 6.8 &  --- &  --- &  --- & 16.8 & 8.3 &  --- &  --- &  --- & 0.0 &  --- &  --- &  --- &  --- &  --- &  --- \\
\\ 
B1925+188 & T & {\bf 19} & $\infty$ & 0 & $\sim$14 & 49 & 7.8 &  --- &  --- &  --- &  --- &  --- &  --- &  --- &  --- &  --- &  --- &  --- &  --- \\
B1925+22 & cT & 27 & $\sim$-8? & +3.3 &  --- & $\sim$6 & 3.6 & 14.5 & 4.8 &  --- & 6.0 & 3.6 & 21.0 & 6.1 &  --- &  --- &  --- & $\sim$21 & 4.8 \\
B1926+18 & cT & 29 & +7.5 & +3.7 &  --- & $\sim$5 & 3.9 & 15.0 & 5.3 &  --- & 6.0 & 4.0 & 0.0 &  --- &  --- &  --- &  --- &  --- &  --- \\
B1927+13 & T & {\bf 90} & $\infty$ & 0 & $\sim$2.8 & $\sim$10 & 5.0 &  --- &  --- & $\sim$2.8 & $\sim$11 & 5.5 &  --- &  --- &  --- &  --- &  --- &  --- &  --- \\
B1929+20 & T? & {\bf 90} & -9 & +6.4 & $\approx$5? & $\sim$10 & 8.1 &  --- &  --- &  --- &  --- &  --- &  --- &  --- &  --- &  --- &  --- &  --- &  --- \\
\end{tabular}
\label{tabA3}
\end{table*}
\begin{table*}
\textbf{Table~\ref{tabA3}.} Emission-Beam Model Geometry (cont'd)
\setlength{\tabcolsep}{3pt} 
\begin{tabular}{lc|ccc|ccccc|ccccc|ccccc}
    \toprule
    Pulsar &  Class & $\alpha$ & $R$ & $\beta$ & $W_c $ & $W_i$ & $\rho_i$ & $W_o$  & $\rho_o$ & $W_c $ & $W_i$ & $\rho_i$ & $W_o$  & $\rho_o$ & $W_c $ & $W_i$ & $\rho_i$ & $W_o$  & $\rho_o$  \\
          &   & (\degr) & (\degr/\degr) & (\degr) & (\degr) & (\degr) & (\degr) & (\degr) & (\degr) & (\degr) & (\degr) & (\degr) & (\degr) & (\degr) & (\degr) & (\degr) & (\degr) & (\degr) & (\degr) \\
    \midrule
    & & \multicolumn{3}{c|}{(1-GHz  Geometry)} & \multicolumn{5}{c|}{(1.4-GHz Beam Sizes)} & \multicolumn{5}{c|}{(327-MHz Beam Sizes)} & \multicolumn{5}{c}{(100-MHz Band Beam Sizes)} \\
    \midrule
    \midrule
B1930+22 & T & {\bf 86} & +8 & +7.6 & 6.5 & $\sim$16 & 11.1 &  --- &  --- &  --- & $\sim$20 & 12.6 &  --- &  --- &  --- & $\sim$33 & 18.2 &  --- &  --- \\
B1930+13 & D? & 60 & $\infty$ & 0 &  --- & 10.5 & 4.5 &  --- &  --- &  --- & 11.6 & 5.0 &  --- &  --- & 25 & 11 &  --- &  --- &  --- \\
B1931+24 & T/M & {\bf 44} & +15 & +2.7 & 3.9 &  --- &  --- & 16.4 & 6.4 & $\sim$5 &  --- &  --- & 16.4 & 6.4 & $\sim$10 &  --- &  --- & 0.0 & 2.7 \\
B1942-00 & D & 35 & -30 & +1.1 &  --- &  --- &  --- & 18.7 & 5.5 &  --- &  --- &  --- & 19.1 & 5.7 &  --- &  --- &  --- & 38.3 & 11.2 \\
B1944+22 & cT & 34 & $\sim$+11? & +2.9 &  --- & $\sim$6 & 3.4 & 14.5 & 5.1 &  --- & $\sim$6 & 3.4 & $\sim$16 & 5.5 &  & $\sim$6 & 3.4 & 14.5 & 4.0 \\
\\ 
B1949+14 & St & {\bf 49} & -18 & +2.4 & 6.2 &  --- &  --- &  --- &  --- & 7.30 &  --- &  --- &  --- &  --- & 11.6 &  --- &  --- &  --- &  --- \\
B1951+32 & Sd?? & 46 & +2.3 & +18.2 &  --- & 28.5 & 21.6 &  --- &  --- &  --- & $\sim$35 & 23.1 &  --- &  --- &  --- & $\sim$40 & 24.4 &  --- &  --- \\
B1953+29 & T & {\bf 65} & $\sim$-3 & -18 & 34.5 & 100 & 44.4 &  --- &  --- & $\sim$43 &  --- & 44.4 &  --- &  --- &  --- &  --- &  --- &  --- &  --- \\
B2000+32 & St & {\bf 90} & -13 & +4.4 & 2.9 & 5.3 & 5.1 &  --- &  --- &  --- & $\sim$35 & 18.0 &  --- &  --- & $\sim$10 &  --- &  --- &  --- &  --- \\
B2002+31 & T & {\bf 45} & $\infty$ & 0 & $\sim$2.4 &  --- &  --- & 11.2 & 3.9 & $\sim$4 &  --- &  --- & 11.5 & 4.0 & $\sim$4 &  --- &  --- & 11.5 & 4.0 \\
\\ 
B2025+21 & cT? & 28 & $\sim$4? & +6.4 &  --- & $\sim$7 & 6.7 & 25.0 & 9.1 &  --- & $\sim$7 & 6.7 & $\sim$25 & 9.1 &  --- &  --- &  --- & 33 & 10.6 \\
B2027+37 & St & {\bf 34} &  --- &  --- & 4.0 &  --- &  --- &  --- &  --- & 7.9 &  --- &  --- &  --- &  --- & $\sim$20 &  --- &  --- &  --- &  --- \\
B2028+22 & cQ & {\bf 50} & -8 & +5.5 & $\approx$4? & $\sim$5? & 5.9 & 12.2 & 7.4 &  --- & $\sim$5? & 5.9 & $\sim$14? & 7.8 &  --- &  --- &  --- & 19.0 & 9.4 \\
B2034+19 & cQ & 47 & +22 & +1.9 &  --- & 6.7 & 3.1 & 9.5 & 4.0 &  --- & 7.6 & 3.4 & 11.3 & 4.6 &  --- & 6.6 & 3.1 & 13.9 & 3.1 \\
B2035+36 & T & {\bf 51} & +20 & +2.2 & 4.0 &  --- &  --- & 17.5 & 7.3 & $\sim$5 &  --- &  --- & 23.5 & 9.5 & $\sim$8 &  --- &  --- & $\sim$42 & 16.7 \\
\\ 
B2044+15 & D & 40 & +11 & +3.4 &  --- &  --- &  --- & 13.0 & 5.5 &  --- &  --- &  --- & 13.9 & 5.7 &  --- &  --- &  --- & 19.4 & 7.3 \\
B2053+21 & D? & 58 & -12 & +4.1 &  --- & 6.1 & 4.8 &  --- &  --- &  --- & 7.8 & 5.3 &  --- &  --- &  --- & $\sim$22 & 10.3 &  --- &  --- \\
B2053+36 & St & {\bf 66} & +8 & +7.0 & 5.7 & 12.4 & 9.1 &  --- &  --- & $\sim$13 &  --- &  --- &  --- &  --- & $\sim$80 &  --- &  --- &  --- &  --- \\
B2113+14 & Sd & 45 & +5 & +8.1 &  --- &  --- &  --- & 7.4 & 8.6 &  --- &  --- &  --- & 7.2 & 8.6 &  --- &  --- &  --- & 16.5 & 10.2 \\
B2122+13 & D & 83 & -27 & +2.1 &  --- &  --- &  --- & 13.0 & 6.8 &  --- &  --- &  --- & 14 & 7.3 &  --- &  --- &  --- & 17.8 & 9.1 \\
\\ 
B2210+29 & M & {\bf 41} & -35 & -1.1 & $\sim$3.7 & $\sim$13 & 4.4 & 17.0 & 5.7 & $\sim$4 & $\sim$15 & 5.0 & $\sim$19 & 6.1 & $\sim$10 &  --- &  --- & 26.2 & 8.6 \\
    \bottomrule
   \end{tabular}
\end{table*}

\twocolumn

\noindent\textit{\textbf{B1900+05}}: We follow ET VI and \citet{W99} in the pulsar as having a core-single profile.  However, more single pulse study is now needed to understand the structure fully.
\vskip 0.1in

\noindent\textit{\textbf{B1900+06}}: Our 1.4-GHz profile as well and those in \citet{W99,JK18} strongly suggest that this must be a core-single profile, and it shows an orderly PPA traverse.  The profile is more complex than a single component, and neither the leading feature nor the possible weak trailing one are well resolved, but a rough inner cone model is possible.  The lower frequency profiles become asymmetric, but scattering seem important only at 100 MHz and below.  
\vskip 0.1in

\begin{figure}
\begin{center}
{\mbox{\includegraphics[height=85mm,angle=-90.]{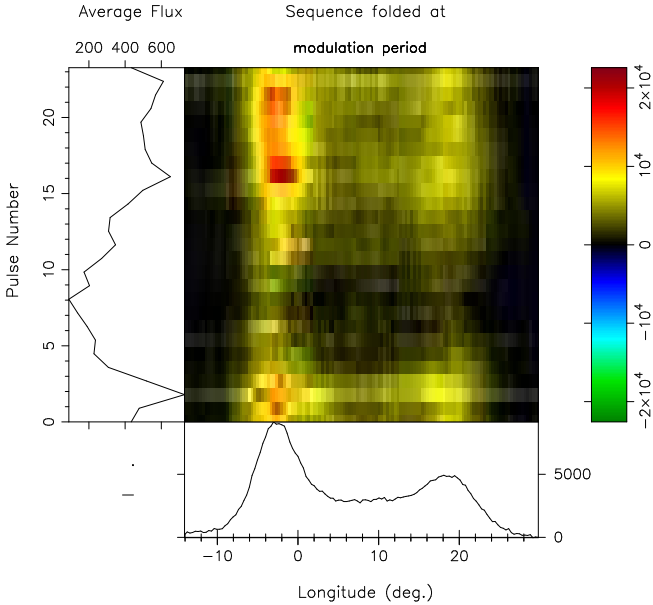}}}
\caption{Pulsar B1901+10 single pulses show a deep 23-$P$ amplitude modulation as shown above in the spulse sequence folded at this interval.}
\hspace{-0.55 cm}
\label{figA105}
\end{center}
\end{figure}

\noindent\textit{\textbf{B1900+01}}: We follow ET VI in regarding the pulsar as primarily being a core feature.  Our strangely shaped 1.4-GHz profile shows a tripartite structure that reflects the usual inner conal outrider structure.   However, drift modulation is detected \citep{W99,Weisberg2004}, apparently in the profile center at 1.4 GHz, so a conal triple structure is also possible.  The \citet{W99} 1.4-GHz profile provides the clearest context of determining the PPA rate.  Our 327-MHz profile shows a prominent scattering ``tail'' making it impossible to distinguish the beam configuration.  The 102-MHz profile \citep{IMS89} is all scattering, and a measurement is given by \citet{kmn+15}.  
\vskip 0.1in
\noindent\textit{\textbf{B1901+10}} exhibits a 23-$P$ amplitude modulation as depicted in the folded pulse sequence in Fig.~\ref{figA105}.  Moreover, the PPA traverse has the expected ``S'' shape.  Therefore we can confidently model the profile with a conal double \textbf{D} beam.  \citet{hulse_taylor} report a larger component separation at 430 MHz---though no discovery profile seems to have been published---so we use an outer cone.
\vskip 0.1in

\noindent\textit{\textbf{B1902--01}}: Scattering is seen in PSR B1902$-$01 at 327 MHz, so we only have the 1.4-GHz to interpret.  Most likely this pulsar has a core-single \textbf{S$_t$} geometry.  The core width implies an $\alpha$ value of 80\degr, and the PPA traverse is disordered, so we assume central sightline. If we are seeing a weak conflated conal component on the trailing side of the profile, the conal width is about 11\degr\ and this is compatible with an inner conal geometry.
\vskip 0.1in

\noindent\textit{\textbf{B1903+07}}:  Taking the long orderly PPA traverse as a guide, we interpret the profile as conal, and given that GL's 606-MHz profile textbf{(the lowest frequency available)} does not seem much wider, we have used an inner cone.   
\vskip 0.1in

\begin{figure*}
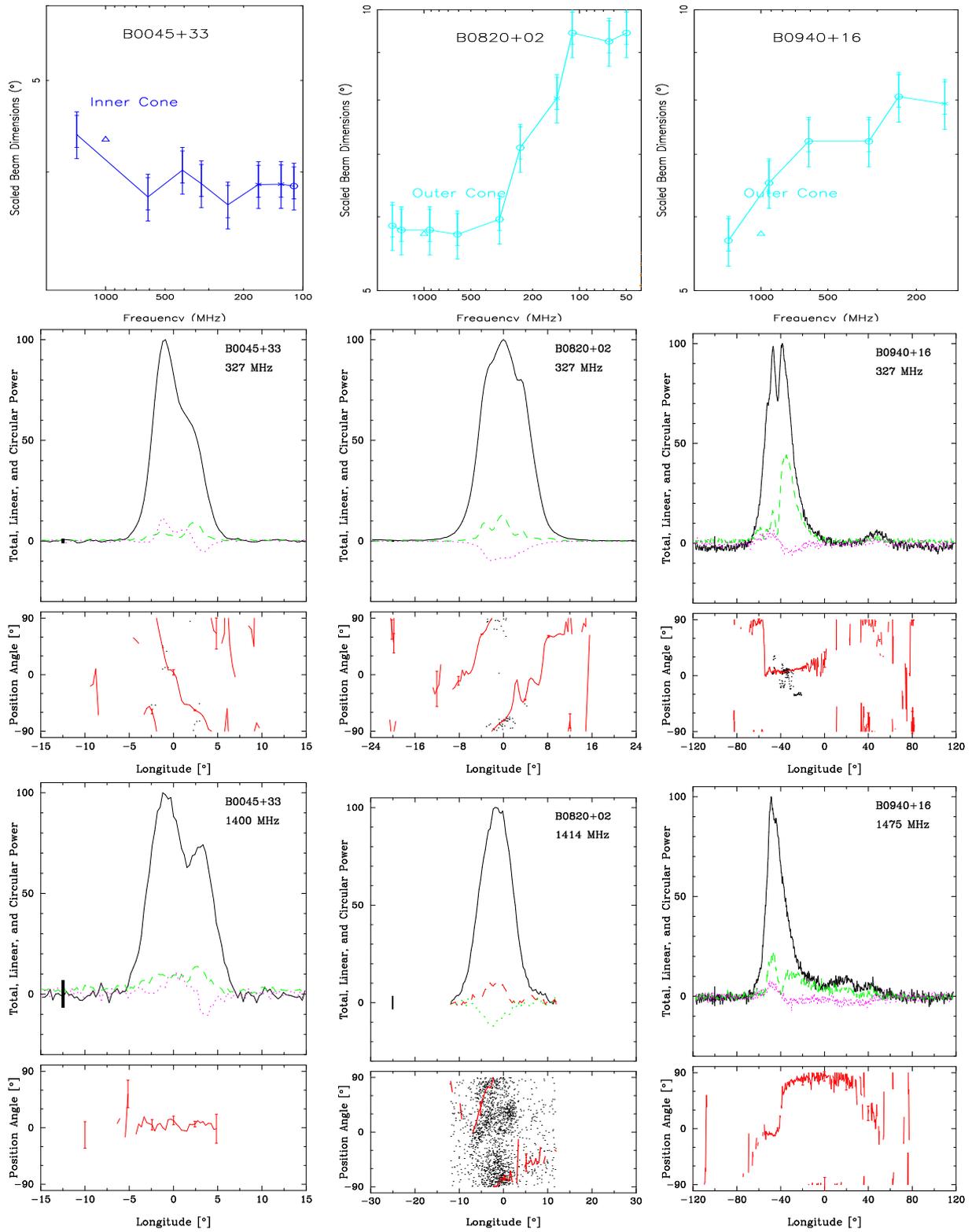

\begin{center}
\begin{tabular}{@{}ll@{}ll@{}}
{\mbox{\includegraphics[width=51mm,height=53mm]{plots/B0045+33_ABmodel.ps}}}& 
{\mbox{\includegraphics[width=51mm,height=53mm]{plots/B0820+02_ABmodel.ps}}}& \ \  
{\mbox{\includegraphics[width=51mm,height=53mm]{plots/B0940+16_ABmodel.ps}}}\\ 
{\mbox{\includegraphics[width=51mm]{plots/B0045+33P_mp.ps}}}& 
{\mbox{\includegraphics[width=51mm]{plots/B0820+02P.ps}}}& \ \ \ 
{\mbox{\includegraphics[width=51mm]{plots/B0940+16P.ps}}}\\ 
{\mbox{\includegraphics[width=51mm]{plots/B0045+33L_mp.ps}}}&
{\mbox{\includegraphics[width=51mm,height=72mm]{plots/d2942048.0820+02l.ps}}}& \ \ \ 
{\mbox{\includegraphics[width=51mm]{plots/B0940+16L.ps}}}\\ 
\end{tabular}

\caption{Scaled beam dimensions and average profiles for PSRs B0045+33, B0820+02 and B0940+16. The scaled beam dimension plots are logarithmic on both axes. Plotted values represent the scaled inner and outer conal beam radii and the core angular width, respectively. The scaled outer and inner conal radii are plotted with blue and cyan lines and the core diameter in red. The nominal values of the three beam dimensions at 1 GHz are shown in each plot by a small triangle. The yellow hatching indicates the average scattering level and the orange hatching indicates 100-MHz scattering times (both where applicable). The top panel in each of the average profiles (second two rows) is the average profile, with the solid line (black) showing the total intensity, the dashed line (green) showing the linear polarization, and dotted (red) line showing the circular polarization. The bottom panels show the polarization angle against longitude. The HR10 1.4-GHz profile has a restricted longitude range and reverses the colour labeling of the linear and circular polarization.} 
\label{figA1}
\end{center}
\end{figure*}

\begin{figure*}
\begin{center}
\begin{tabular}{@{}ll@{}ll@{}}
{\mbox{\includegraphics[width=51mm]{plots/B1534+12_ABmodel.ps}}}&
{\mbox{\includegraphics[width=51mm]{plots/B1726-00_ABmodel.ps}}}& \ \ \ 
{\mbox{\includegraphics[width=51mm]{plots/B1802+03_ABmodel.ps}}}\\
{\mbox{\includegraphics[width=51mm]{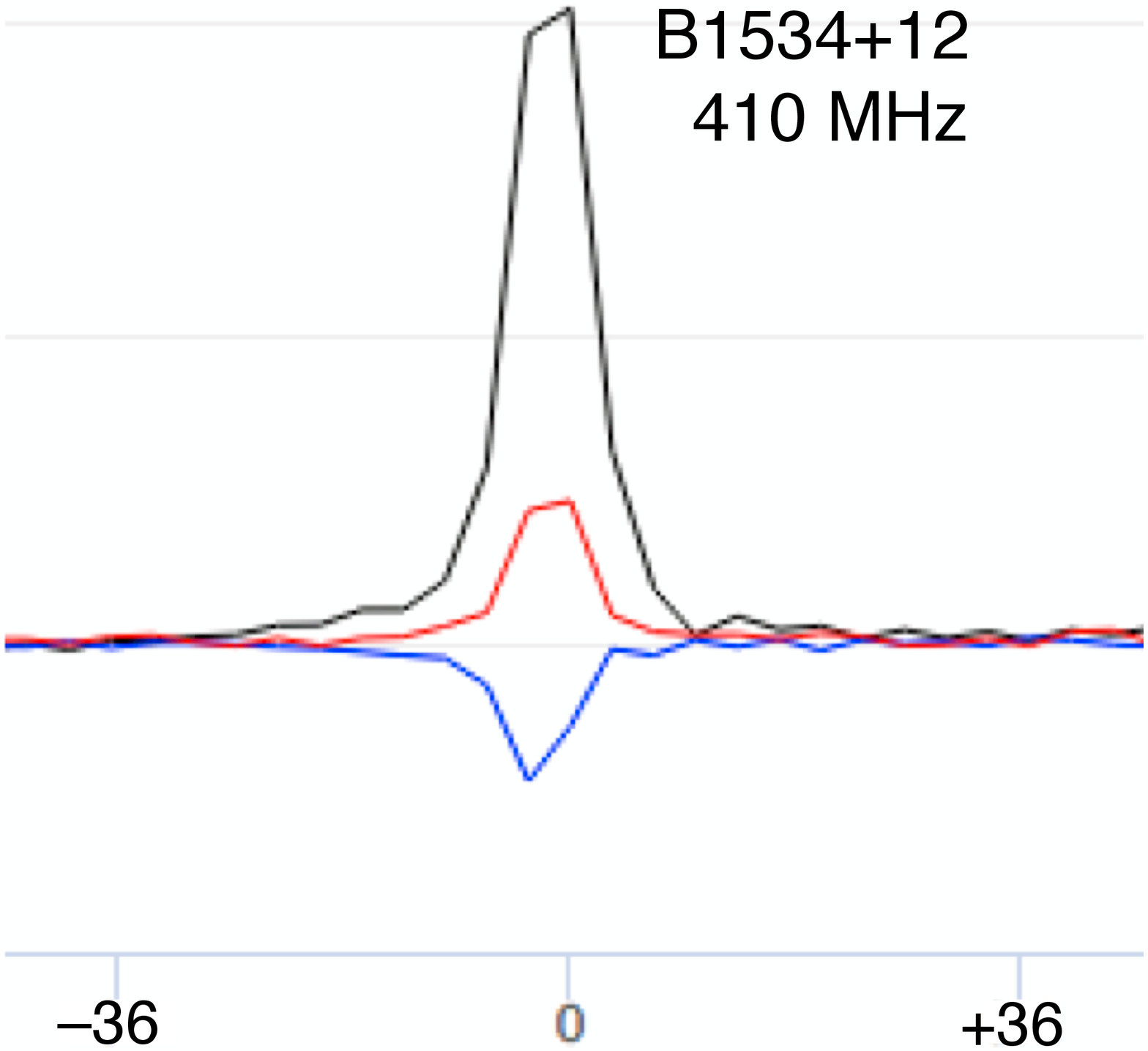}}}& 
{\mbox{\includegraphics[width=51mm]{plots/B1726-00P.ps}}}&  \ \ \
{\mbox{\includegraphics[width=51mm]{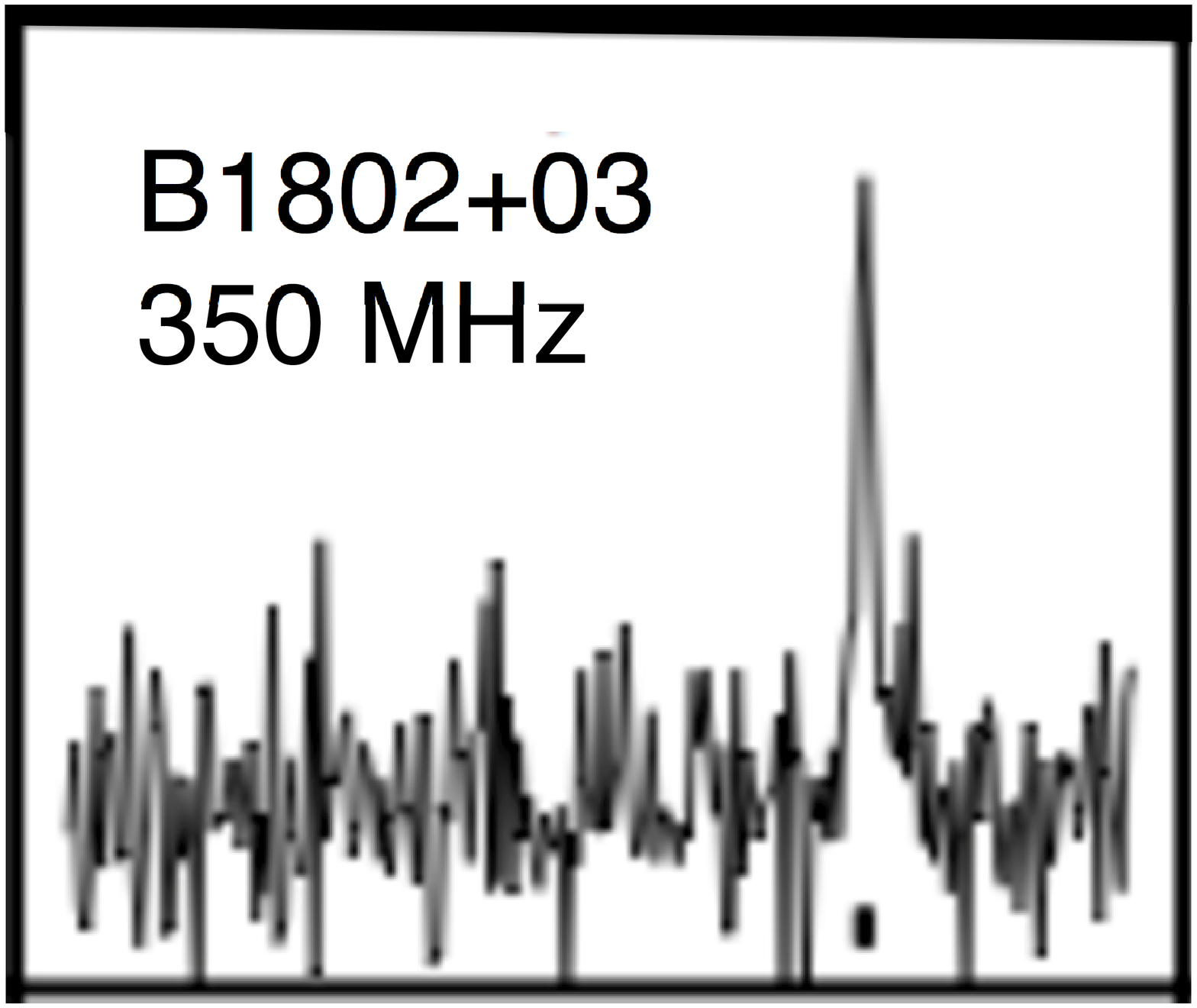}}}\\
{\mbox{\includegraphics[width=51mm]{plots/PQB1534+12.55637la.ps}}}&
{\mbox{\includegraphics[width=51mm]{plots/B1726-00L.ps}}}& \ \ \
{\mbox{\includegraphics[width=51mm]{plots/B1802+03L.ps}}}\\ 
\end{tabular}
\caption{Scaled beam dimensions and average profiles for PSRs B1534+12 (410-MHz profile from EPN Database; 36\degr\ tick marks), B1726+00 and B1802+03/J1805+0306 (full period, 350-MHz profile from \citet{mcewen}) as in Fig~\ref{figA1}.} 
\label{figA2}
\end{center}
\end{figure*}

\begin{figure*}
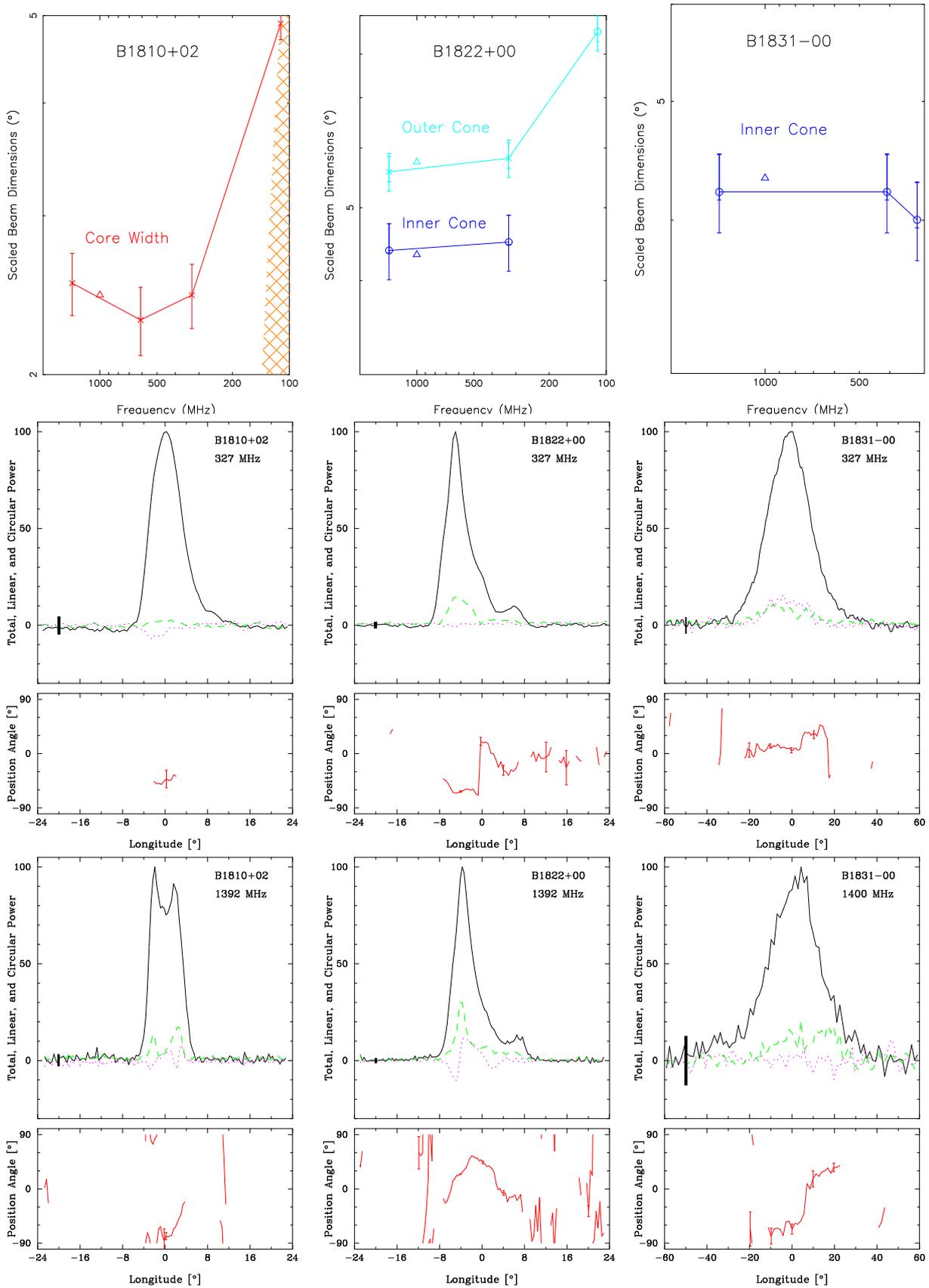

\begin{center}
\begin{tabular}{@{}ll@{}ll@{}}
{\mbox{\includegraphics[width=51mm]{plots/B1810+02_ABmodel.ps}}}& 
{\mbox{\includegraphics[width=51mm]{plots/B1822+00_ABmodel.ps}}}& \ \ \ 
{\mbox{\includegraphics[width=51mm]{plots/B1831-00_ABmodel.ps}}}\\
{\mbox{\includegraphics[width=51mm]{plots/B1810+02P.ps}}}&
{\mbox{\includegraphics[width=51mm]{plots/B1822+00P.ps}}}& \ \ \ 
{\mbox{\includegraphics[width=51mm]{plots/PQB1831-00.53377ap.ps}}}\\
{\mbox{\includegraphics[width=51mm]{plots/B1810+02L.ps}}}& 
{\mbox{\includegraphics[width=51mm]{plots/B1822+00L.ps}}}& \ \ \ 
{\mbox{\includegraphics[width=51mm]{plots/PQB1831-00.52735la.ps}}}& \ \ \ \ \ \
\end{tabular}
\caption{Scaled beam dimensions and average profiles for PSRs B1810+02, B1822+00 and B1831--00 in Fig~\ref{figA1}.}
\label{figA3}
\end{center}
\end{figure*}

\begin{figure*}
\begin{center}
\begin{tabular}{@{}ll@{}ll@{}}
{\mbox{\includegraphics[width=51mm]{plots/B1848+04_ABmodel.ps}}}&
{\mbox{\includegraphics[width=51mm]{plots/B1853+01_ABmodel.ps}}}& \ \ \ 
{\mbox{\includegraphics[width=51mm]{plots/B1854+00_ABmodel.ps}}}\\ 
{\mbox{\includegraphics[width=60mm,height=75mm]{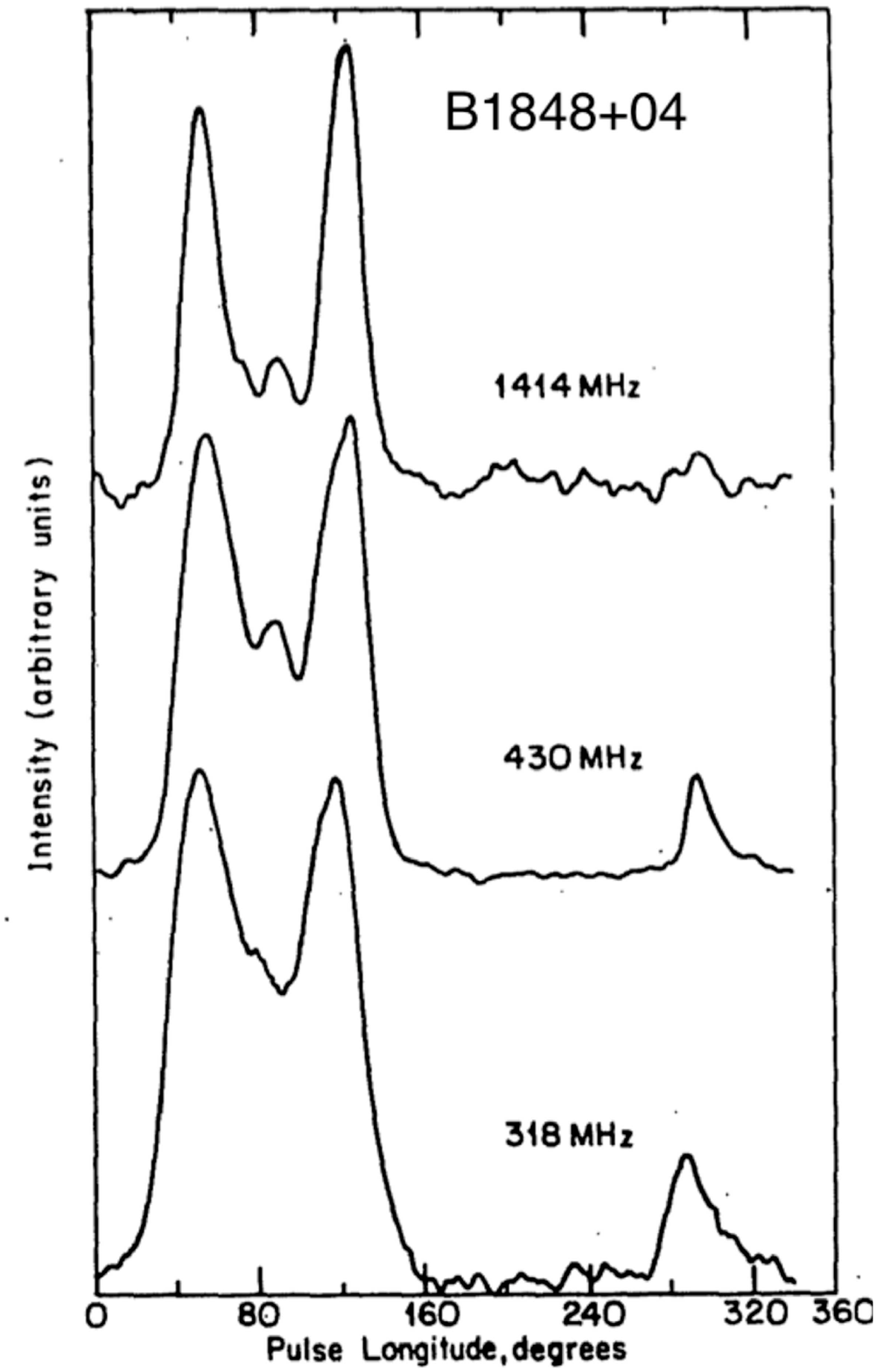}}}&
{\mbox{\includegraphics[width=51mm]{plots/B1853+01P.ps}}}& \ \ \ 
{\mbox{\includegraphics[width=51mm]{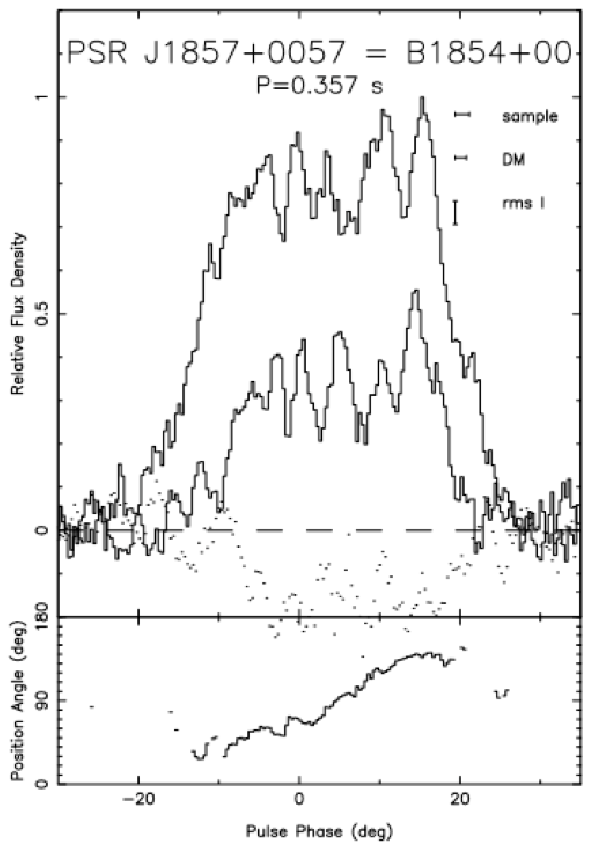}}}\\ 
{\mbox{\includegraphics[width=51mm]{plots/B1848+04L.ps}}}&
{\mbox{\includegraphics[width=51mm]{plots/B1853+01L.ps}}}& \ \ \ 
{\mbox{\includegraphics[width=51mm]{plots/B1854+00L.ps}}}\\ 
\end{tabular}
\caption{Scaled beam dimensions and average profiles for PSRs B1848+04 (profiles from \citet{boriakoff}), B1853+01 and  B1854+00/J1857+0057 (full-period, 430-MHz profile from \citet{Weisberg2004})as in Fig~\ref{figA1}.}
\label{figA4}
\end{center}
\end{figure*}

\begin{figure*}
\begin{center}
\begin{tabular}{@{}ll@{}ll@{}}
{\mbox{\includegraphics[width=51mm]{plots/B1859+01_ABmodel.ps}}}&
{\mbox{\includegraphics[width=51mm]{plots/B1859+03_ABmodel.ps}}}& \ \ \ 
{\mbox{\includegraphics[width=51mm]{plots/B1859+07_ABmodel.ps}}}\\
{\mbox{\includegraphics[width=51mm,height=75mm]{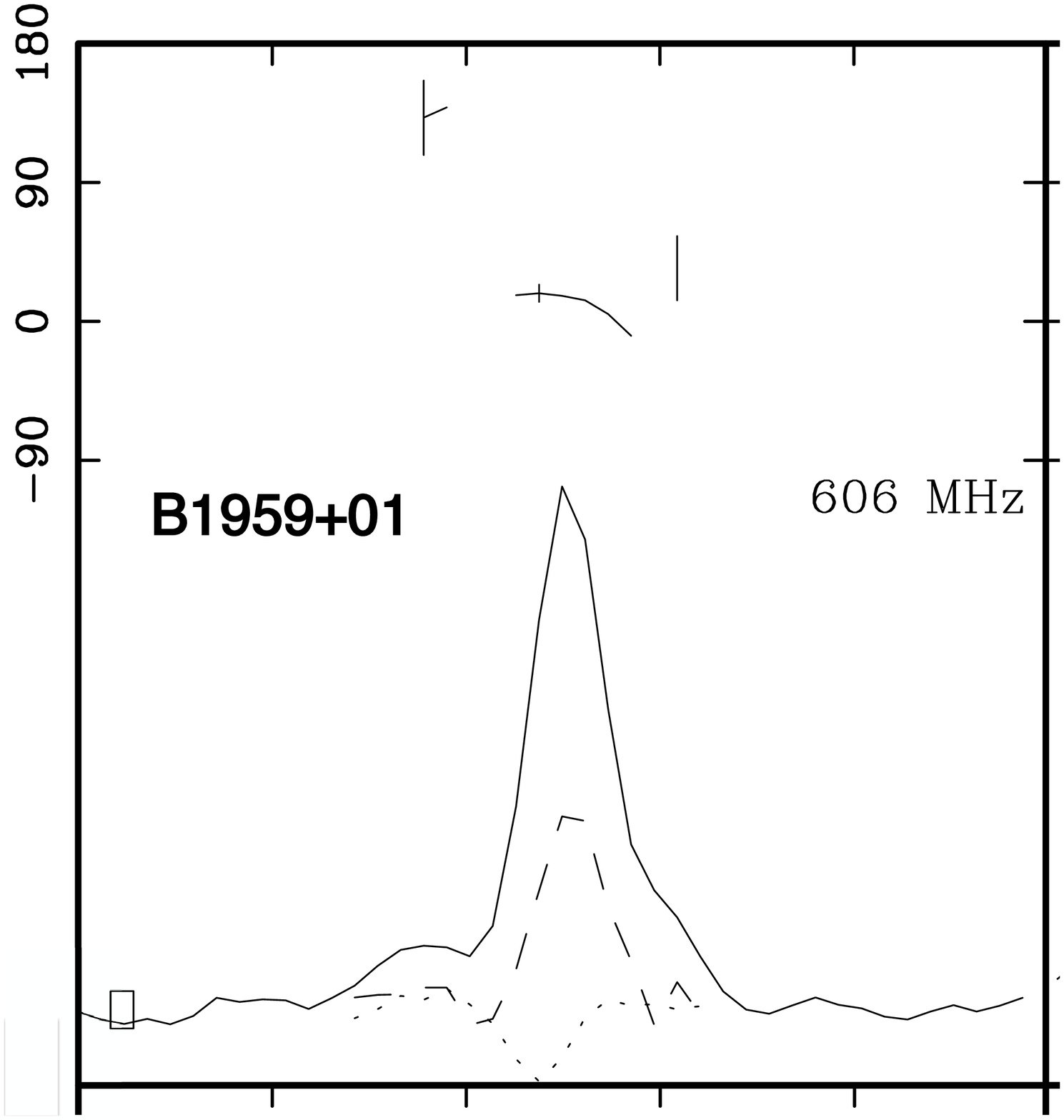}}}&
{\mbox{\includegraphics[width=48mm,height=65mm]{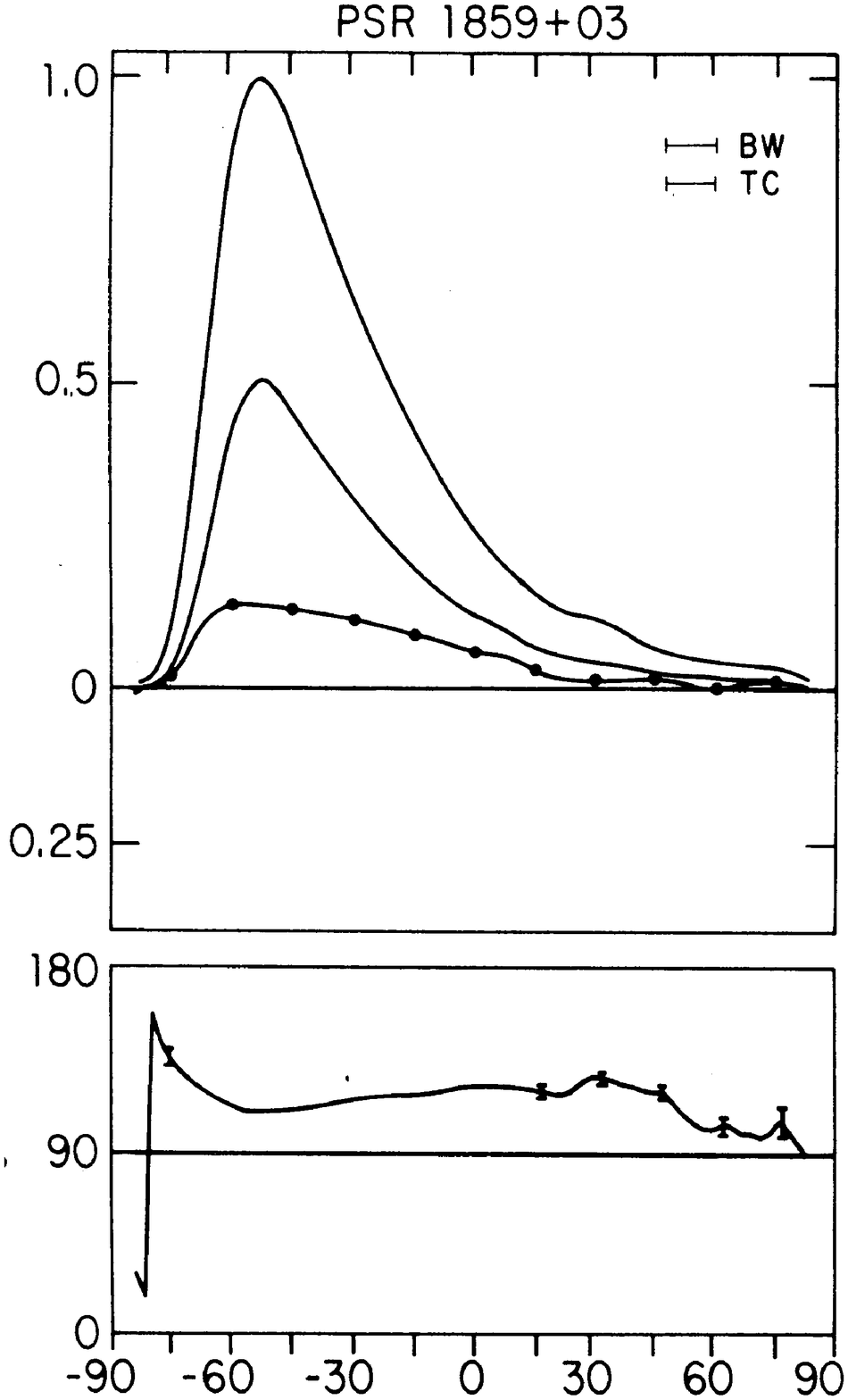}}}& \ \ \ 
{\mbox{\includegraphics[width=51mm,height=75mm]{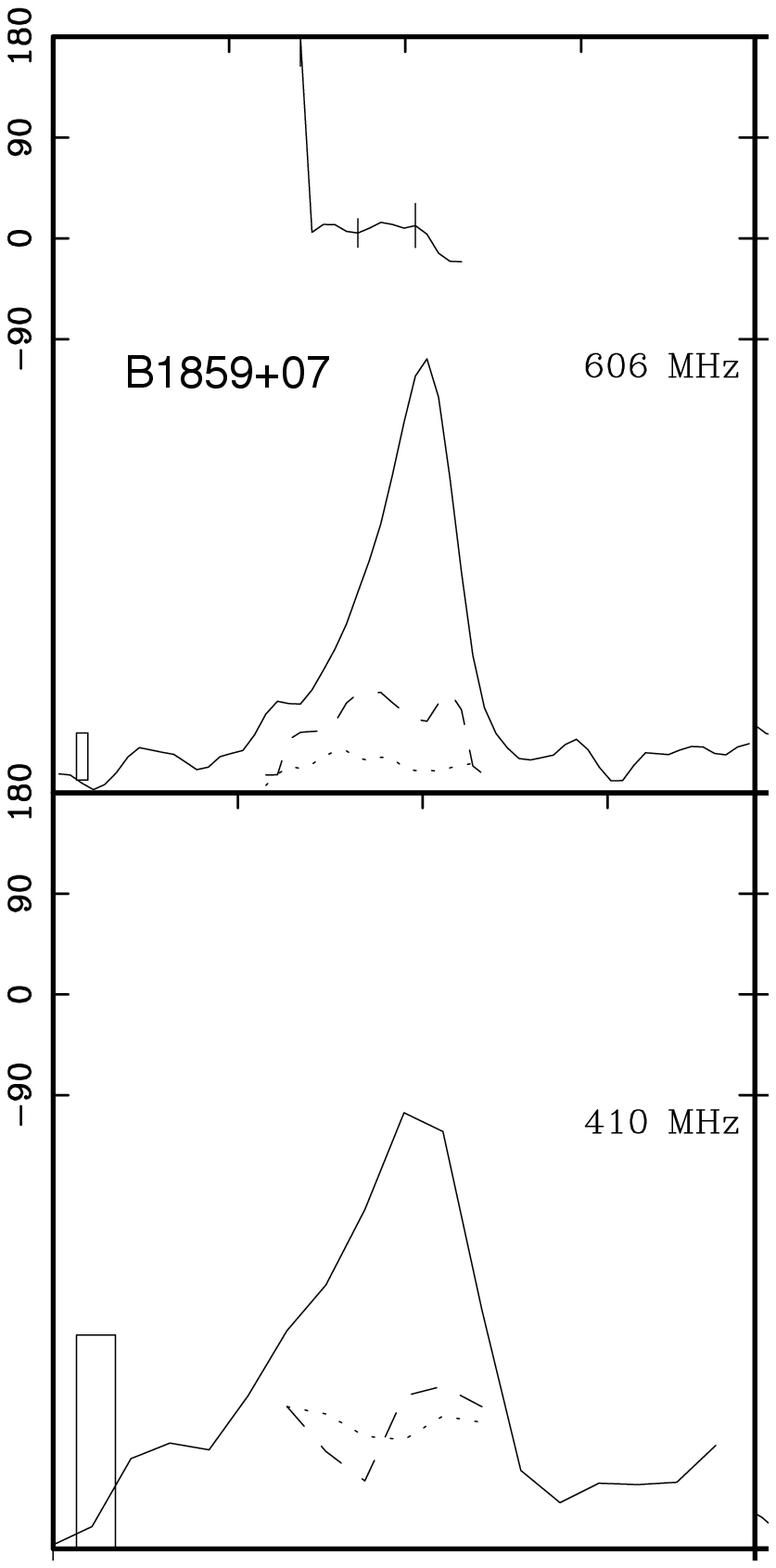}}}\\
{\mbox{\includegraphics[width=51mm]{plots/PQB1859+01.54842la.ps}}}& 
{\mbox{\includegraphics[width=51mm]{plots/PQB1859+03.56768la.ps}}}& \ \ \ 
{\mbox{\includegraphics[width=51mm]{plots/PQB1859+07.57121la.ps}}}\\ 
\end{tabular}
\caption{Scaled beam dimensions and average profiles for PSRs B1859+01 (606-MHz profile from GL98, B1859+03 and B1859+07 (profiles from GL98 with 20\degr\ tick marks) as in Fig~\ref{figA1}.}
\label{figA5}
\end{center}
\end{figure*}

\begin{figure*}
\begin{center}
\begin{tabular}{@{}ll@{}ll@{}}
{\mbox{\includegraphics[width=51mm]{plots/B1900+05_ABmodel.ps}}}&
{\mbox{\includegraphics[width=51mm]{plots/B1900+06_ABmodel.ps}}}&  \ \ \ 
{\mbox{\includegraphics[width=51mm]{plots/B1900+01_ABmodel.ps}}}\\
{\mbox{\includegraphics[height=51mm,width=65mm,angle=90.]{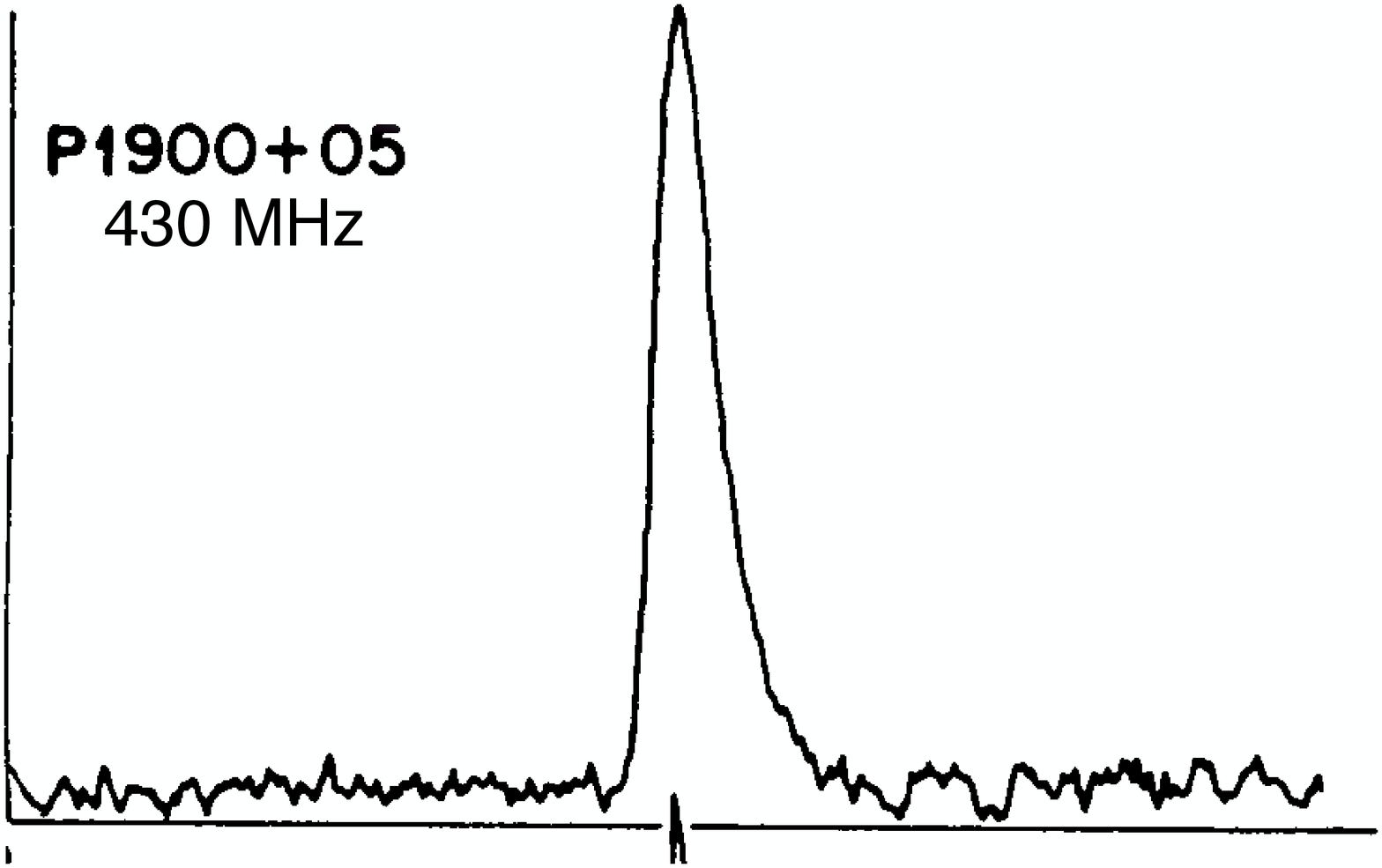}}}& \ \ \ 
{\mbox{\includegraphics[height=45mm,width=65mm,angle=90.]{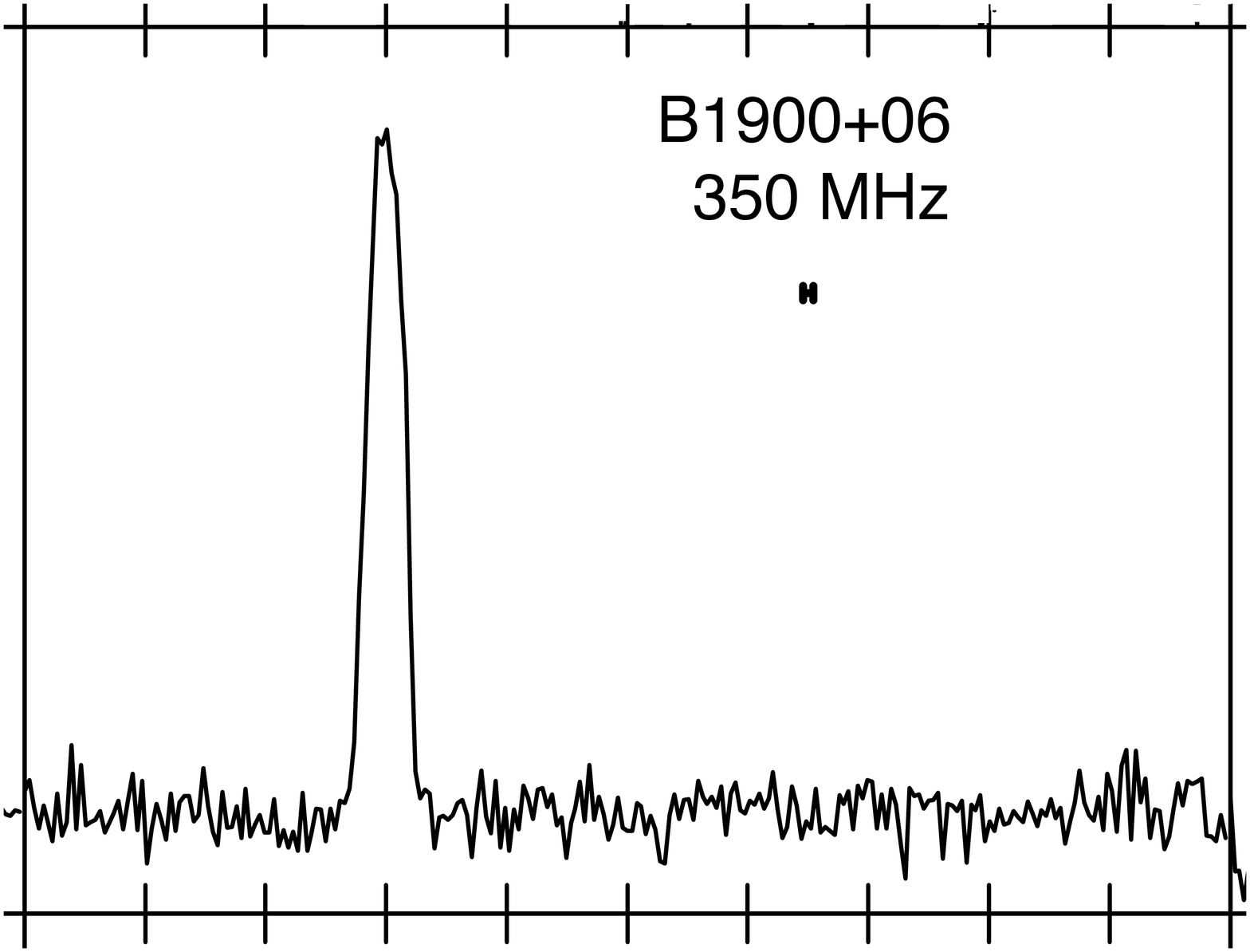}}}&  \ \ \ 
{\mbox{\includegraphics[width=51mm]{plots/B1900+01P.ps}}}\\
{\mbox{\includegraphics[width=51mm]{plots/PQB1900+05.54842la.ps}}}& \ \ \ 
{\mbox{\includegraphics[width=51mm]{plots/PQB1900+06.55633la.ps}}}& \ \ \ 
{\mbox{\includegraphics[width=51mm]{plots/B1900+01L.ps}}}\\ 
\end{tabular}
\caption{Scaled beam dimensions and average profiles for PSRs B1900+05 (430-MHz full period profile from \citet{GR78}, B1900+06/J1902+0615 (350-MHz full period profile from \citet{mcewen}) and B1900+01 as in Fig~\ref{figA1}.}
\label{figA6}
\end{center}
\end{figure*}

\begin{figure*}
\begin{center}
\begin{tabular}{@{}ll@{}ll@{}}
{\mbox{\includegraphics[width=51mm]{plots/B1901+10_ABmodel.ps}}}&
{\mbox{\includegraphics[width=51mm]{plots/B1902-01_ABmodel.ps}}}& \ \ \ 
{\mbox{\includegraphics[width=51mm]{plots/B1903+07_ABmodel.ps}}}\\
{\mbox{\includegraphics[width=51mm]{plots/PQB1901+10.53777ap.ps}}}&
{\mbox{\includegraphics[width=51mm]{plots/B1902-01P.ps}}}& \ \ \   
{\mbox{\includegraphics[width=51mm]{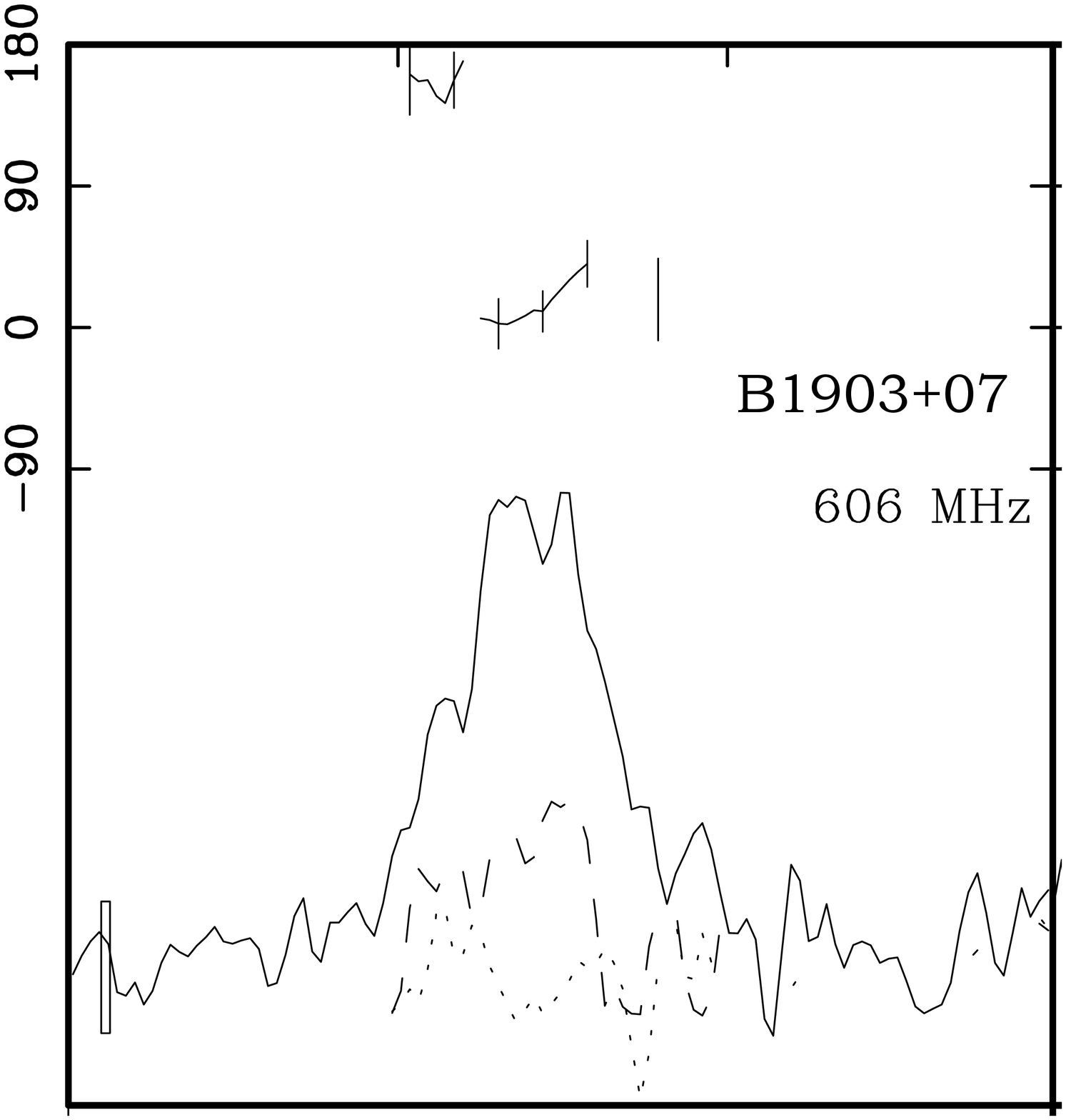}}}\\
{\mbox{\includegraphics[width=51mm]{plots/PQB1901+10.56563la.ps}}}&
{\mbox{\includegraphics[width=51mm]{plots/B1902-01L.ps}}}& \ \ \ 
{\mbox{\includegraphics[width=51mm]{plots/PQB1903+07.57115la.ps}}}\\ 
\end{tabular}
\caption{Scaled beam dimensions and average profiles for PSRs B1901+10, B1902-01 and B1903+07 (150\degr\ profile from GL98)as in Fig~\ref{figA1}.}
\label{figA7}
\end{center}
\end{figure*}

\begin{figure*}
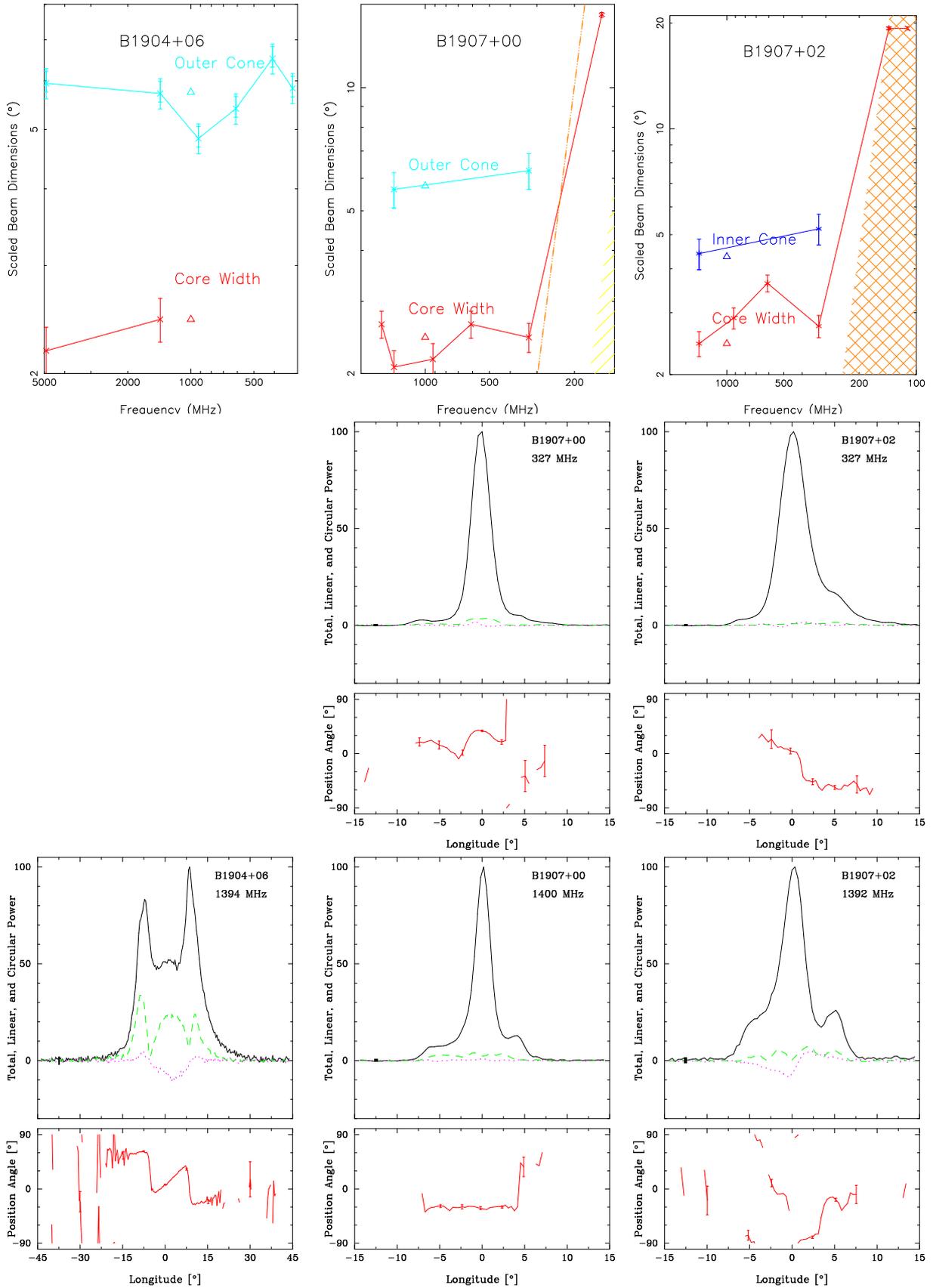

\begin{center}
\begin{tabular}{@{}ll@{}ll@{}}
{\mbox{\includegraphics[width=51mm]{plots/B1904+06_ABmodel.ps}}}&
{\mbox{\includegraphics[width=51mm]{plots/B1907+00_ABmodel.ps}}}& \ \ \ 
{\mbox{\includegraphics[width=51mm]{plots/B1907+02_ABmodel.ps}}}\\
& 
{\mbox{\includegraphics[width=51mm]{plots/B1907+00P.ps}}}&  \ \ \ 
{\mbox{\includegraphics[width=51mm]{plots/B1907+02P.ps}}}\\
{\mbox{\includegraphics[width=51mm]{plots/PQB1904+06.57004la.ps}}}&
{\mbox{\includegraphics[width=51mm]{plots/B1907+00L.ps}}}& \ \ \ 
{\mbox{\includegraphics[width=51mm]{plots/B1907+02L.ps}}}\\
\end{tabular}
\caption{Scaled beam dimensions and average profiles for PSRs B1904+06, B1907+00 and B1907+02 and as in Fig~\ref{figA1}.}
\label{figA8}
\end{center}
\end{figure*}

\begin{figure*}
\begin{center}
\begin{tabular}{@{}ll@{}ll@{}}
{\mbox{\includegraphics[width=51mm]{plots/B1907+10_ABmodel.ps}}}&
{\mbox{\includegraphics[width=51mm]{plots/B1907+03_ABmodel.ps}}}& \ \ \
{\mbox{\includegraphics[width=51mm]{plots/B1907+12_ABmodel.ps}}}\\
{\mbox{\includegraphics[width=51mm]{plots/B1907+10P.ps}}}&
{\mbox{\includegraphics[width=51mm]{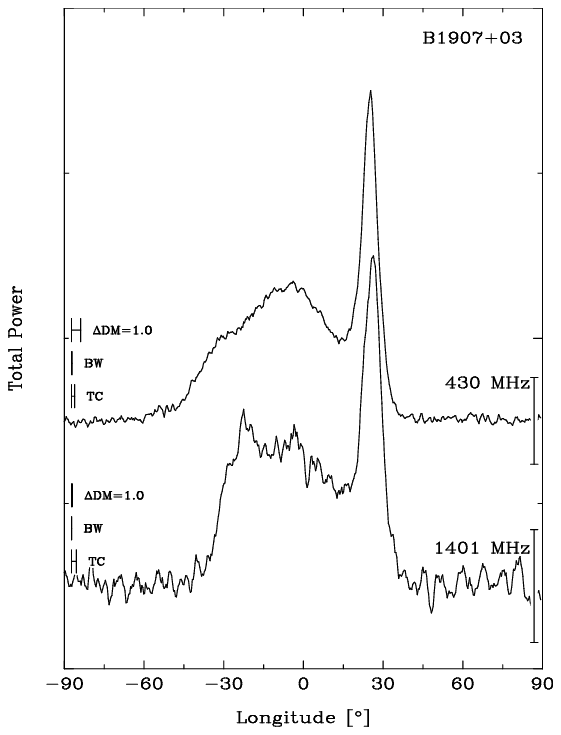}}}& \ \ \
{\mbox{\includegraphics[width=51mm]{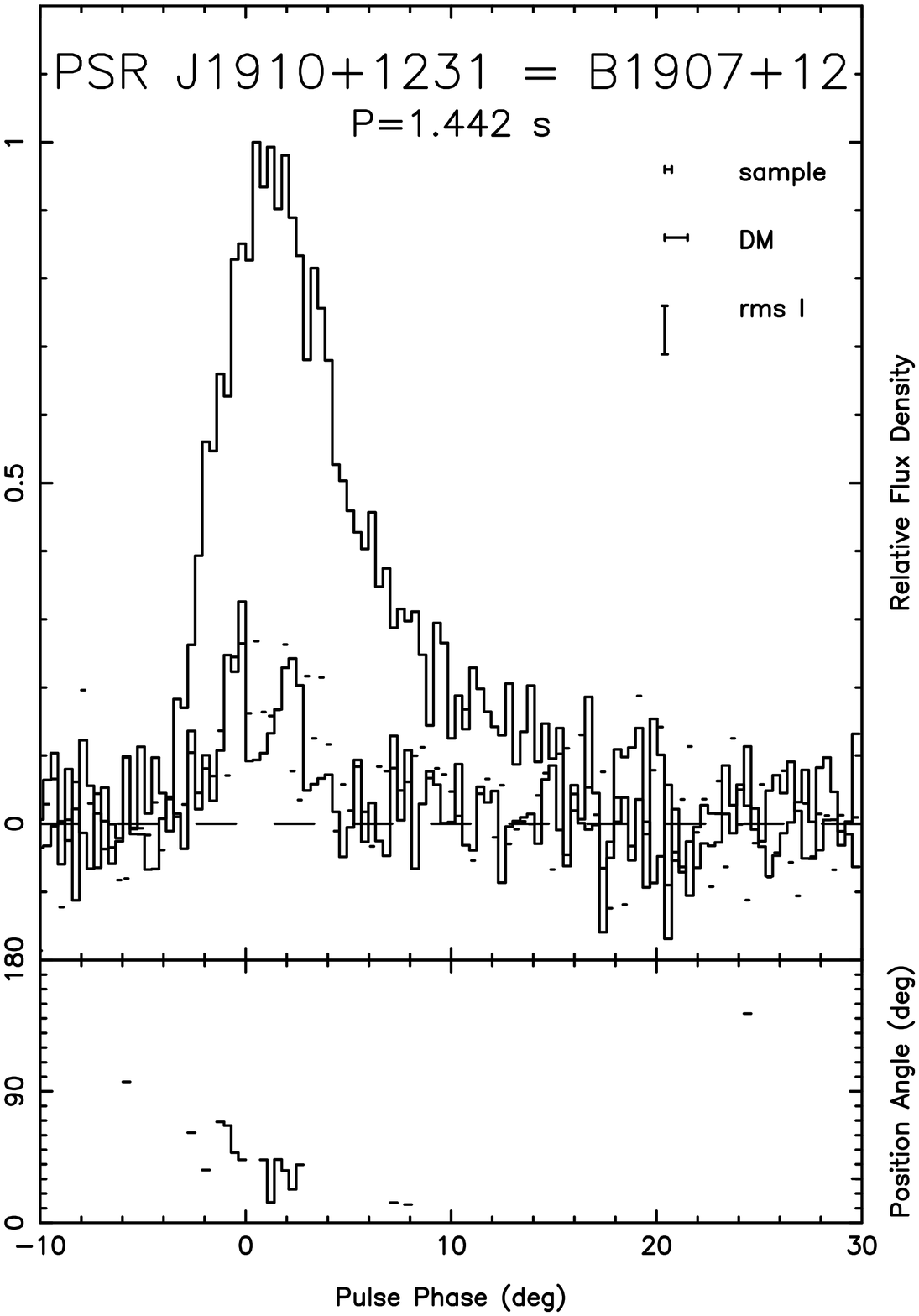}}}\\
{\mbox{\includegraphics[width=51mm]{plots/B1907+10L.ps}}}&
{\mbox{\includegraphics[width=51mm]{plots/B1907+03L.ps}}}& \ \ \
{\mbox{\includegraphics[width=51mm]{plots/B1907+12L.ps}}}\\
\end{tabular}
\caption{Scaled beam dimensions and average profiles for PSRs B1907+10, B1907+03 (time-aligned 430/1401-MHz profiles from HR10) and B1907+12 as in Fig~\ref{figA1}.}
\label{figA9}
\end{center}
\end{figure*}

\begin{figure*}
\begin{center}
\begin{tabular}{@{}ll@{}ll@{}}
{\mbox{\includegraphics[width=51mm]{plots/B1911+09_ABmodel.ps}}}&
{\mbox{\includegraphics[width=51mm]{plots/B1911+13_ABmodel.ps}}}& \ \ \
{\mbox{\includegraphics[width=51mm]{plots/B1911+11_ABmodel.ps}}}\\
{\mbox{\includegraphics[height=50mm,width=55mm]{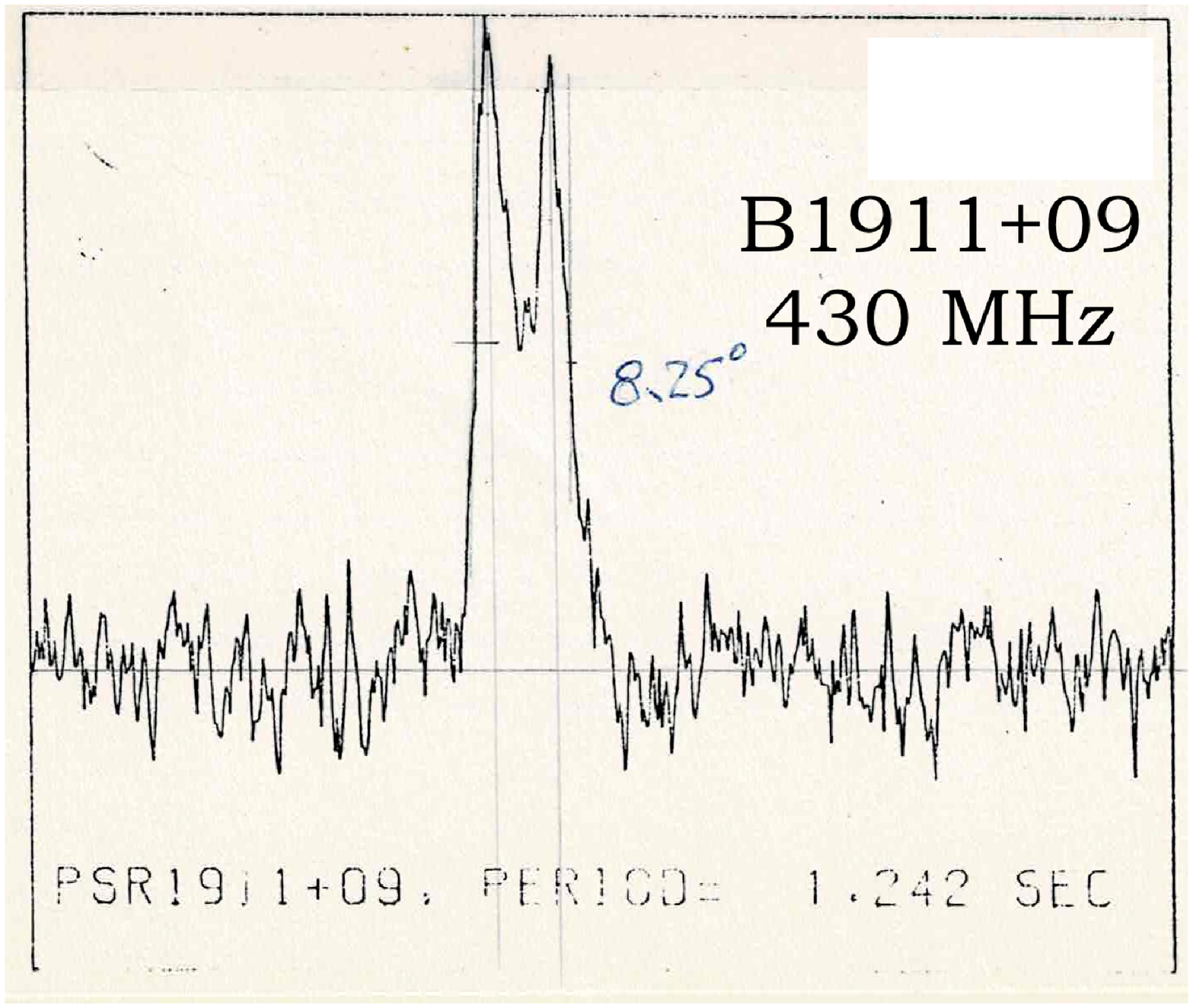}}}&
{\mbox{\includegraphics[height=51mm,width=50mm]{plots/B1911+13P.ps}}}& \ \ \
{\mbox{\includegraphics[height=51mm,width=60mm,angle=90.]{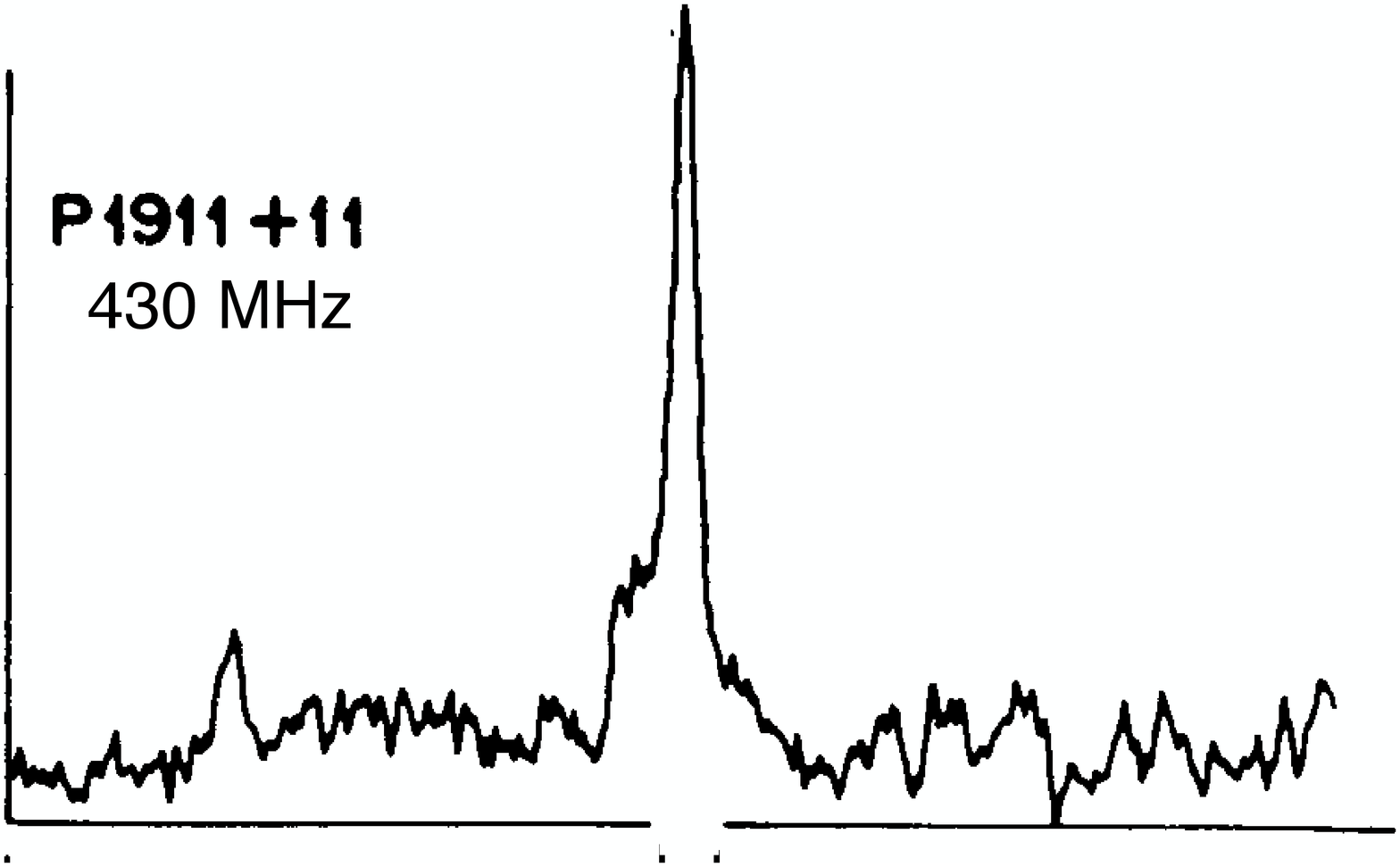}}}\\
{\mbox{\includegraphics[width=51mm]{plots/PQB1911+09.56769la.ps}}}&
{\mbox{\includegraphics[width=51mm]{plots/B1911+13L.ps}}}& \ \ \
{\mbox{\includegraphics[width=51mm]{plots/PQB1911+11.52738la.ps}}}\\
\end{tabular}
\caption{Scaled beam dimensions and average profiles for PSRs B1911+09 (150\degr/ profile from \citet{w80}), B1911+13 and B1911+11 as in Fig~\ref{figA1}.}
\label{figA10}
\end{center}
\end{figure*}

\begin{figure*}
\begin{center}
\begin{tabular}{@{}ll@{}ll@{}}
{\mbox{\includegraphics[width=51mm]{plots/B1913+10_ABmodel.ps}}}&
{\mbox{\includegraphics[width=51mm]{plots/B1913+16_ABmodel.ps}}}& \ \ \ 
{\mbox{\includegraphics[width=51mm]{plots/B1913+167_ABmodel.ps}}}\\
{\mbox{\includegraphics[height=51mm,angle=90]{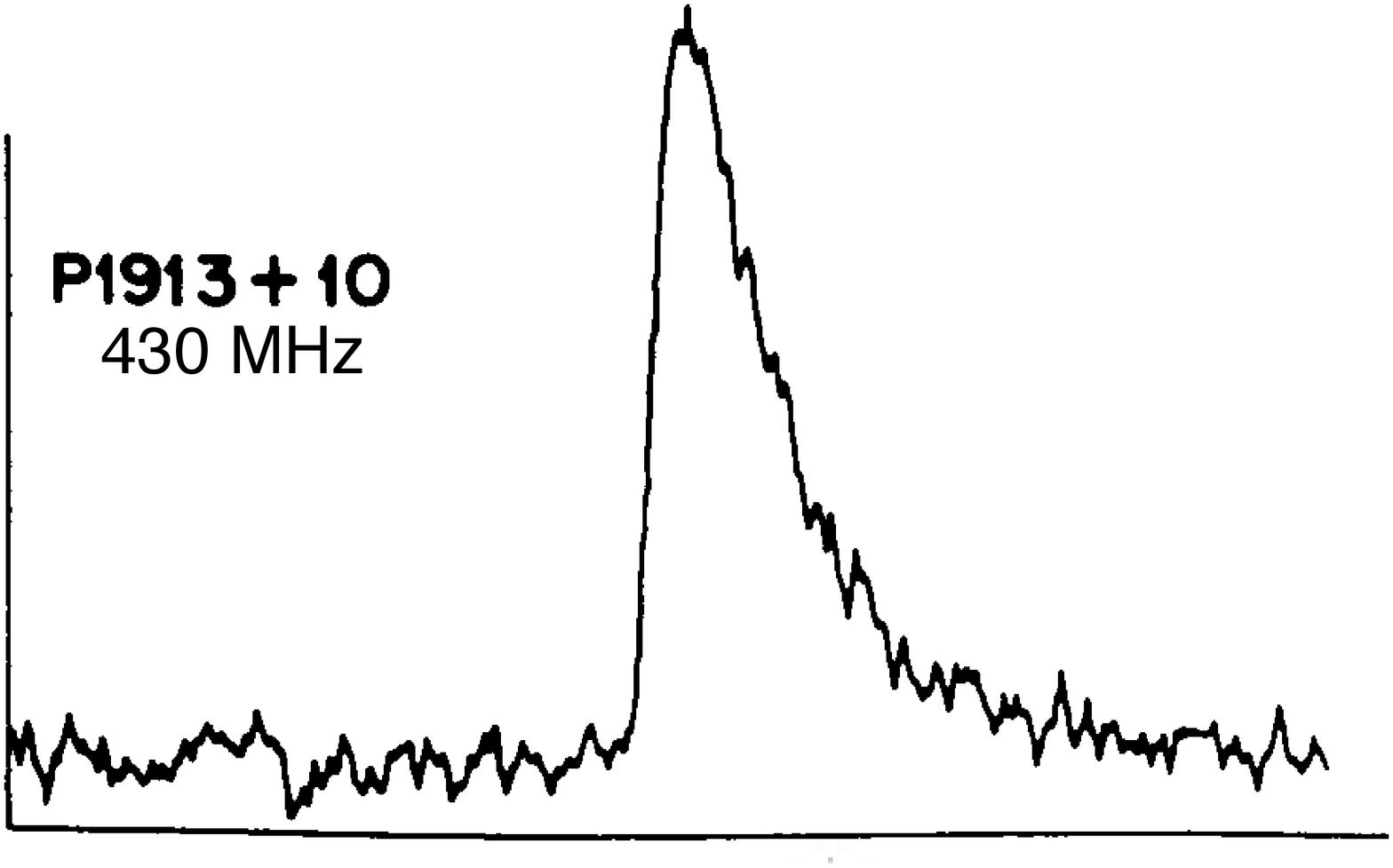}}}&
{\mbox{\includegraphics[width=51mm,height=70mm]{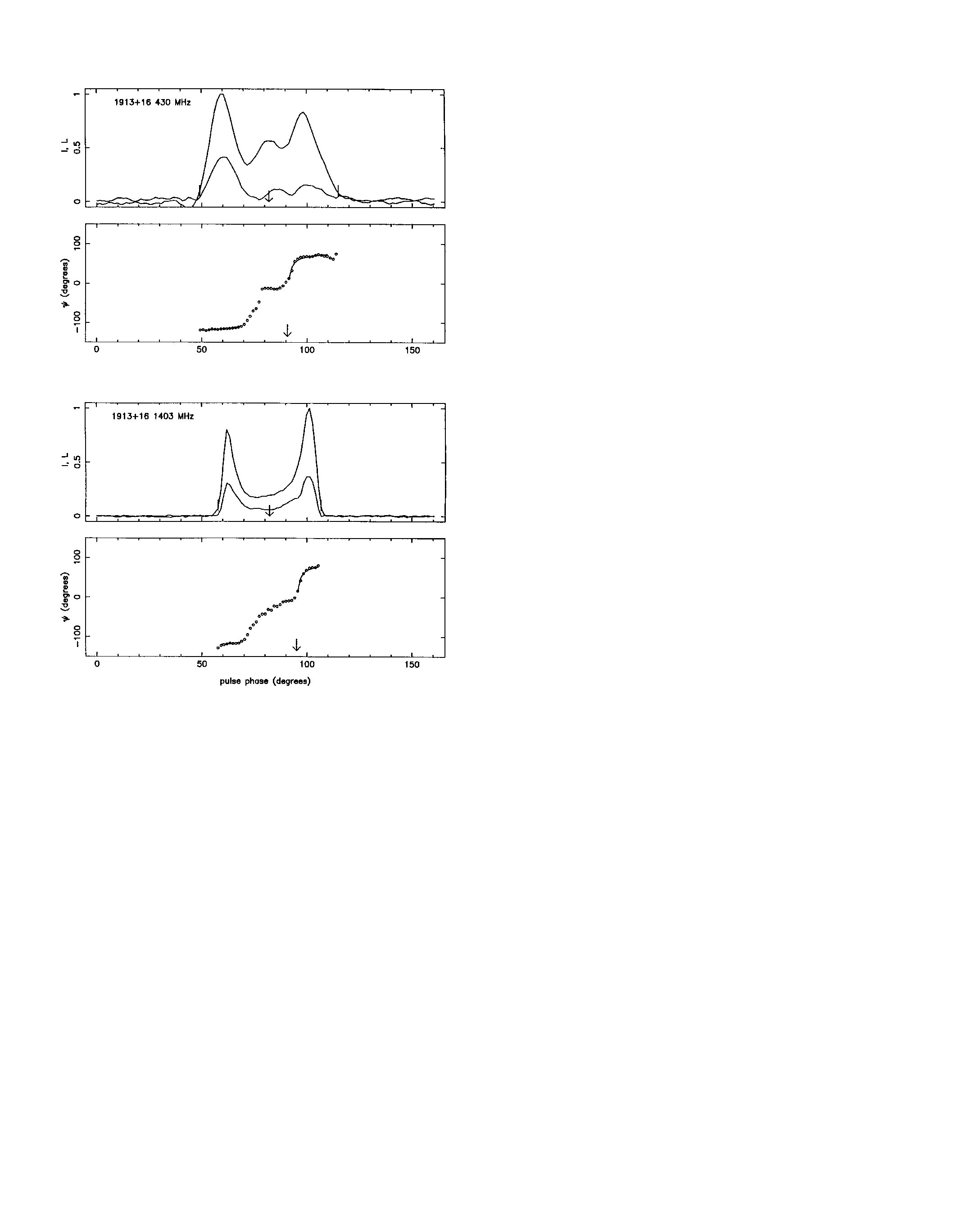}}}& \ \ \ 
{\mbox{\includegraphics[width=51mm]{plots/B1913+167P.ps}}}\\
{\mbox{\includegraphics[width=51mm]{plots/PQB1913+10.54538la.ps}}}&
{\mbox{\includegraphics[width=51mm]{plots/PQB1913+16.56171la.ps}}}& \ \ \
{\mbox{\includegraphics[width=51mm]{plots/B1913+167L.ps}}}\\
\end{tabular}
\caption{Scaled beam dimensions and average profiles for PSRs B1913+10 (430-MHz profile from \citet{GR78}), B1913+16 (profiles from \citet{Blaskiewic} and B1913+167 as in Fig~\ref{figA1}.}
\label{figA11}
\end{center}
\end{figure*}

\begin{figure*}
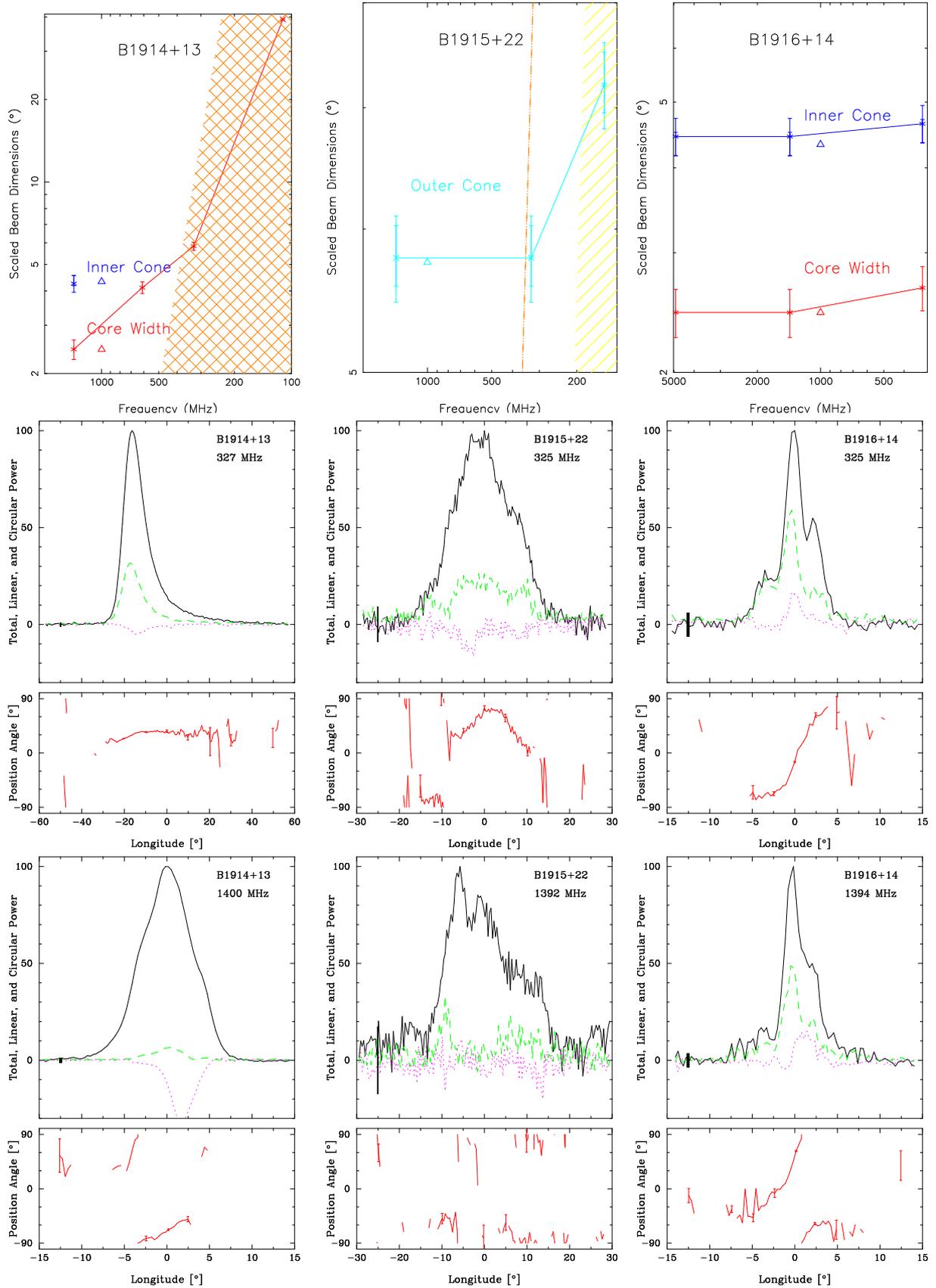

\begin{center}
\begin{tabular}{@{}ll@{}ll@{}}
{\mbox{\includegraphics[width=51mm]{plots/B1914+13_ABmodel.ps}}}&
{\mbox{\includegraphics[width=51mm]{plots/B1915+22_ABmodel.ps}}}& \ \ \ 
{\mbox{\includegraphics[width=51mm]{plots/B1916+14_ABmodel.ps}}}\\
{\mbox{\includegraphics[width=51mm]{plots/B1914+13P}}}&
{\mbox{\includegraphics[width=51mm]{plots/B1915+22P}}}& \ \ \ 
{\mbox{\includegraphics[width=51mm]{plots/B1916+14P.ps}}}\\
{\mbox{\includegraphics[width=51mm]{plots/B1914+13L.ps}}}&
{\mbox{\includegraphics[width=51mm]{plots/B1915+22L.ps}}}& \ \ \ 
{\mbox{\includegraphics[width=51mm]{plots/B1916+14L.ps}}}\\
\end{tabular}
\caption{Scaled beam dimensions and average profiles for PSRs B1914+13, B1915+22 and B1916+14 as in Fig~\ref{figA1}.}
\label{figA12}
\end{center}
\end{figure*}

\begin{figure*}
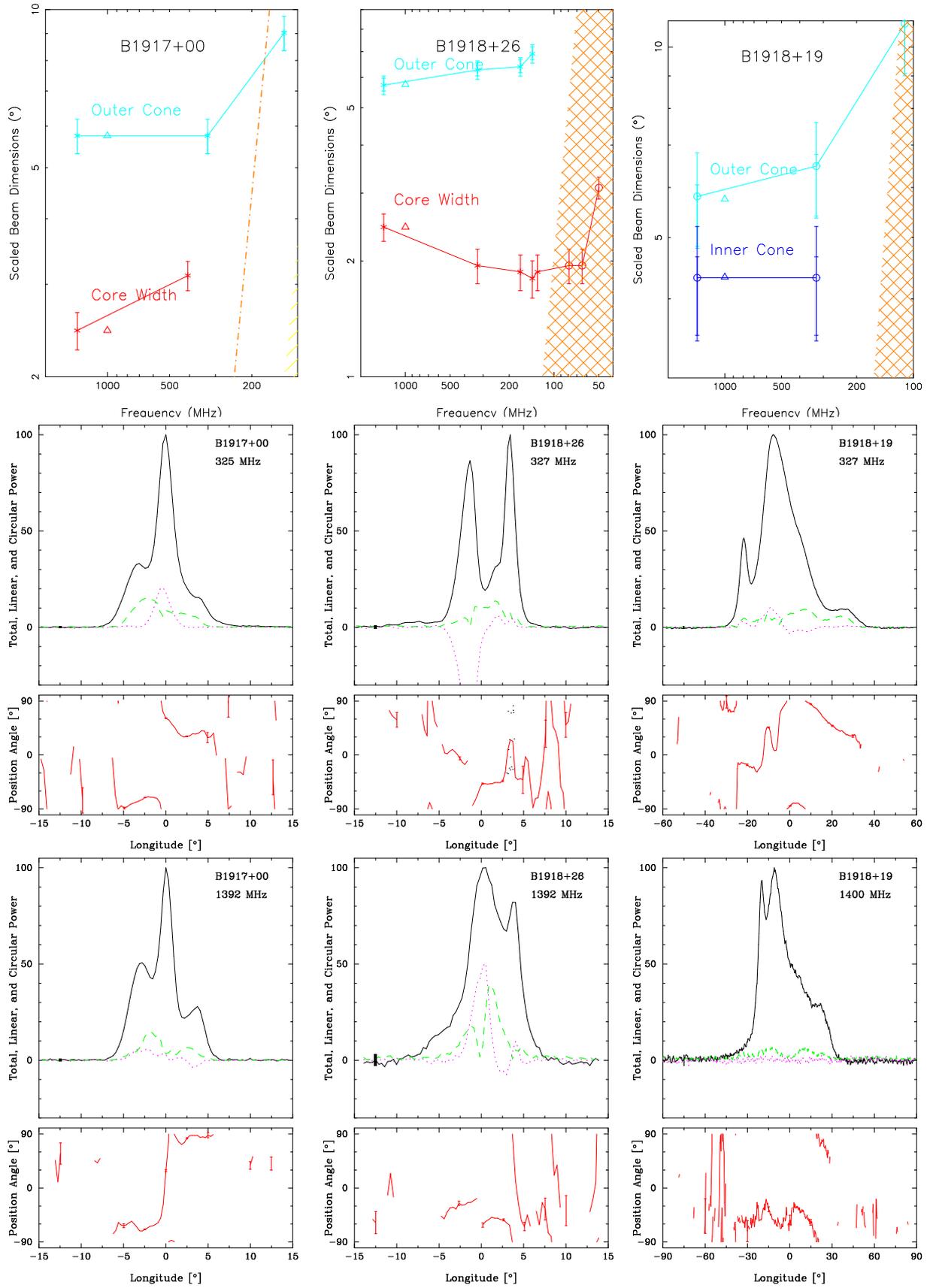

\begin{center}
\begin{tabular}{@{}ll@{}ll@{}}
{\mbox{\includegraphics[width=51mm]{plots/B1917+00_ABmodel.ps}}}&
{\mbox{\includegraphics[width=51mm]{plots/B1918+26_ABmodel.ps}}}& \ \ \ 
{\mbox{\includegraphics[width=51mm]{plots/B1918+19_ABmodel.ps}}}\\
{\mbox{\includegraphics[width=51mm]{plots/B1917+00P.ps}}}&
{\mbox{\includegraphics[width=51mm]{plots/B1918+26P.ps}}}& \ \ \ 
{\mbox{\includegraphics[width=51mm]{plots/B1918+19P.ps}}}\\
{\mbox{\includegraphics[width=51mm]{plots/B1917+00L.ps}}}&
{\mbox{\includegraphics[width=51mm]{plots/B1918+26L.ps}}}& \ \ \ 
{\mbox{\includegraphics[width=51mm]{plots/B1918+19L.ps}}} \\
\end{tabular}
\caption{Scaled beam dimensions and average profiles for PSRs B1917+00, B1918+26 and B1918+19 as in Fig~\ref{figA1}.}
\label{figA13}
\end{center}
\end{figure*}

\begin{figure*}
\begin{center}
\begin{tabular}{@{}ll@{}ll@{}}
{\mbox{\includegraphics[width=51mm]{plots/B1919+14_ABmodel.ps}}}&
{\mbox{\includegraphics[width=51mm]{plots/B1919+20_ABmodel.ps}}}& \ \ \ 
{\mbox{\includegraphics[width=51mm]{plots/B1920+20_ABmodel.ps}}}\\
{\mbox{\includegraphics[width=51mm]{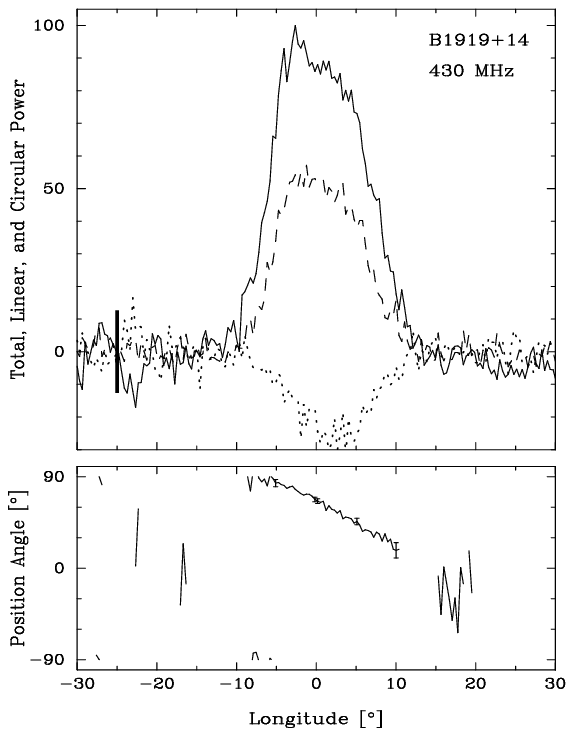}}}&
{\mbox{\includegraphics[width=51mm]{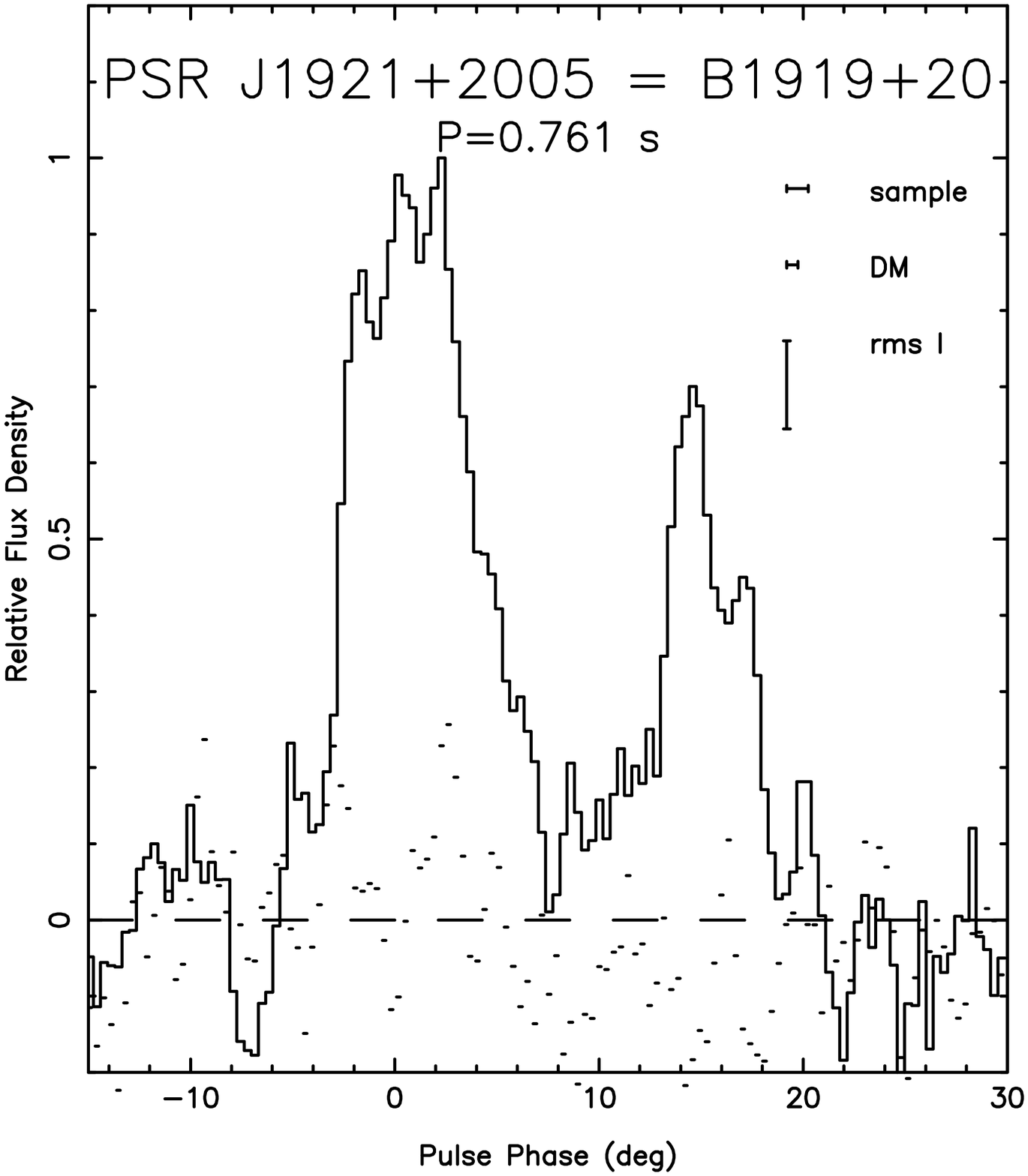}}}& \ \ \ 
{\mbox{\includegraphics[width=51mm]{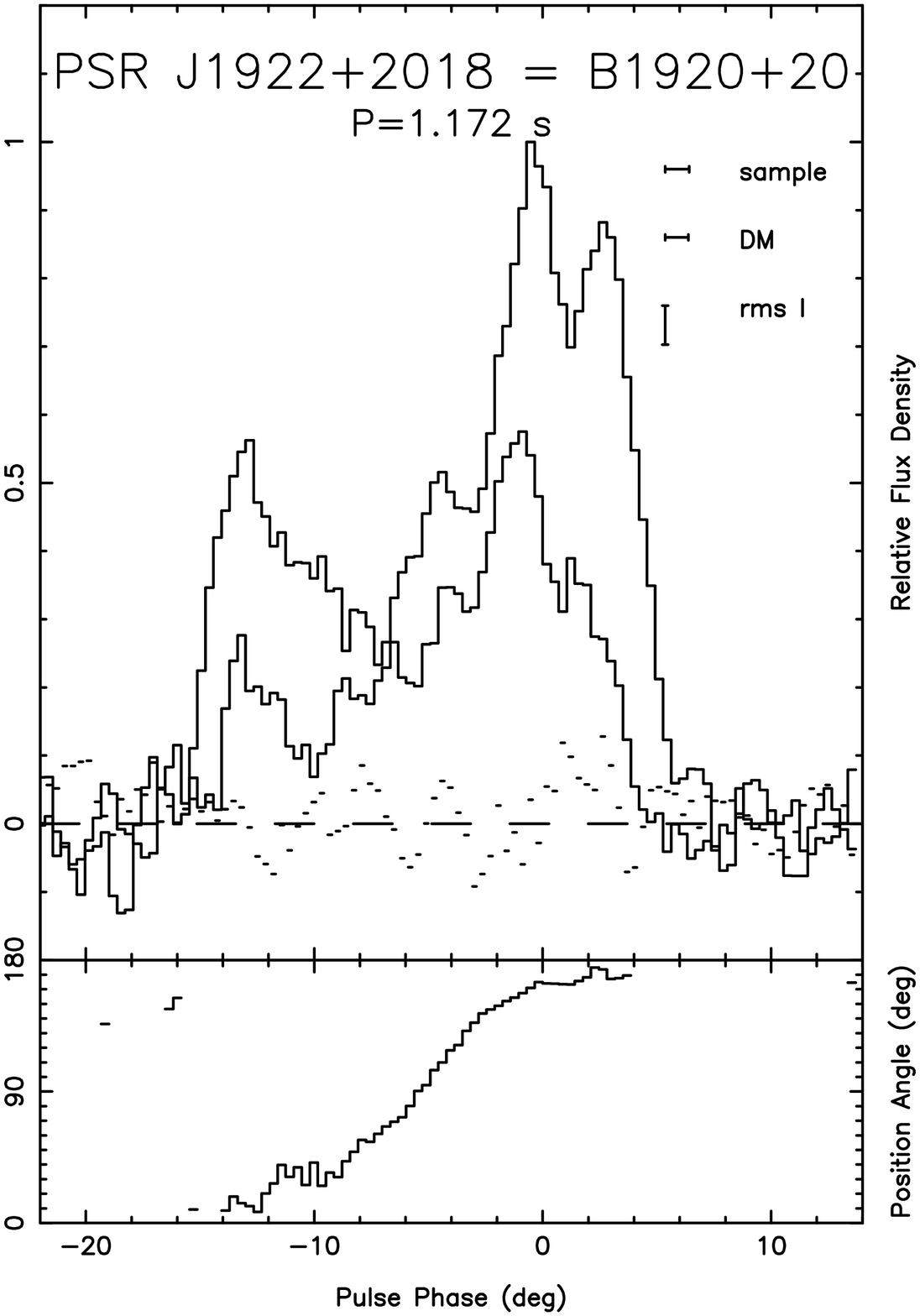}}}\\
{\mbox{\includegraphics[width=51mm]{plots/B1919+14L.ps}}}&
{\mbox{\includegraphics[width=51mm]{plots/PQB1919+20.56419la.ps}}}& \ \ \  
{\mbox{\includegraphics[width=51mm]{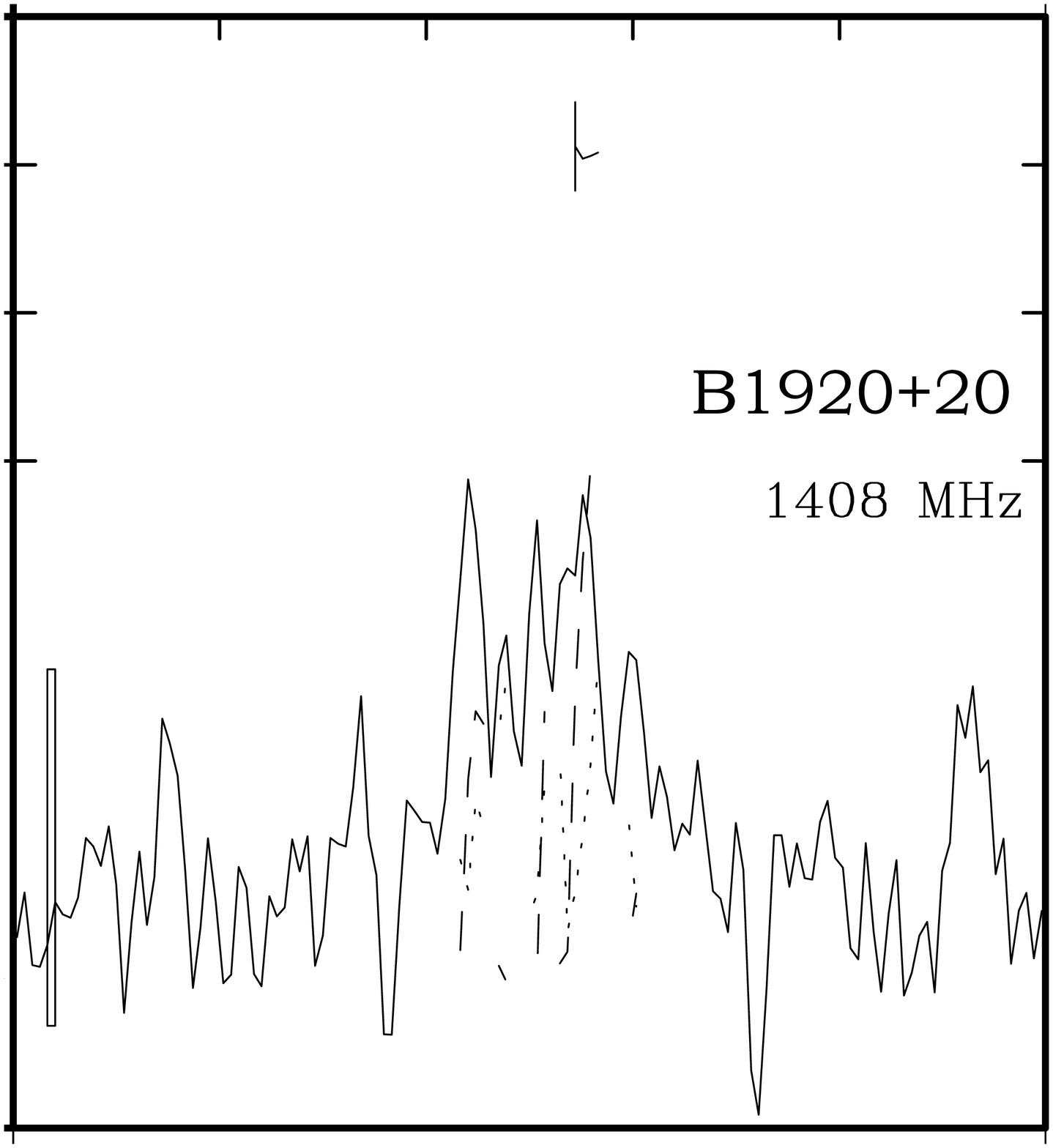}}}\\ 
\end{tabular}
\caption{Scaled beam dimensions and average profiles for PSRs B1919+14 (430-MHz profile from HR10), B1919+20, B1920+20 (430-MHz profile from W04; 50\degr\ 1.4-GHz from GL98) as in Fig~\ref{figA1}.}
\label{figA14}
\end{center}
\end{figure*}

\begin{figure*}
\begin{center}
\begin{tabular}{@{}ll@{}ll@{}}
{\mbox{\includegraphics[width=51mm]{plots/B1920+21_ABmodel.ps}}}&
{\mbox{\includegraphics[width=51mm]{plots/B1921+17_ABmodel.ps}}}&  \ \ \ 
{\mbox{\includegraphics[width=51mm]{plots/B1924+14_ABmodel.ps}}}\\
{\mbox{\includegraphics[width=51mm]{plots/B1920+21P.ps}}}&
{\mbox{\includegraphics[width=51mm]{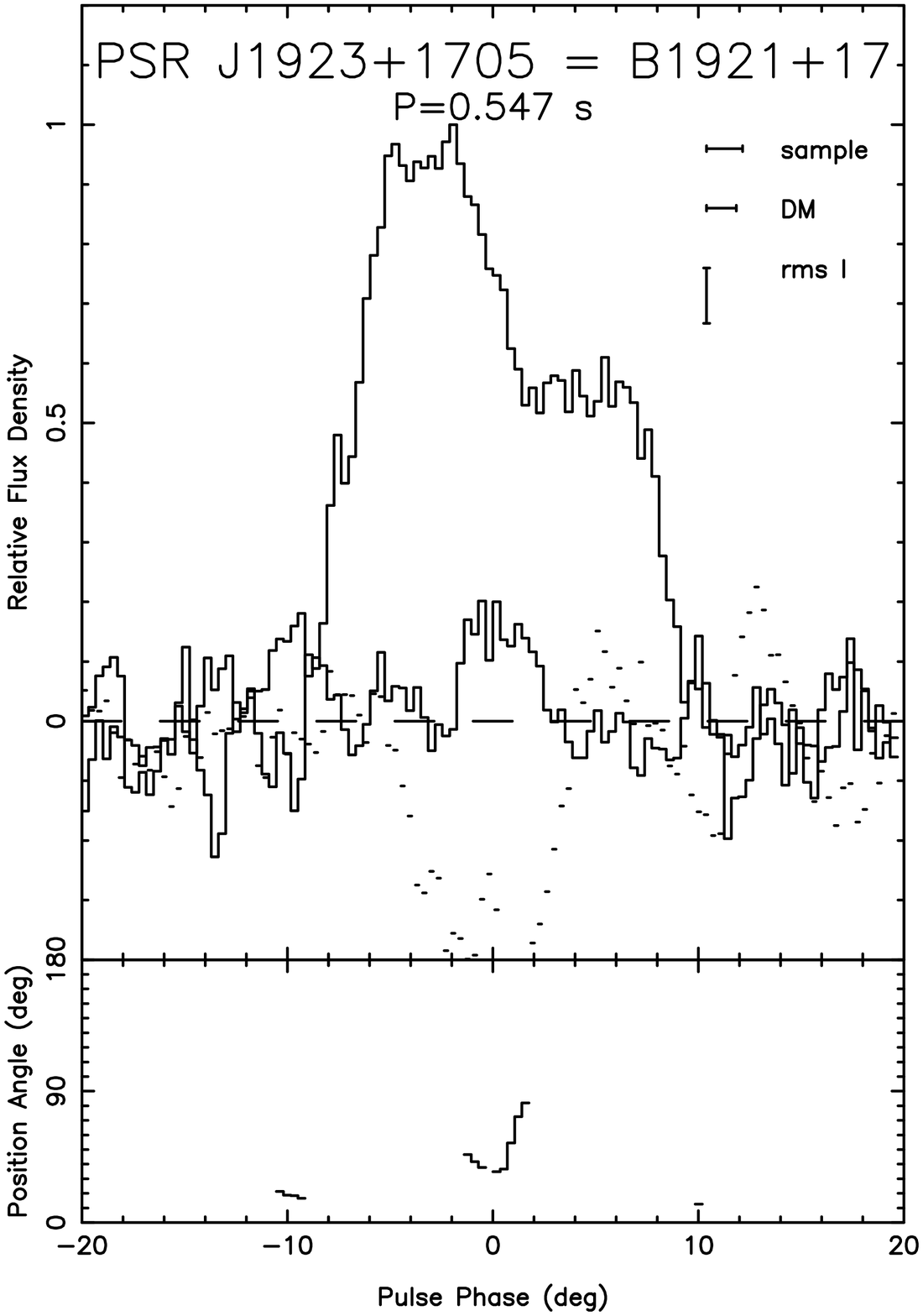}}}& \ \ \  
{\mbox{\includegraphics[width=51mm]{plots/B1924+14P.ps}}}\\
{\mbox{\includegraphics[width=51mm]{plots/B1920+21L.ps}}}& 
{\mbox{\includegraphics[width=51mm]{plots/PQB1921+17.56768la.ps}}}& \ \ \ 
{\mbox{\includegraphics[width=51mm]{plots/B1924+14L.ps}}}\\
\end{tabular}
\caption{Scaled beam dimensions and average profiles for PSRs B1920+21, B1921+17 (430-MHz profile from W04) and B1924+14 as in Fig~\ref{figA1}.}
\label{figA15}
\end{center}
\end{figure*}

\begin{figure*}
\begin{center}
\begin{tabular}{@{}ll@{}ll@{}}
{\mbox{\includegraphics[width=51mm]{plots/B1924+16_ABmodel.ps}}}&
{\mbox{\includegraphics[width=51mm]{plots/B1925+18_ABmodel.ps}}}& \ \ \ 
{\mbox{\includegraphics[width=51mm]{plots/B1925+22_ABmodel.ps}}}\\
{\mbox{\includegraphics[width=51mm]{plots/B1924+16P.ps}}}&
{\mbox{\includegraphics[width=51mm]{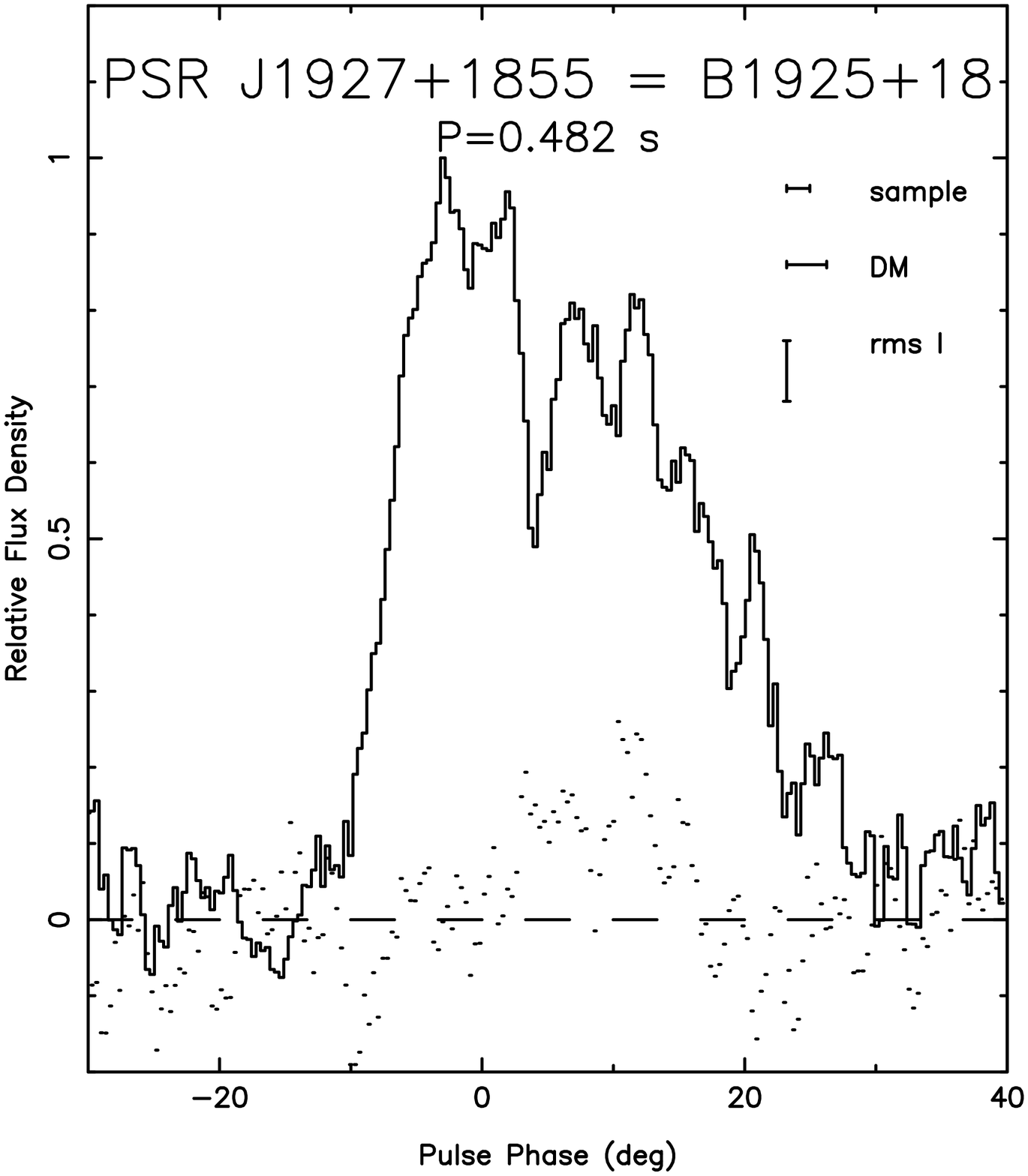}}}& \ \ \ 
{\mbox{\includegraphics[width=51mm]{plots/B1925+22P.ps}}}\\
{\mbox{\includegraphics[width=51mm]{plots/B1924+16L.ps}}}&
{\mbox{\includegraphics[width=51mm]{plots/PQB1925+18.56769la.ps}}}& \ \ \ 
{\mbox{\includegraphics[width=51mm]{plots/B1925+22L.ps}}}\\
\end{tabular}
\caption{Scaled beam dimensions and average profiles for PSRs B1924+16, B1925+18 (430-MHz profile from W04) and B1925+22 (1,4-GHz profile from \citet{W99}) as in Fig~\ref{figA1}.}
\label{figA16}
\end{center}
\end{figure*}

\begin{figure*}
\begin{center}
\begin{tabular}{@{}ll@{}ll@{}}
{\mbox{\includegraphics[width=51mm]{plots/B1926+18_ABmodel.ps}}}&
{\mbox{\includegraphics[width=51mm]{plots/B1927+13_ABmodel.ps}}}& \ \ \ 
{\mbox{\includegraphics[width=51mm]{plots/B1929+20_ABmodel.ps}}}\\
{\mbox{\includegraphics[width=51mm]{plots/PQB1926+18.53967p.ps}}}&
{\mbox{\includegraphics[width=51mm]{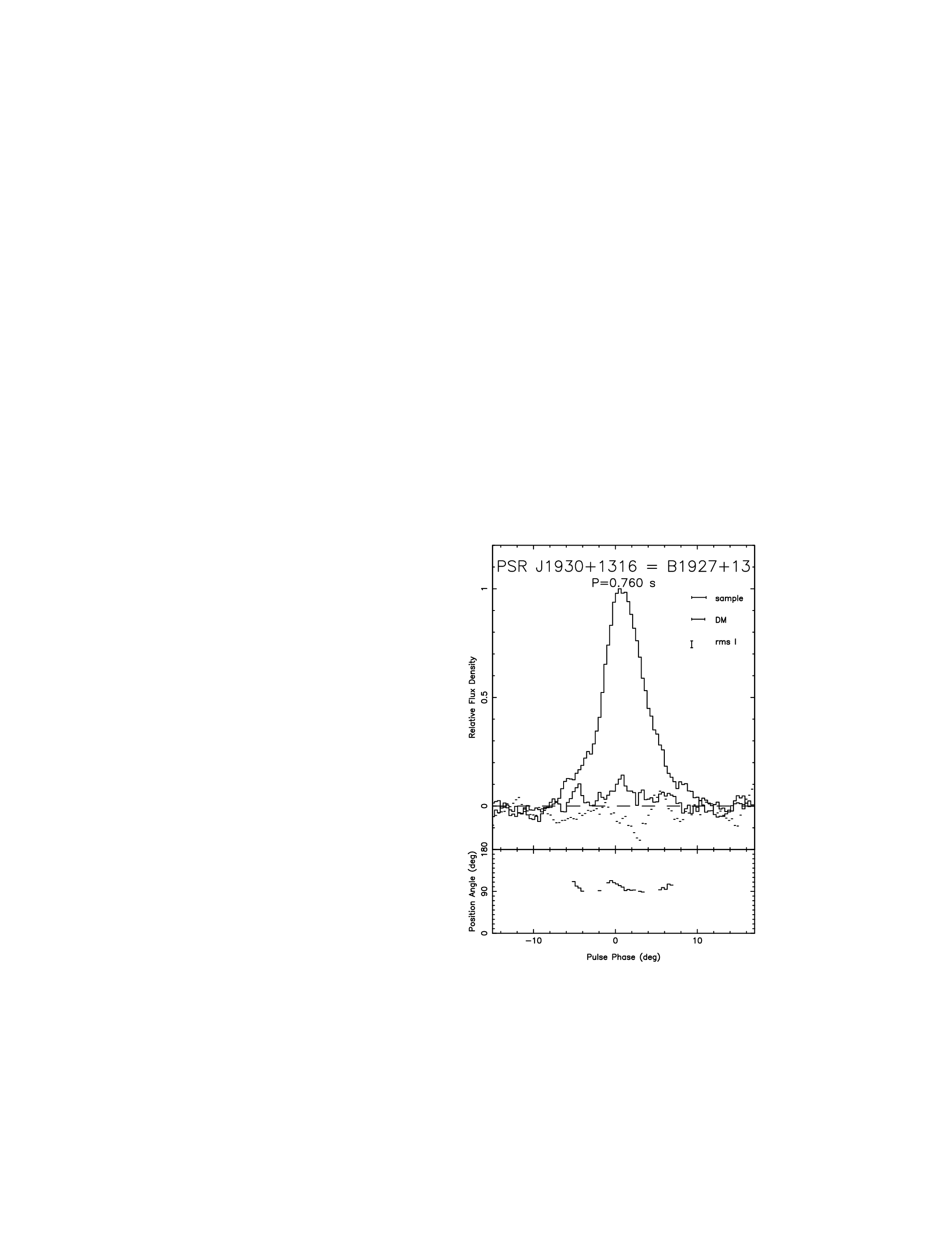}}}& \ \ \ 
{\mbox{\includegraphics[width=51mm]{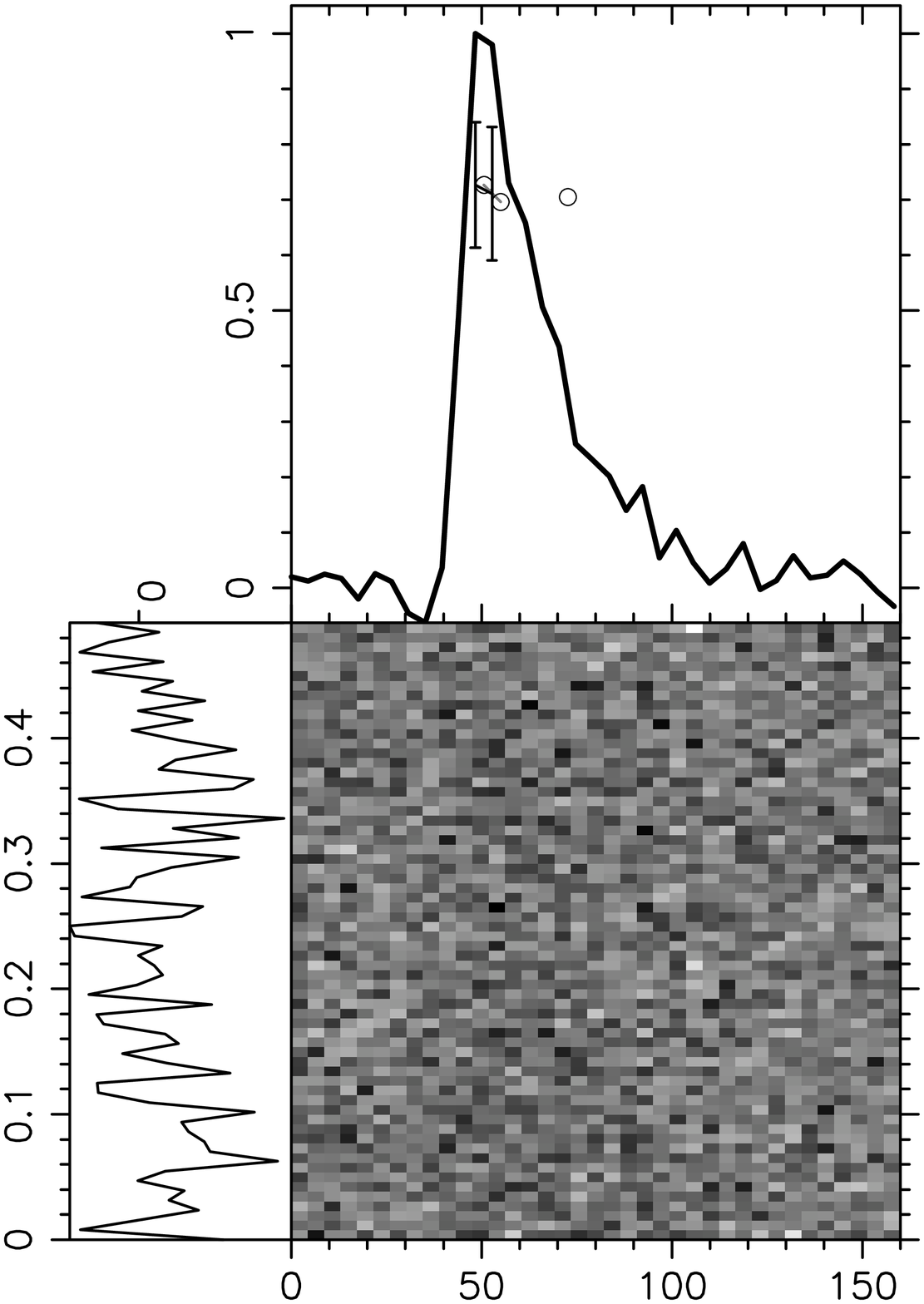}}}\\ 
{\mbox{\includegraphics[width=51mm]{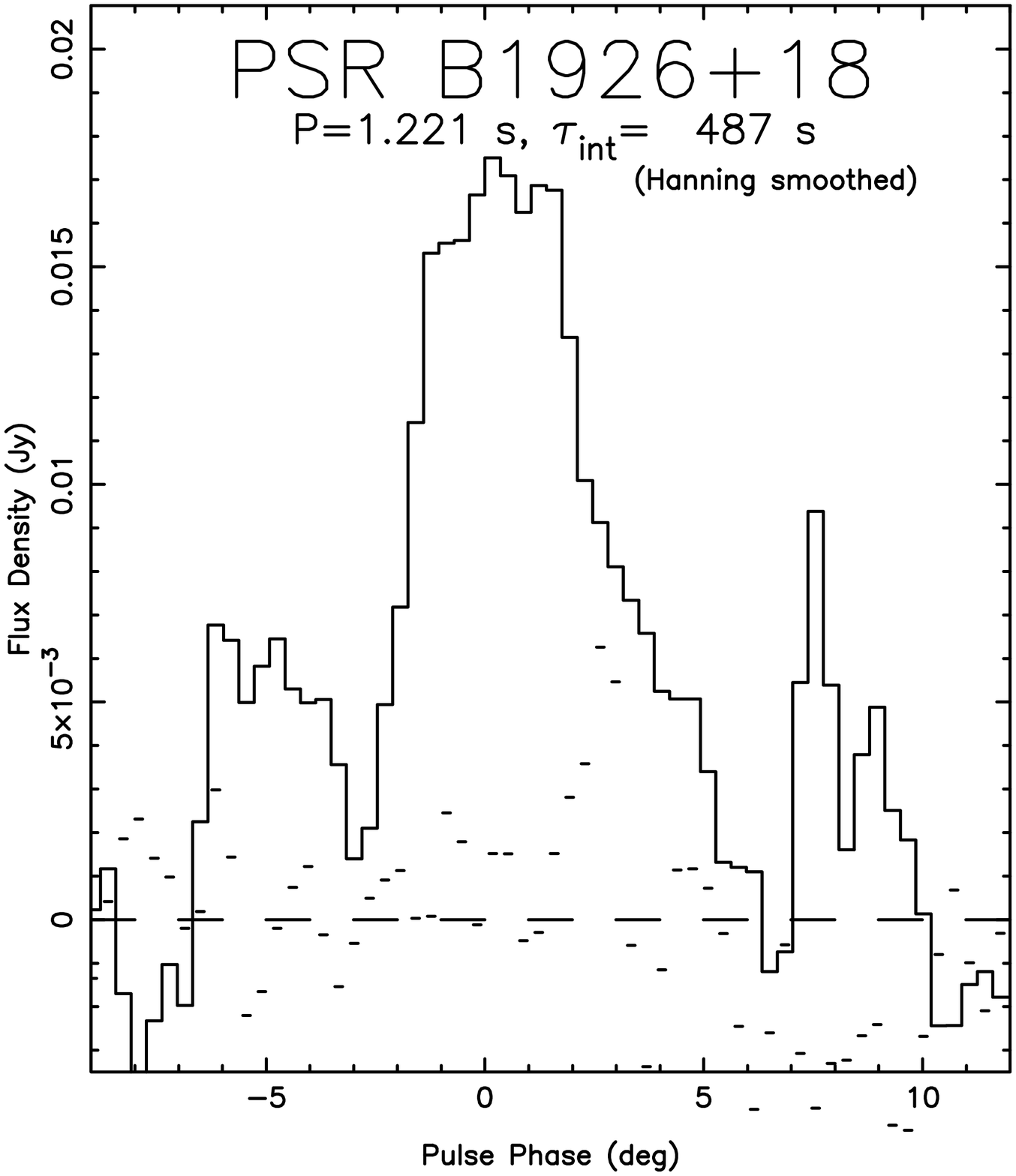}}}&
{\mbox{\includegraphics[width=51mm]{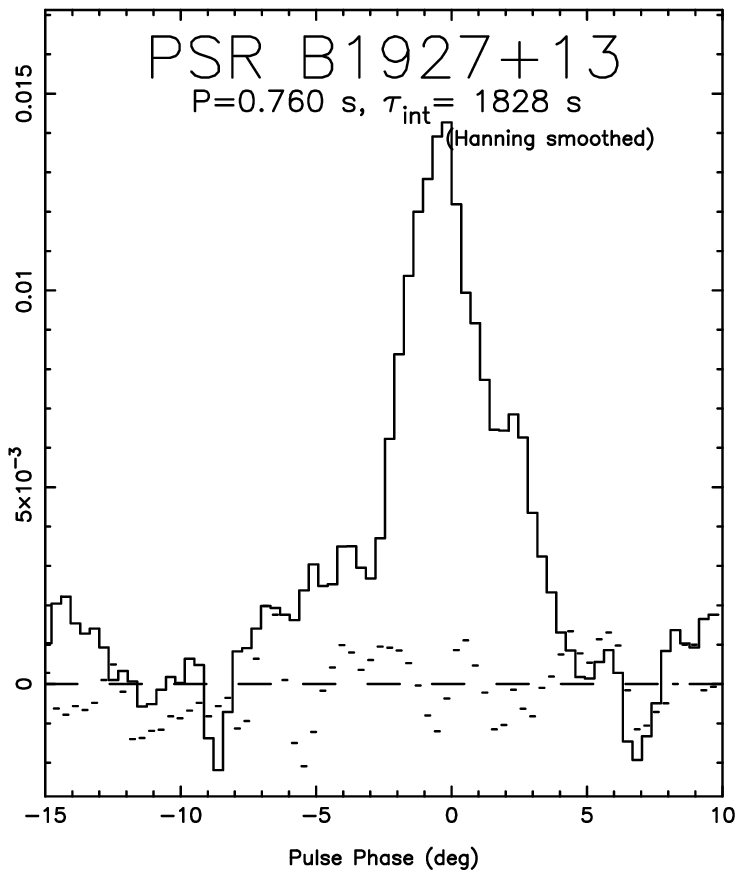}}}& \ \ \ 
{\mbox{\includegraphics[width=51mm]{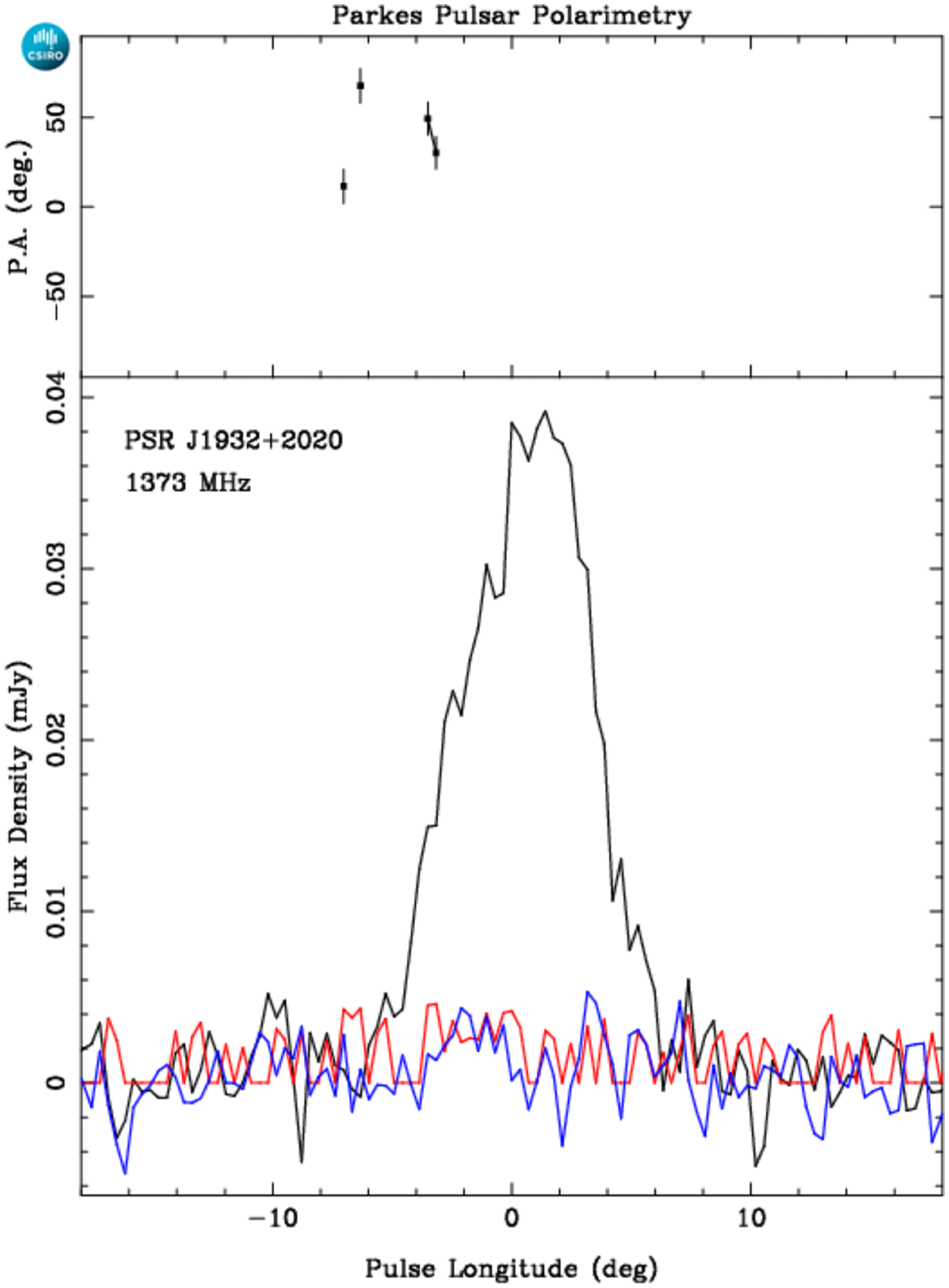}}}\\
\end{tabular}
\caption{Scaled beam dimensions and average profiles for PSRs B1926+18, B1927+13 (profiles from \citet{W99,Weisberg2004} and B1929+20 [327-MHz and 1.4-GHz profiles from \citet{Weltevrede2007,JK18}] as in Fig~\ref{figA1}.}
\label{figA17}
\end{center}
\end{figure*}

\begin{figure*}
\begin{center}
\begin{tabular}{@{}ll@{}ll@{}}
{\mbox{\includegraphics[width=51mm]{plots/B1930+22_ABmodel.ps}}}&
{\mbox{\includegraphics[width=51mm]{plots/B1930+13_ABmodel.ps}}}& \ \ \ 
{\mbox{\includegraphics[width=51mm]{plots/B1931+24_ABmodel.ps}}}\\
{\mbox{\includegraphics[width=51mm]{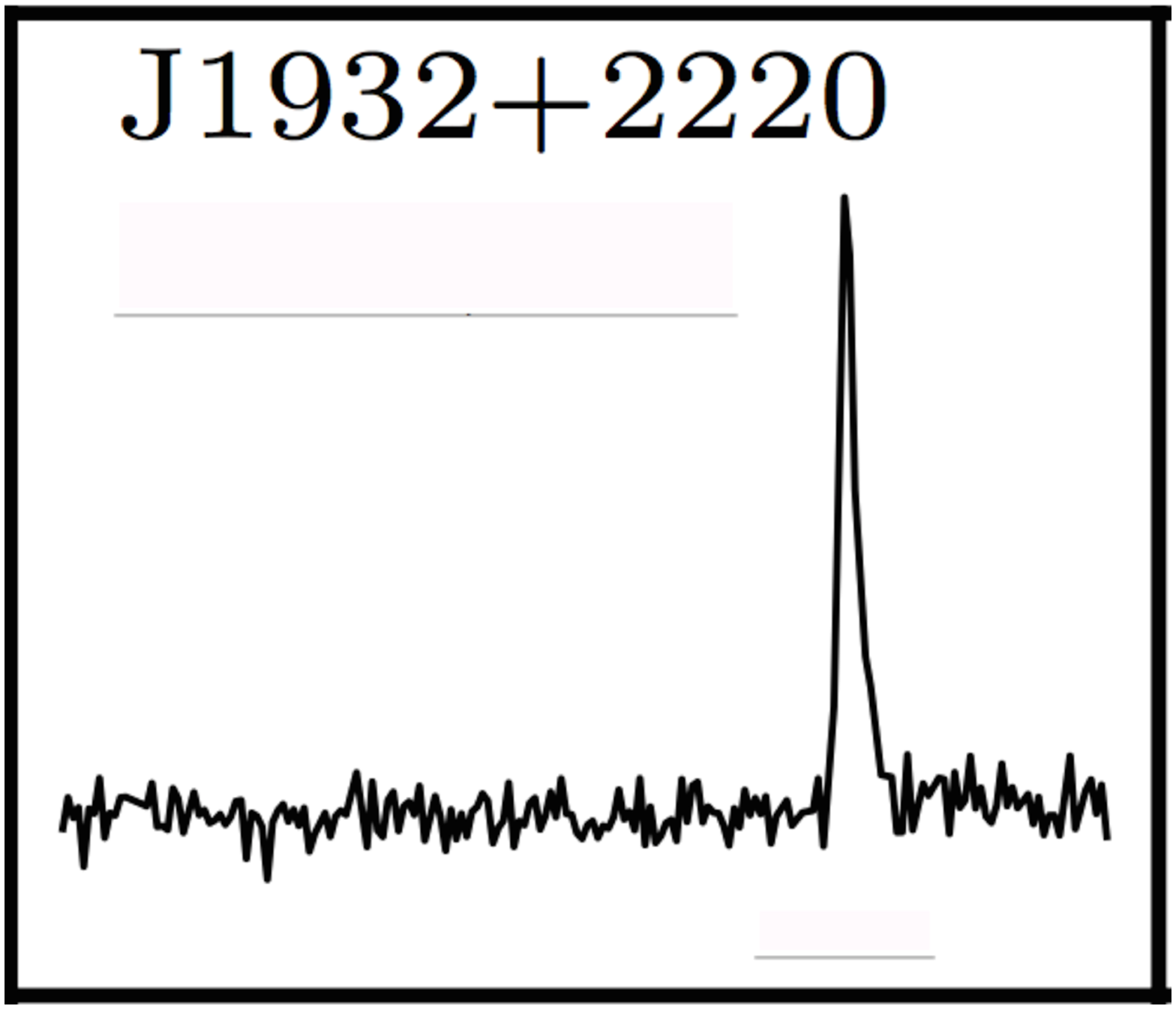}}}&
{\mbox{\includegraphics[width=51mm]{plots/B1930+13P.ps}}}& \ \ \ 
{\mbox{\includegraphics[width=51mm]{plots/B1931+24P.ps}}}\\ 
{\mbox{\includegraphics[width=51mm]{plots/B1930+22L.ps}}}&
{\mbox{\includegraphics[width=51mm]{plots/B1930+13L.ps}}}& \ \ \ 
{\mbox{\includegraphics[width=51mm]{plots/B1931+24L.ps}}}\\
\end{tabular}
\caption{Scaled beam dimensions and average profiles for PSRs B1930+22/J1932+2220 (350-MHz profile from \citet{mcewen}, B1930+13 and B1931+24 as in Fig~\ref{figA1}.}
\label{figA18}
\end{center}
\end{figure*}

\begin{figure*}
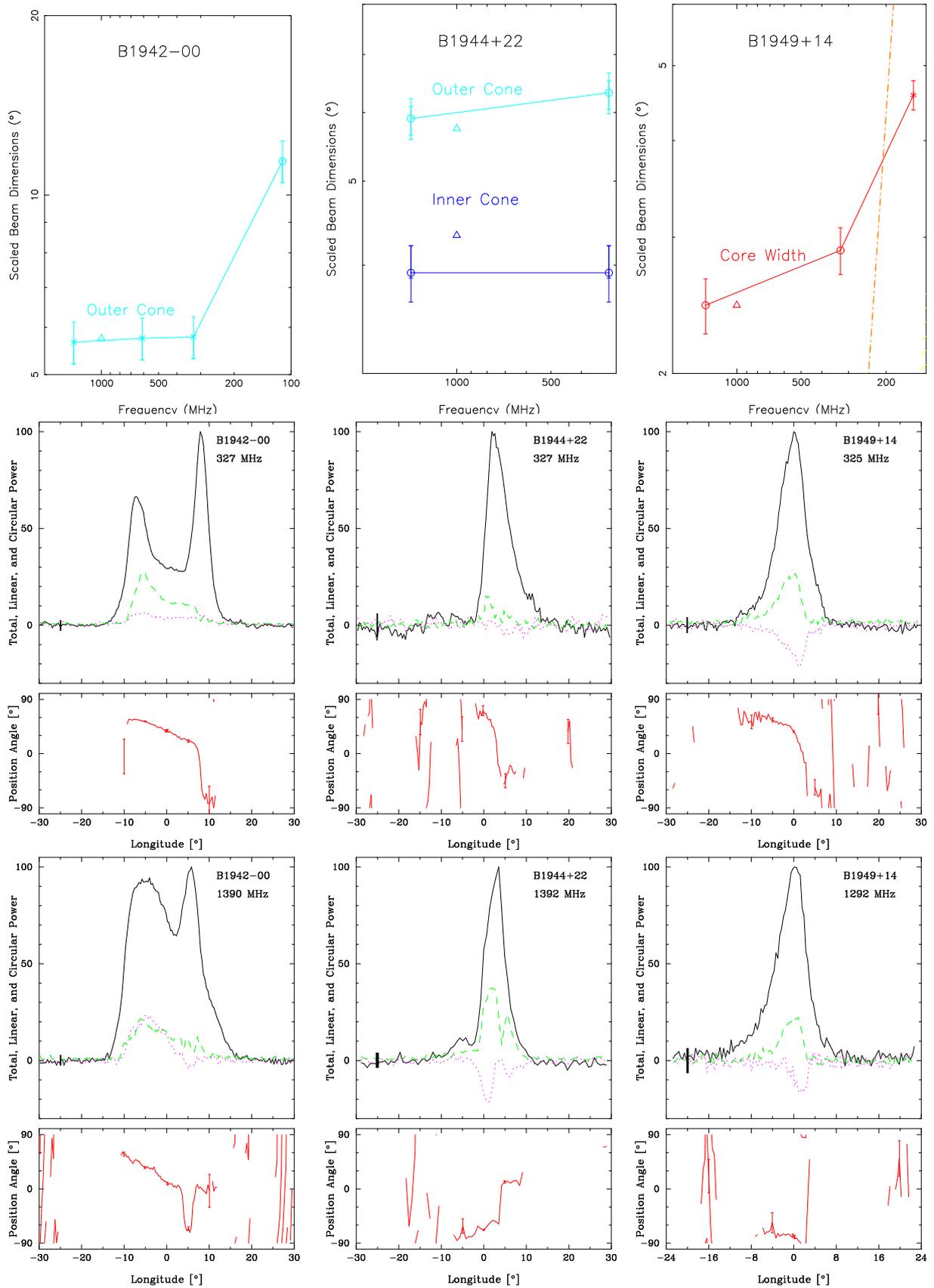

\begin{center}
\begin{tabular}{@{}ll@{}ll@{}}
{\mbox{\includegraphics[width=51mm]{plots/B1942-00_ABmodel.ps}}}&
{\mbox{\includegraphics[width=51mm]{plots/B1944+22_ABmodel.ps}}}& \ \ \  
{\mbox{\includegraphics[width=51mm]{plots/B1949+14_ABmodel.ps}}}\\ 
{\mbox{\includegraphics[width=51mm]{plots/B1942-00P.ps}}}&
{\mbox{\includegraphics[width=51mm]{plots/B1944+22P.ps}}}& \ \ \ 
{\mbox{\includegraphics[width=51mm]{plots/B1949+14P.ps}}}\\
{\mbox{\includegraphics[width=51mm]{plots/B1942-00L.ps}}}& 
{\mbox{\includegraphics[width=51mm]{plots/B1944+22L.ps}}}& \ \ \ 
{\mbox{\includegraphics[width=51mm]{plots/B1949+14L.ps}}}\\
\end{tabular}
\caption{Scaled beam dimensions and average profiles for PSRs B1942--00, B1944+22 and B1949+14 as in Fig~\ref{figA1}.}
\label{figA19}
\end{center}
\end{figure*}

\begin{figure*}
\begin{center}
\begin{tabular}{@{}ll@{}ll@{}}
{\mbox{\includegraphics[width=51mm]{plots/B1951+32_ABmodel.ps}}}&
{\mbox{\includegraphics[width=51mm]{plots/B1953+29_ABmodel.ps}}}& \ \ \ 
{\mbox{\includegraphics[width=51mm]{plots/B2000+32_ABmodel.ps}}}\\
{\mbox{\includegraphics[width=50mm]{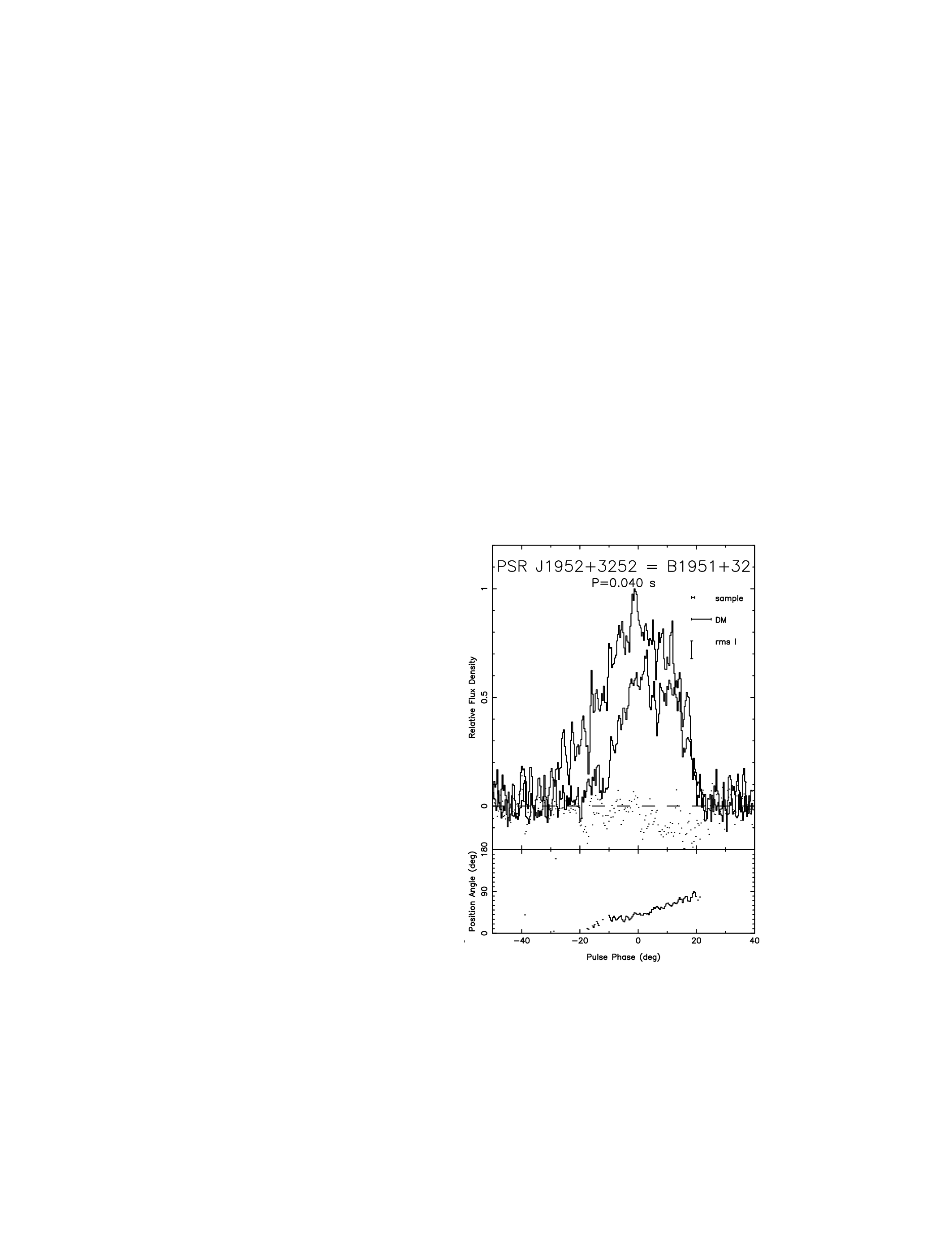}}}&
{\mbox{\includegraphics[width=51mm]{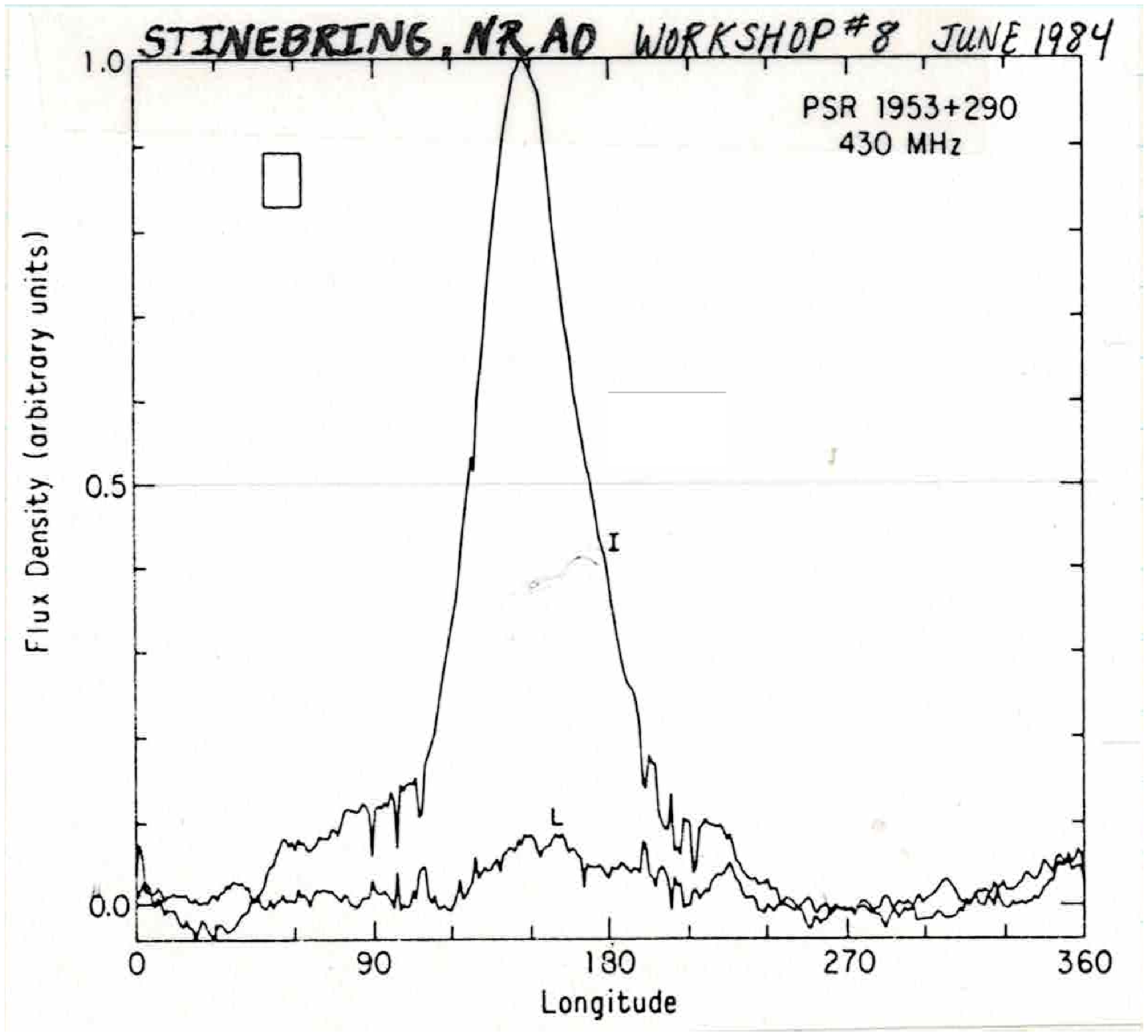}}}& \ \ \ 
{\mbox{\includegraphics[width=51mm]{plots/B2000+32P.ps}}}\\
{\mbox{\includegraphics[width=51mm]{plots/B1951+32L.ps}}}&
{\mbox{\includegraphics[width=51mm]{plots/PQB1953+29.56585la.ps}}}& \ \ \ 
{\mbox{\includegraphics[width=51mm]{plots/B2000+32L.ps}}}\\
\end{tabular}
\caption{Scaled beam dimensions and average profiles for PSRs B1951+32/J1952+3252 (430-MHz profile from W04), B1953+29 (430-MHz profile from \citet{sbc+84}) and B2000+32 as in Fig~\ref{figA1}.}
\label{figA20}
\end{center}
\end{figure*}

\begin{figure*}
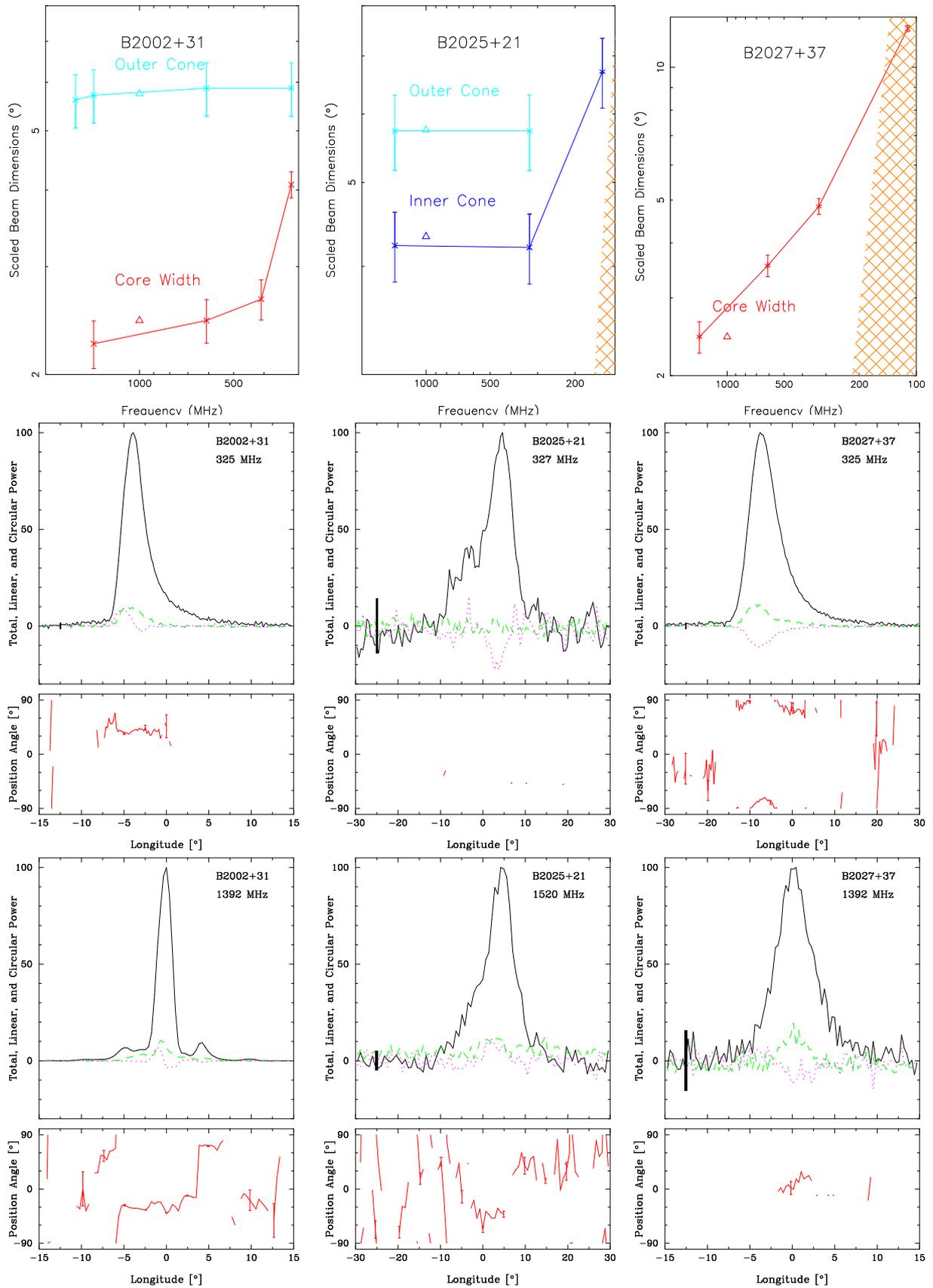

\begin{center}
\begin{tabular}{@{}ll@{}ll@{}}
{\mbox{\includegraphics[width=51mm]{plots/B2002+31_ABmodel.ps}}}&
{\mbox{\includegraphics[width=51mm]{plots/B2025+21_ABmodel.ps}}}& \ \ \ 
{\mbox{\includegraphics[width=51mm]{plots/B2027+37_ABmodel.ps}}}\\
{\mbox{\includegraphics[width=51mm]{plots/B2002+31P.ps}}}&
{\mbox{\includegraphics[width=51mm]{plots/B2025+21P.ps}}}& \ \ \  
{\mbox{\includegraphics[width=51mm]{plots/B2027+37P.ps}}}\\
{\mbox{\includegraphics[width=51mm]{plots/B2002+31L.ps}}}&
{\mbox{\includegraphics[width=51mm]{plots/B2025+21L.ps}}}& \ \ \ 
{\mbox{\includegraphics[width=51mm]{plots/B2027+37L.ps}}}\\
\end{tabular}
\caption{Scaled beam dimensions and average profiles for PSRs B2002+31, B2025+21 and B2027+37 as in Fig~\ref{figA1}.}
\label{figA21}
\end{center}
\end{figure*}

\begin{figure*}
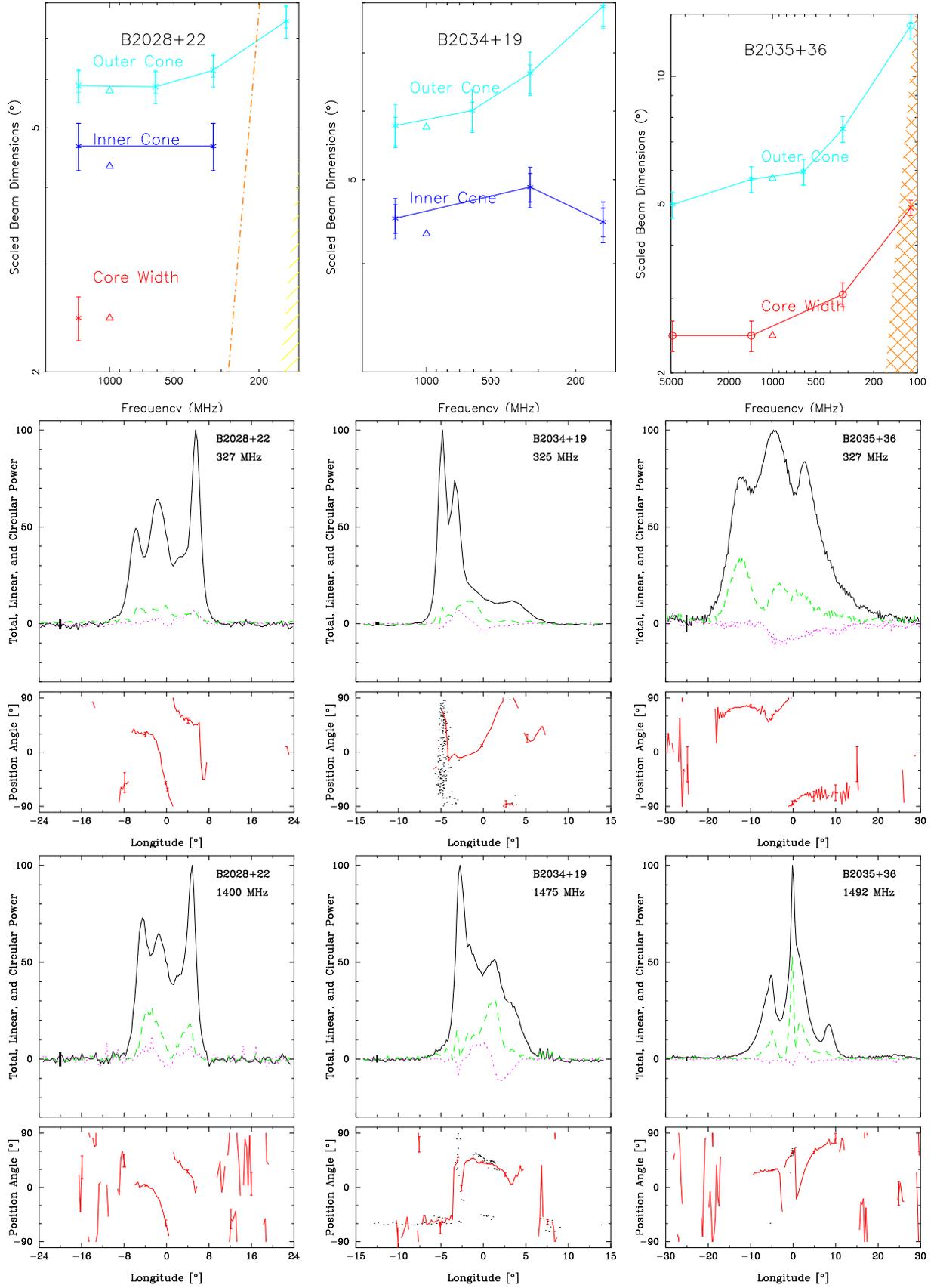

\begin{center}
\begin{tabular}{@{}ll@{}ll@{}}
{\mbox{\includegraphics[width=51mm]{plots/B2028+22_ABmodel.ps}}}&
{\mbox{\includegraphics[width=51mm]{plots/B2034+19_ABmodel.ps}}}& \ \ \ 
{\mbox{\includegraphics[width=51mm]{plots/B2035+36_ABmodel.ps}}}\\
{\mbox{\includegraphics[width=51mm]{plots/B2028+22P.ps}}}&
{\mbox{\includegraphics[width=51mm]{plots/B2034+19P.ps}}}& \ \ \ 
{\mbox{\includegraphics[width=51mm]{plots/B2035+36P.ps}}}\\
{\mbox{\includegraphics[width=51mm]{plots/B2028+22L.ps}}}&
{\mbox{\includegraphics[width=51mm]{plots/B2034+19L.ps}}}& \ \ \ 
{\mbox{\includegraphics[width=51mm]{plots/B2035+36L.ps}}}\\
\end{tabular}
\caption{Scaled beam dimensions and average profiles for PSRs B2028+22, B2034+19 and B2035+36 as in Fig~\ref{figA1}.}
\label{figA22}
\end{center}
\end{figure*}

\begin{figure*}
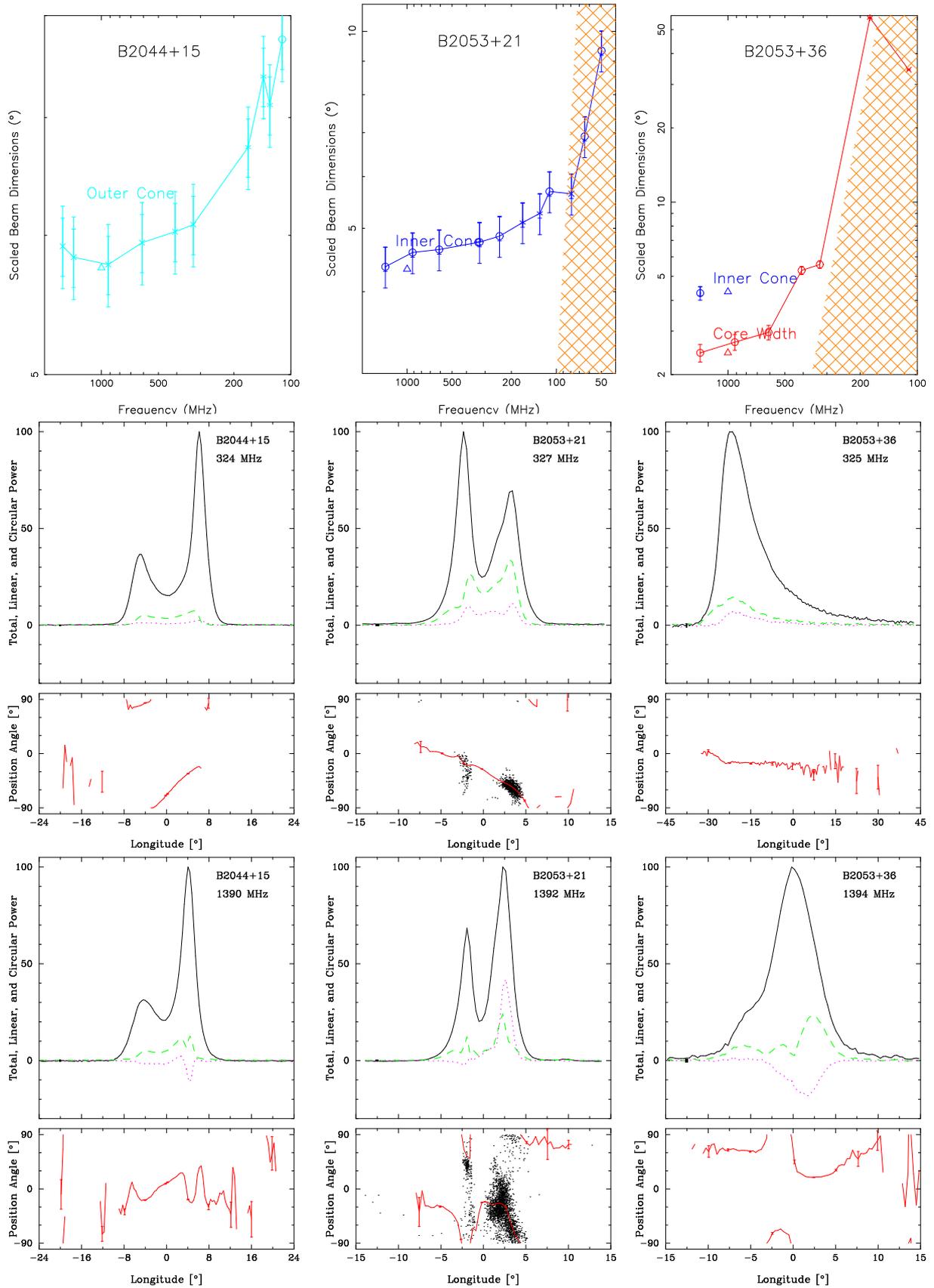

\begin{center}
\begin{tabular}{@{}ll@{}ll@{}}
{\mbox{\includegraphics[width=51mm]{plots/B2044+15_ABmodel.ps}}}&
{\mbox{\includegraphics[width=51mm]{plots/B2053+21_ABmodel.ps}}}& \ \ \ 
{\mbox{\includegraphics[width=51mm]{plots/B2053+36_ABmodel.ps}}}\\
{\mbox{\includegraphics[width=51mm]{plots/B2044+15P.ps}}}&
{\mbox{\includegraphics[width=51mm]{plots/B2053+21P.ps}}}& \ \ \ 
{\mbox{\includegraphics[width=51mm]{plots/B2053+36P.ps}}}\\
{\mbox{\includegraphics[width=51mm]{plots/B2044+15L.ps}}}&
{\mbox{\includegraphics[width=51mm]{plots/B2053+21L.ps}}}& \ \ \ 
{\mbox{\includegraphics[width=51mm]{plots/B2053+36L.ps}}}\\
\end{tabular}
\caption{Scaled beam dimensions and average profiles for PSRs B2044+15, B2053+21 and B2053+36 as in Fig~\ref{figA1}.}
\label{figA23}
\end{center}
\end{figure*}

\begin{figure*}
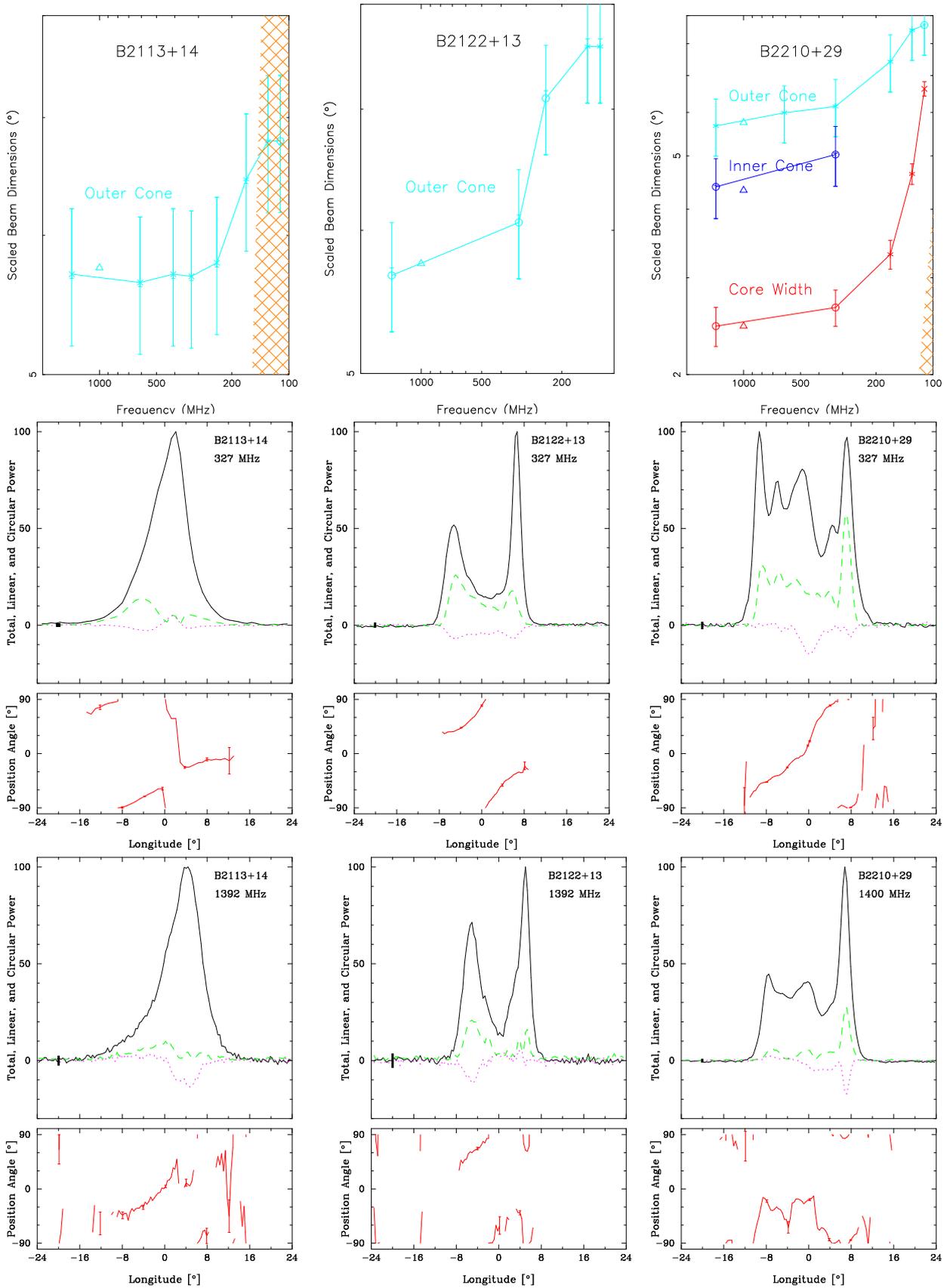

\begin{center}
\begin{tabular}{@{}ll@{}ll@{}}
{\mbox{\includegraphics[width=51mm]{plots/B2113+14_ABmodel.ps}}}&
{\mbox{\includegraphics[width=51mm]{plots/B2122+13_ABmodel.ps}}}& \ \ \ 
{\mbox{\includegraphics[width=51mm]{plots/B2210+29_ABmodel.ps}}}\\
{\mbox{\includegraphics[width=51mm]{plots/B2113+14P.ps}}}& 
{\mbox{\includegraphics[width=51mm]{plots/B2122+13P.ps}}}& \ \ \ 
{\mbox{\includegraphics[width=51mm]{plots/B2210+29P.ps}}}\\
{\mbox{\includegraphics[width=51mm]{plots/B2113+14L.ps}}}& \ \ \ 
{\mbox{\includegraphics[width=51mm]{plots/B2122+13L.ps}}}& \ \ \ 
{\mbox{\includegraphics[width=51mm]{plots/B2210+29L.ps}}}\\
\end{tabular}
\caption{Scaled beam dimensions and average profiles for B2113+14, PSRs B2113+14 and B2210+29 as in Fig~\ref{figA1}.}
\label{figA24}
\end{center}
\end{figure*}

\begin{figure}
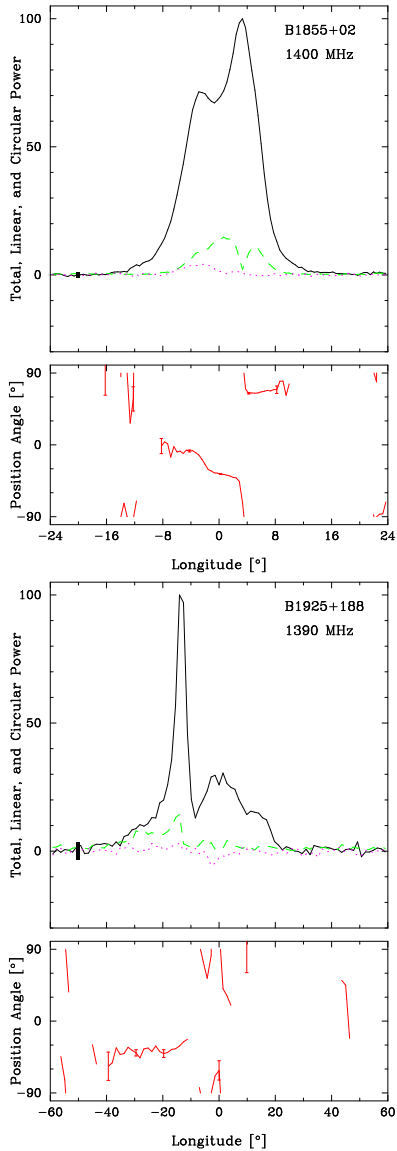

\begin{center}
\begin{tabular}{@{}ll@{}ll@{}}
{\mbox{\includegraphics[width=51mm]{plots/PQB1855+02.52739la.ps}}} \\
{\mbox{\includegraphics[width=51mm]{plots/PQB1925+188.56769la.ps}}} \\
\end{tabular}
\caption{Pulsars with available observations only at 1.4 GHz}
\label{figA30}
\end{center}
\end{figure}

\noindent\textit{\textbf{B1904+06}} has a classic core-cone triple \textbf{T} profile with a well defined PPA traverse as studied by \citet{W99} as well as \citet{GL98} and \citet{JK18}.  However, none of GL98's lower frequency profiles show the evolution clearly, and their width estimates used here with attempted corrections are probably far off.  We model the conal profile as an outer cone, and find that the core would have a width of about 9\degr\ which is quite plausible---and this is confirmed qualitatively by the \citet{von_thesis} 4.9-GHz EPN profile which shows the core clearly---however, the observations is very poorly resolved and no meaningful correction seems possible.  We thus presume that the 4.9-GHz profile dimensions are similar to those as 1.4 GHz.
\vskip 0.1in

\noindent\textit{\textbf{B1906+09}}:  The pulsar may well have a wide conal double profile, but none of the observations are sensitive enough to define either the widths or PPA rate.  The best \citep{Weisberg2004,GR78} show two components, whereas the GL98 detections are so marginal that only the brighter first component is shown.  Following ET VI (which had similar complaints) this is almost certainly a conal double structure but without more than a guess at the PPA rate a quantitative mode is meaningless.   
\vskip 0.1in

\noindent\textit{\textbf{B1907+00}}: The profiles show a clear triple \textbf{T} structure, and we model it with a core-cone triple beam geometry. 
Little can be inferred from the profile regarding the PPA rate, so we assume a central sightline passage.   
\vskip 0.1in
\noindent\textit{\textbf{B1907+02}}: In \citet{rankin1993a}, this pulsar was classified as a core single, but in that we see conal outriders at 327 MHz, it is more likely a triple \textbf{T} profile.  We model it as such and find that the MM10 detection at 111 MHz is highly scattered.  

\begin{figure}
\begin{center}
\includegraphics[height=85mm,angle=-90.]{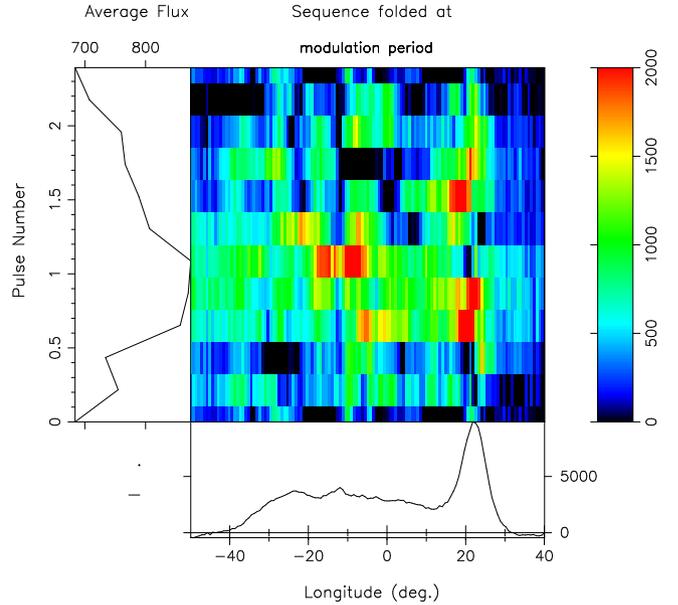}
\caption{Pulsar B1907+03 pulse sequence folded at a 2.4-$P$ modulation with the unvarying ``base'' removed.  Different sections of the profile are illuminated over the cycle as is usual for conal modulations.  The resolution across the pulsar's broad profile is unusually course with only 11 bins displayed.
}
\label{figA105a}
\end{center}
\end{figure}
\vskip 0.1in
\noindent\textit{\textbf{B1907+03}}: This pulsar's profile is modulated such that different regions are bright at different phases in its cycle supporting the conal triple or quadruples identification. 
Fig~\ref{figA105a} shows a 2.4-$P$ modulation, but in other intervals a 9.6-$P$ cycle is seen.  This shows that the pulsar has a conal triple or quadruple \textbf{T/Q} profile, but the overall poor quality of many profiles make it difficult to see the evolution of its inner structure clearly.  
\vskip 0.1in

\noindent\textit{\textbf{B1907+10}}: The profile of PSR B1907+10 is difficult to interpret.  Earlier ET VI regarded it as having a core \textbf{S$_t$} with only the leading conal outrider visible.  \citet{WSE07}'s fluctuation spectrum shows a feature at about 15-$P$ (as does our own), but this seems to reflect the amplitude modulation of short bursts of pulses, as there is no hint of phase modulation.  The core may overlie the trailing outrider at 1.4 GHz (as we surmise), and the ``tail'' at 327 MHz is not scattering, so may be conal.  Therefore we model it with a triple \textbf{T} geometry. The 143-MHz LOFAR observation shows a clear scattering ``tail''---but the MM10 profile does not, so it is not clear how to interpret it.
\vskip 0.1in

\noindent\textit{\textbf{B1907+12}}: The available profiles \citep{sgg+95,Weisberg2004} of this weak pulsar including our own hardly justify a model, both because the PPA rate is poorly determined and because scattering is increasingly prominent below 1 GHz.  Nonetheless, our narrow 1.4-GHz profile with some antisymmetric $V$ suggests core  flanked by conal emission.  The inner conal model relies on a poor guess at the PPA rate. MM10 detect the scattered profile at 111 MHz.
\vskip 0.1in

\noindent\textit{\textbf{B1911+09}}'s emission alternates between bursts of 50-100 pulses with null or weak pulse intervals of similar length.   An outer cone model is used because the \citet{hulse_taylor} discovery paper gives a significantly larger width. 
\vskip 0.1in

\noindent\textit{\textbf{B1911+13}}: PSR B1911+13 has a core-outer cone triple \textbf{T} geometry.  Only one scattered feature is visible in MM10's 100-MHz profile, and we have no way to distinguish whether this is core, cone or both.  The somewhat broader 4.9-GHz profile is probably so because of stronger conal outriders, but it cannot be usefully measured.  
\vskip 0.1in

\noindent\textit{\textbf{B1911+11}} is definitely conal with a deep cycle of 19.7-$P$ amplitude modulation in both components.  The poor \citet{GL98} meter wavelength profiles show no large increase in profile width, so we use an inner cone model.   
\vskip 0.1in

\begin{figure}
\begin{center}
\includegraphics[height=85mm,angle=-90.]{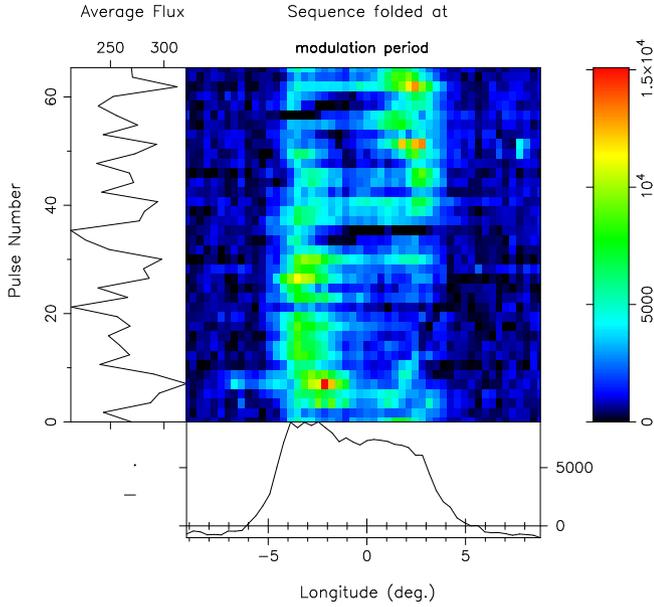}
\caption{Pulsar B1913+167 exhibits a strong 65.4-$P$ modulation.  Remarkably, it represents an alternation between bright emission in the leading and trailing conal components.  And to make it more interesting, it seems that there is core emission at the onset of each part of the cycle.  The main panel shows the modulation-folded pulse sequence, the lower one the average profile and the lefthand one the amplitudes through the cycle.}
\label{figA104}
\end{center}
\end{figure}
\noindent\textit{\textbf{B1913+167}}: The ET VI study regarded the pulsar as having a \textbf{T} profile; however, a fluctuation spectrum shows a strong 65.4-$P$ feature.  Interestingly, this is produced by a regular alternating pattern of emission in the leading and trailing conal components with core emission at the beginning of both intervals, as shown in Fig~\ref{figA104}.  No scattering time value is available.  
\vskip 0.1in
\noindent\textit{\textbf{B1913+10}}:  We follow \citet{rankin1993b} as well as \citet{W99} in regarding this pulsar's geometry as reflecting a core beam without evidence of conal emission.  At lower frequencies, \citet{GL98} show that scattering sets in, and the EPN 5-GHz profile shows no hint of conal outriders.  Moreover the PPA traverse is disorderly, so only an $\alpha$ value can be estimated from the observation.  At no frequency are conal outriders clearly discernible, but the increased width at 4.9 GHz \citep{kkwj98} may be their result.
\vskip 0.1in

\noindent\textit{\textbf{B1913+16}}:  This is the famous first Binary Pulsar, and we follow the analysis in table 4 of \citet{rankin17}.  The profile is a core-cone triple \textbf{T} but the core is conflated with the conal emission at 1.4 GHz.  
\vskip 0.1in

\begin{figure}
\begin{center}
\includegraphics[width=85mm,angle=0.]{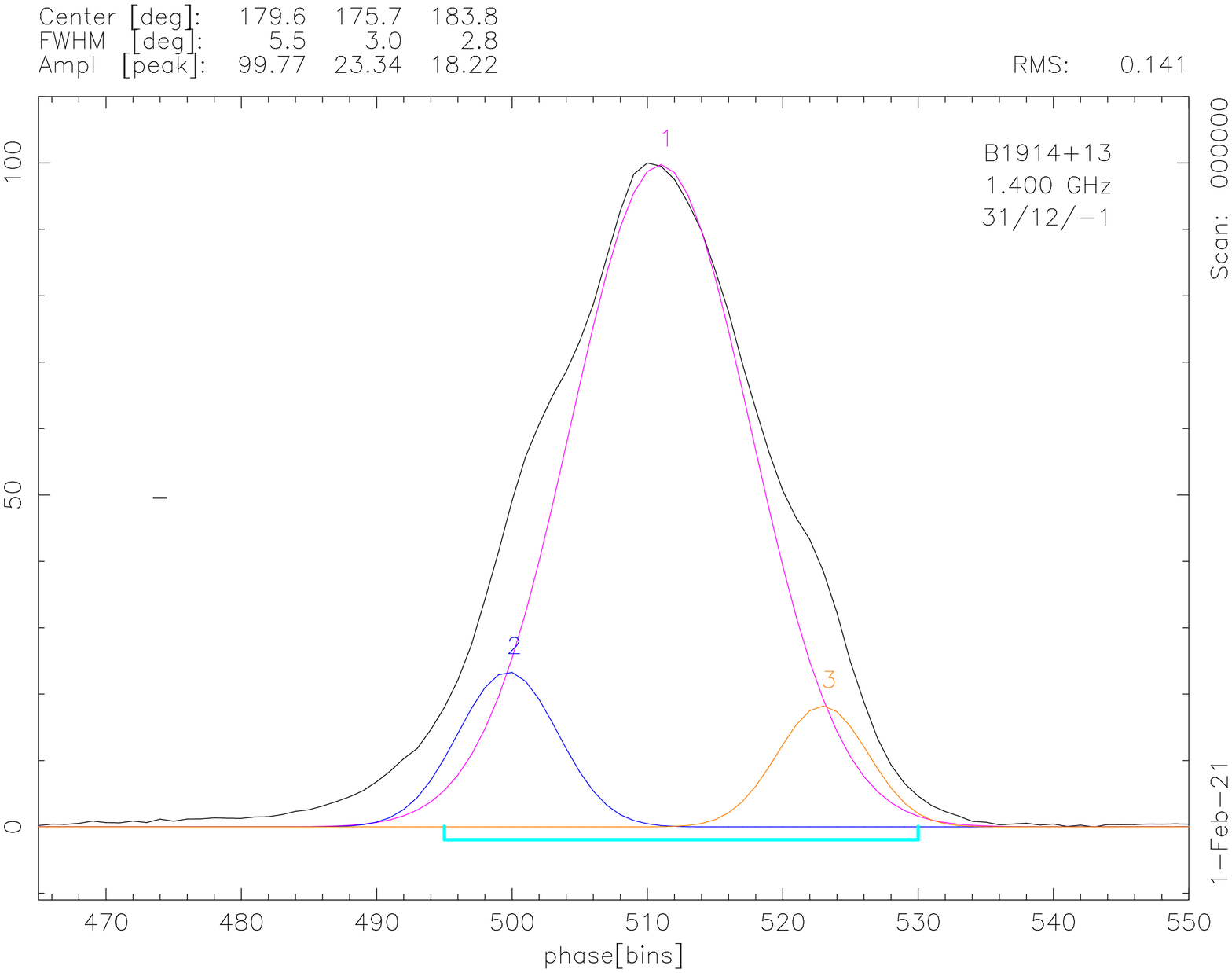}
\caption{Gaussian-component fitting of B1914+13 at 1.4 GHz. The three components are highly conflated, but they show in the inflections on the leading and trailing edges of the profile.}
\label{fig105a}
\end{center}
\end{figure}

\noindent\textit{\textbf{B1914+13}} A conflated core-inner cone triple \textbf{T} profile is present in this pulsar at 1.4 GHz, as clearly seen from the Gaussian component fitting shown in Fig~\ref{fig105a}.  We estimate the core at about 5\degr. By 327 MHz we see substantial scattering, and we assume the power is mostly that of the core.
\vskip 0.1in

\noindent\textit{\textbf{B1915+22}}: Both the 1.4-GHz and 149-MHz profiles are asymmetric with the leading part much stronger than the trailing---a common property of c\textbf{T} profiles.  However, the 327-MHz profile form is so different than the others that no interpretation is possible.  All three profiles show hints of both an inner and outer cone, but there is no hint of regular modulation in the fluctuation spectra. We model it nominally as having an outer cone, any core/cone structure remains unclear.
\vskip 0.1in

\noindent\textit{\textbf{B1916+14}}: The core-cone triple structure at 1.4 GHz has been exhibited by \citet{Blaskiewic} and \citet{W99} as well as ourselves.  The \citet{von_thesis} 4.9-GHz exhibits structures similar to those at 1.4 GHz, and we mirror them in the model.  We follow the geometry in ET VI down to 327 MHz.  
\vskip 0.1in

\noindent\textit{\textbf{B1917+00}}: A core-cone \textbf{T} profile is seen in this pulsar at higher frequencies but GL98 shows that the first component becomes more dominant at meter wavelengths.  Therefore we interpret the 135-MHz detection as having a width of 16\degr\ or more.
\vskip 0.1in

\noindent\textit{\textbf{B1918+26}}: This pulsar exhibits a triple—or perhaps \textbf{M} where what may be the trailing inner cone at 327 MHz is conflated with the outer cone. The leading conal component(s) are very weak, and this structure is echoed in the LOFAR profile. The \citet{KTSD23} profiles permit the core width to be measured down to 50 MHz, where it begins to suffer from scattering (and detect it at 35 MHz), but the conal features are buried in the noise.  The 327-MHz core is narrower than the polar cap width, perhaps incomplete with only negative Stokes $V$. Single pulses show no periodicity but a different emission pattern in the core and trailing component.  $R$ cannot be determined accurately, but --11\degr/\degr\ is a plausible estimate.  The sign of V is different in our two observations, but the GL and W99 profiles also seem inconsistent.  
\vskip 0.1in

\noindent\textit{\textbf{B1918+19}}: \citet{hankins1987} show that the pulsar has three modes and a double-cone profile structure.  We thus model it using a conal quadruple c\textbf{Q} geometry (rather than the c\textbf{T} envisioned in ET VI).  This geometry is explored in \citet{RankinWright+13} and \citet{IMS89} provide a profile at 102 MHz.  A scattering time scale has been measured by \citet{kmn+15}.
\vskip 0.1in

\begin{figure}
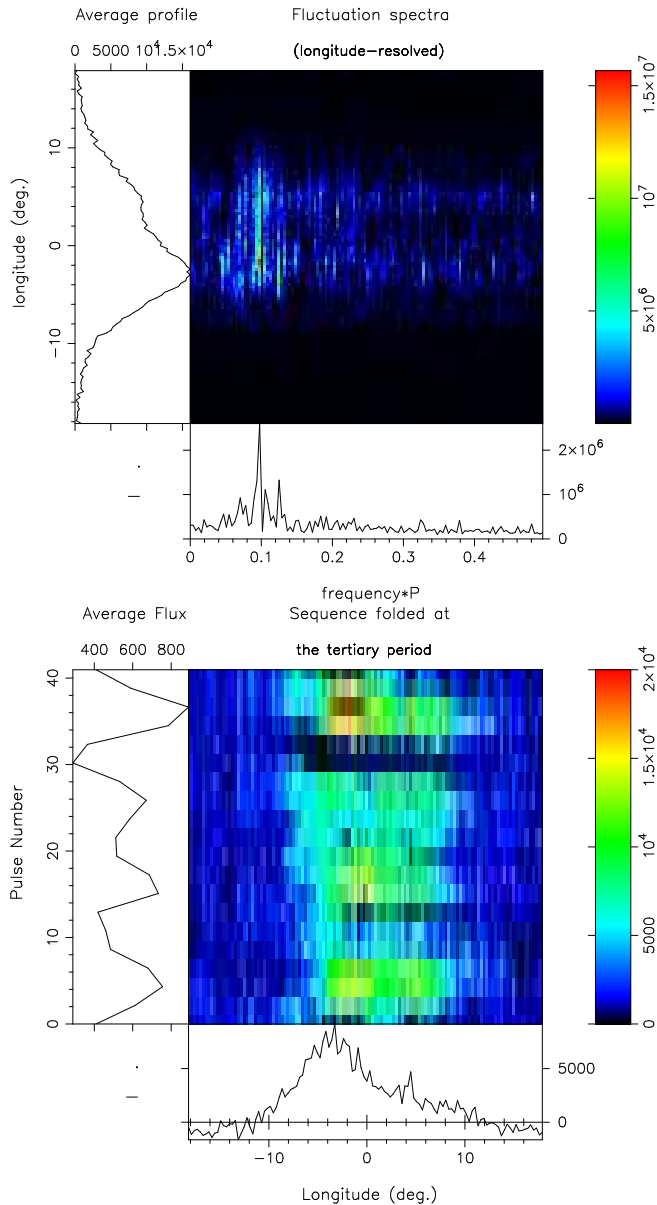

\begin{center}
\includegraphics[height=85mm,angle=-90.]{plots/PQB1919+14.56768_lrf_257-512.ps}
\includegraphics[height=85mm,angle=-90.]{plots/PQB1919+14.56768_modx4_19bin_257-512.ps}
\caption{Pulsar B1919+14 shows an unusually coherent phase and amplitude modulation, here 10.24-$P$ as shown in the upper display.  Note also the sidebands spaced at 1/4 this frequency in this 256-pulse interval.  This may correspond to a stable pattern of 4 ``beamlets'' as shown in the lower display, that may in turn correspond to a 4-element carousel pattern in the manner of the 20-element configuration of pulsar B0943+10 \citep{DR99,DR01}}.
\label{figA106}
\end{center}
\end{figure}
\noindent\textit{\textbf{B1919+14}}: There is no question that the profile is conal as its pulses show a strong 10-period fluctuation feature as shown in Fig~\ref{figA106}.  Two conflated components are seen at both 1.4 GHz and 430 MHz (HR10), and the profile width seems constant down to 102 MHz (MM10).  We thus model it with a \textbf{S$_d$} geometry. 
\vskip 0.1in

\begin{figure}
\begin{center}
{\mbox{\includegraphics[height=85mm,angle=-90.]{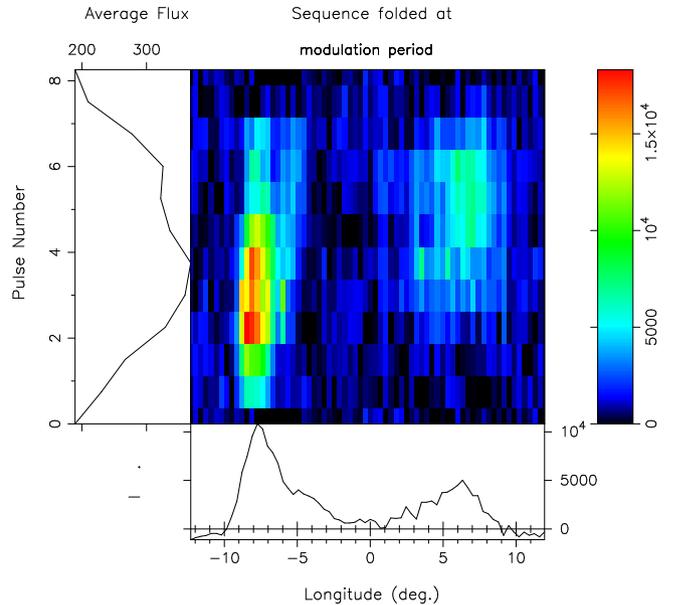}}}
\caption{Pulsar B1919+20 single pulses show a 8.3-$P$ amplitude modulation as shown above in the pulse sequence folded at this interval over the first 256 pulses.}
\hspace{-0.55 cm}
\label{figA107}
\end{center}
\end{figure}

\noindent\textit{\textbf{B1919+20}} shows us clearly that the emission is conal in Fig.~\ref{figA107} and our 327-MHz as well as the 774-MHz \citet{han2009} profiles with somewhat larger widths suggest that it is probably an outer cone.  
\vskip 0.1in

\noindent\textit{\textbf{B1920+20}}: The \citet{Weisberg2004} 430-MHz profile shows 3 or 4 components with widths and spacings suggesting a double conal structure, only the leading inner conal feature lacks definition.  Even a possible core with about the right width is hinted at, and the PPA traverse is very well defined.  The pulsar seems to have a steep spectrum as it is barely detected in the GL98 1.4-GHz profile, but this does give a rough estimate of its overall width.  We thus model the beam geometry with an outer conal model but it may have a conal quadruple c\textbf{Q} or even \textbf{M} structure.  
\vskip 0.1in

\noindent\textit{\textbf{B1920+21}}: This closely-spaced and conflated triple profile could be interpreted as either a core-cone \textbf{T} or conal triple c\textbf{T}.  However, \citet{WES06,WSE07}, and this paper find that the pulsar has a flat fluctuation spectrum, strongly suggesting that the bright feature is a core component.  At LOFAR frequencies, the profile, is mostly core emission and shows significant scattering, so nothing can be said about its intrinsic width here. 
\vskip 0.1in

\noindent\textit{\textbf{B1921+17}} gives no hints from its weak fluctuation spectra.  The profile suggests three features, and the PPA traverse seems very steep.  \citet{hulse_taylor} report a width a little larger that that at 1.4 GHz, but we suspect this was poor resolution, and we model the pulsar as having an inner cone.  If there is a central core component, its width would need to be about 3.85\degr, and such a width does seem plausible looking at the profile.  The geometry does not seem to square with a conal triple configuration.  
\vskip 0.1in

\noindent\textit{\textbf{B1924+14}}: This pulsar shows a very usual outer conal double \textbf{D} profile.  The profile may show scattering even at 327 MHz, and the narrower one in MM10 seems incompatible with the higher frequencies and is probably misleading. The suggestion of a more central conflated component at 1.4 GHz is due to a single pulse of radar RFI.   
\vskip 0.1in

\noindent\textit{\textbf{B1924+16}}: Following ET VI and ET IX the pulsar seems to have a classic core-single profile with little width increase down to 327-MHz and no clearly discernible conal outriders---although no observation is available above 1.6 GHz.  Only red noise is seen in the fluctuation spectrum \citep{WES06}.  
\vskip 0.1in

\noindent\textit{\textbf{B1925+18}} shows a strong 6.4-$P$ amplitude modulation in certain intervals, so the emission clearly seems conal.  The poor GL 610-MHz may have a larger width, so we model it as outer conal.  
\vskip 0.1in

\noindent\textit{\textbf{B1925+188}}'s single pulses show three separate regions of emission, and the fluctuation spectra have a 50-pulse peak reflecting that bright groups of pulses occur at intervals with nulls or weak pulses in between.  The \citet{han2009} 774-MHz profile has a similar form and width to our 1.4-GHz one (see Fig.~\ref{figA30}), so an inner cone model seems appropriate.  If the central core feature has a very plausible 14\degr\  width, then a triple \textbf{T} geometry works perfectly.
\vskip 0.1in

\noindent\textit{\textbf{B1925+22}}: Our 1.4-GHz profile shows a clear triple structure, but that of \citet{W99} is so different as to suggest modal activity.  The 327-MHz profile is broader and has a rough conal double form.  We model the profiles using a core-cone triple geometry.  No scattering timescale value is available.  
\vskip 0.1in

\begin{figure}
\begin{center}
\includegraphics[height=85mm,angle=-90.]{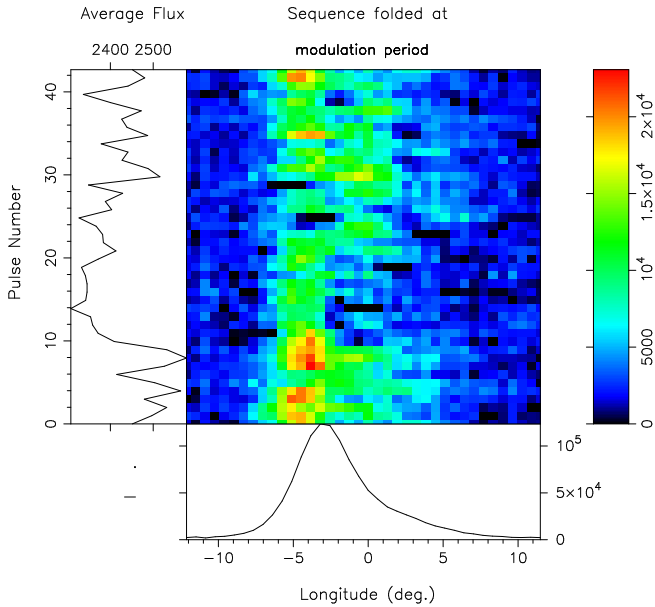}
\caption{Pulsar B1930+22 shows a fairly coherent modulation with a 42.7-$P$ period.  The 4000-pulse sequence here folded on this cycle shows that the modulation is both deep and persistent.  Note the 3-4 distinct regions that are illuminated throughout the cycle.  Here the unvarying ``base'' has been removed removed.}
\label{figA108}
\end{center}
\end{figure}

\noindent{\textbf{B1926+18}}: The \citet{W99} apparently has three components; whereas, no clear triplicity is seen in our 327-MHz profile.  Probably this is compatible with the moding discovered early by \citet{Ferguson+81} that produces strong variations in the strength of the central component relative to the peripheral ones. We model the beam system using a conal triple c\textbf{T} structure, while bearing in mind the moding behavior suggest a central core.  However, we have tilted to its being conal given that the pulsar is not very energetic.  The profile triplicity is hardly apparent in our 327-MHz profile and is thus most similar to the above authors' C mode---and our observation is not of high enough S/N to explore which modes might be mixed in this 1023-pulse sequence.
\vskip 0.1in

\noindent{\textbf{B1927+13}}: ET VI listed the pulsar as probably having a core single geometry, but lack of a PPA rate frustrated any strong conclusion on the basis of the poor observations then available.  The best now available are those of \citet{W99,Weisberg2004}, but neither provides an $R$ estimate.  However, their quality permits us to conjecture that each shows a strong core component flanked by a conal ``outrider'' pair.  We model the geometry assuming a central sightline traverse, and we find that an inner conal configuration results.  No scattering measurement is available. 
\vskip 0.1in

\noindent{\textbf{B1929+20/J1932+2220}} was studied by GL98 but several of the profiles seem less well resolved than that of \citet{JK18}.  There is no good determination of the PPA rate.  The profile seems to be primarily core radiation, perhaps with some conal outriders at high frequency, but any structure is obliterated at low frequency by scattering as seen in the \citet{WSE07} profile at 327 MHz---a scattering timescale of 19$\pm$1 ms at 327 MHz \citep{kmn+15}, indicating that the pulsar is fully scattered out in the 100-MHz band.  
\vskip 0.1in

\noindent\textit{\textbf{B1930+22}}: This energetic fast pulsar was studied in ET IX \citep{ETIX}  at 1.4 GHz as a ``partial cone'' and found to have sporadic emission on the trailing (and perhaps on the leading) side of the bright component.  Its short rotation period and large DM make it difficult to observe at meter wavelengths---and in any case it is clearly scattered---so there is little reliable evidence regarding the profile's evolution.  Here we find it also has a regular cycle of about 43 rotation periods as we can see in Fig.~\ref{figA108}, during which different parts of its profile are activated.  Such orderly modulation is unusual in a pulsar with such a large $\dot E$, may well not be conal ``drift'' modulation, and surely deserves further detailed study.  The width of the sporadic emission is roughly compatible with that of an inner cone, so we model the pulsar with an inner cone triple \textbf{T} geometry.  At 102 MHz MM10's profile seems too narrow to reflect the remainder of a ``scattered out" response.

\begin{figure}
\begin{center}
\includegraphics[height=85mm,angle=-90.]{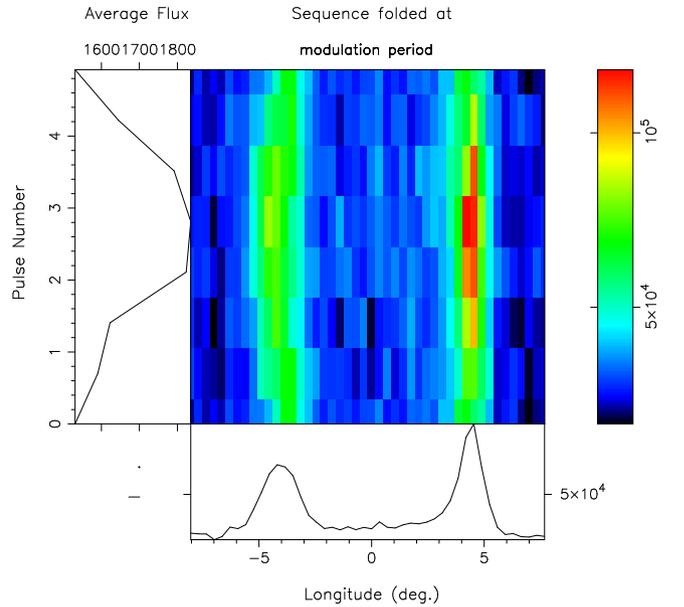}
\caption{Pulsar B1930+13 has a 4.92-$P$ modulation common to both components that is part phase and part amplitude as shown in the display.  Here we see the first 512 pulses folded at this period with the unvarying ``base'' removed.}
\label{figA109}
\end{center}
\end{figure}
\vskip 0.1in

\noindent\textit{\textbf{B1930+13}}: This pulsar seems to be a \textbf{D} profile with some filling of the profile interior at 327 MHz and below.  The modulation-folded display in Fig.~\ref{figA109} leaves not doubt that the emission is primarily conal.  The polarimetry gives little indication of $R$ but suggests that $\beta$ is close to 0.  $t_{scatt}$ is expected to be some 12\degr\ at 149 MHz which may account for its broad form.
\vskip 0.1in

\noindent\textit{\textbf{B1931+24}}: The pulsar, famous for its intermittent character \citep{Kramer06}, has a core-cone triple, or perhaps five-component profile.  We model it using a triple \textbf{T} geometry because the putative trailing components cannot be distinguished.  The PPA rate is well defined and consistent at both frequencies.  No scattering measurement is available.  
\begin{figure}
\begin{center}
\includegraphics[height=84mm,angle=-90.]{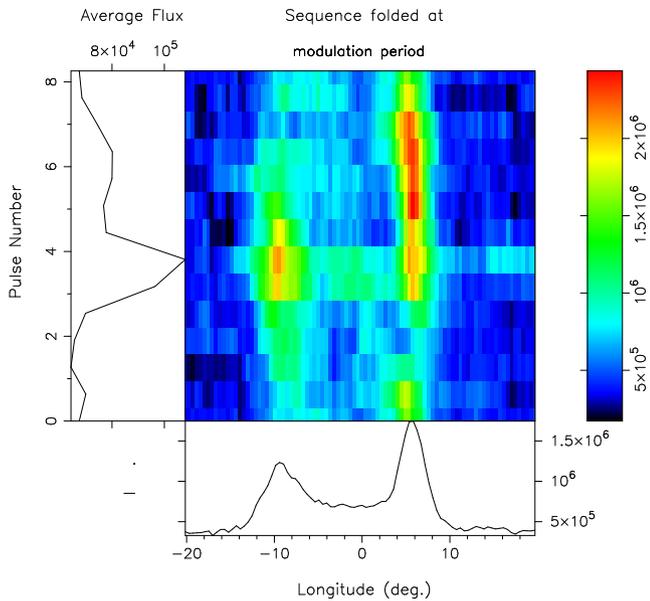}
\caption{Pulsar B1942--00 shows an 8.26-$P$ periodicity common to both components that appears as pure phase modulation.  Here we see the first 512 pulses folded at this period.}
\label{figA110}
\end{center}
\end{figure}
\vskip 0.1in
\noindent\textit{\textbf{B1942--00}}: PSR B1942$-$00 was classified as an inner conal double in ET VI, and we see in Fig~\ref{figA110} that the modulation is an entirely conal one.  However, the broader detection of MM10 at 111 MHz, suggests an outer cone with little scattering here. 
\vskip 0.1in

\noindent\textit{\textbf{B1944+22}}: Our 1.4-GHz profile has a triple form, and the central feature almost looks core-like; however, it seems more likely that all three components are conal within a conal triple c\textbf{T} structure.  The leading component is well separated; see also the \citet{W99,Weisberg2004} observations.  The PPA rate is not well determined, but some profiles suggest a value around +11\degr/\degr\  \citep{ETIX}.  \citet{kmn+15} provide a scattering timescale value that is too small to appear on the model plot, showing that the 327-MHz profile structure is intrinsic. 
\vskip 0.1in

\noindent\textit{\textbf{B1949+14}}: This pulsar apparently has an \textbf{S$_t$} profile with a suggestion of weak conflated conal components or some scattering as possibly responsible for the increasing width at lower frequencies.  Its emission comes in irregular bursts of about 100 pulses duration.
\vskip 0.1in

\noindent\textit{\textbf{B1951+32}}:  This 40-ms pulsar gives few hints about the character of its emission, and it is too weak to study its individual pulses.  Like many MSPs, its profile changes little across a wide band, although its relatively strong $B_{surf}$ puts it among the normal pulsar population.  It has a single broad component and a linear PPA traverse, so we have modeled it as having a conal single beam geometry, however without much conviction that this gives any full picture of its configuration. 
\vskip 0.1in

\noindent\textit{\textbf{B1953+29}}: This is the second MSP once known as ``Boriakov's Pulsar.  It shows a clear core-cone \textbf{T} structure as well as an interpulse.  Again we follow the geometrical analysis in table 4 of \citet{rankin17} together with the polarimetry of \citet{sbc+84}.  
\vskip 0.1in

\begin{figure}
\begin{center}
\includegraphics[height=85mm,angle=-90.]{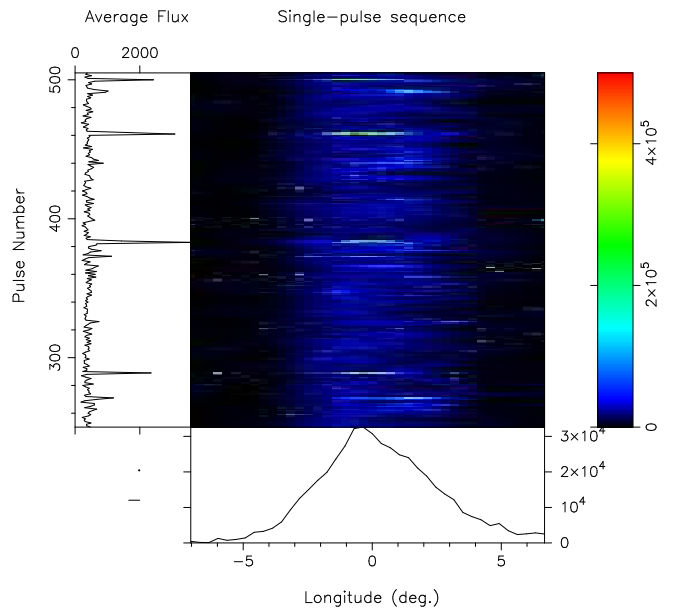}
\caption{Pulsar B2000+32 emits very strong single pulses against a background of very weak ones as shown in the 256-pulse display.  Some or many of these are probably so-called ``giant'' pulses.}
\label{figA111}
\end{center}
\end{figure}

\noindent\textit{\textbf{B2000+32}}: A weak interpulse with a nearly 180\degr\ spacing (that we do not show) is seen in this pulsar, arguing that the object is a two-pole interpulsar. Its emission is overall weak at both frequencies but punctuated by occasional very strong subpulses, many of which probably qualify as ``giants'' as shown in Fig~\ref{figA111}---such that it may be likened to `RRAT'' modulation \citep[\eg][]{keane2011}.  The profiles remain single over the entire band until distorted by scattering.  Our 1.4-GHz profile is typical as being too wide to be entirely core for an orthogonal geometry, and breaks suggest a three-part structure, and so we model it. The PPA rate is difficult to discern, and the model value corresponds to an inner cone, but were it about --9\degr/\degr\ as suggested by many of the GL98 profiles, an outer conal model would results.  The 4.9-GHz profile \citep{kkwj98} has similar dimensions and probably structure.  The 327-MHz profile shows some scattering \citep{kmn+15}, and the MM10 profile cannot be interpreted and compared because scattering accounts for the full width of the 111-MHz profile.  
\vskip 0.1in

\noindent\textit{\textbf{B2002+31}}: Profiles have a symmetrical core-cone triple \textbf{T} structure with some $\pm V$ signature down to 600 MHz---and we so model it following ET VI.  Scattering \citep{kmn+15} distorts the profiles at lower frequencies.  Unsurprisingly, \citet{WES06} find a flat fluctuation spectrum. A scattering timescale has been measured by \citet{kmn+15}.
\vskip 0.1in

\noindent\textit{\textbf{B2025+21}}: PSR B2025+21 seems to have a conal triple c\textbf{T} profile, but the P-band and LOFAR profiles cannot be measured well enough to confirm this, nor is a reliable PPA sweep rate discernible.  Therefore, only the outer cone is modeled, and this assuming a central sightline traverse.  $t_{scatt}$ is expected to be some 8\degr\ at 149 MHz.
\vskip 0.1in

\noindent\textit{\textbf{B2027+37}}:  This is another pulsar with core profiles and no clear evidence of conal emission.  Scattering seems an issue at 327 MHz and surely in MM10's 111-MHz profile.
\vskip 0.1in

\noindent\textit{\textbf{B2028+22}}: Both profiles show four features, and we model the pulsar as having a conal quadruple c\textbf{Q} beam system.  The LOFAR profile is too poor to show both cones, so we measure only the outer one.  No scattering measurement has been reported.
\vskip 0.1in

\noindent\textit{\textbf{B2034+19}}: Four components are present in this pulsar at LOFAR frequencies: the first two very close together with the second stronger than the first; the trailing two are much weaker. This pulsar has been well studied and shows modulation and nulling periodicities \citep{rankin2017_nulling}. 
The latter two components are conflated at 327 MHz, while all four can be distinguished at at 1.4 GHz.  We follow the above paper in modeling this pulsar as having a conal quadruple c\textbf{Q} beam configuration.  The extreme symmetry between its leading and trailing halves is similar to several other pulsars with this beam structure.
\vskip 0.1in

\noindent\textit{\textbf{B2035+36}}: This pulsar seems to be a fine example of a core/outer cone triple \textbf{T} configuration.  We find no discernible periodicities in the pulse trains; its emission comes in strong occasional subpulses against a weak background with intervals of stronger leading and trailing emission.  Our 1.4-GHz core seems strangely asymmetric, but the W99 profile affords a more reliable estimate of about 4.0\degr.  The PPA rate is well determined, and the MM10 profile, though noisy, seems to afford reliable width estimates.  
\vskip 0.1in

\noindent\textit{\textbf{B2044+15}}: The 18-$P$ drift feature detected at both frequencies by \citet{WES06,WSE07} confirm that the pulsar has a standard outer conal \textbf{D} profile.
\vskip 0.1in

\begin{figure}
\begin{center}
\includegraphics[height=85mm,angle=-90.]{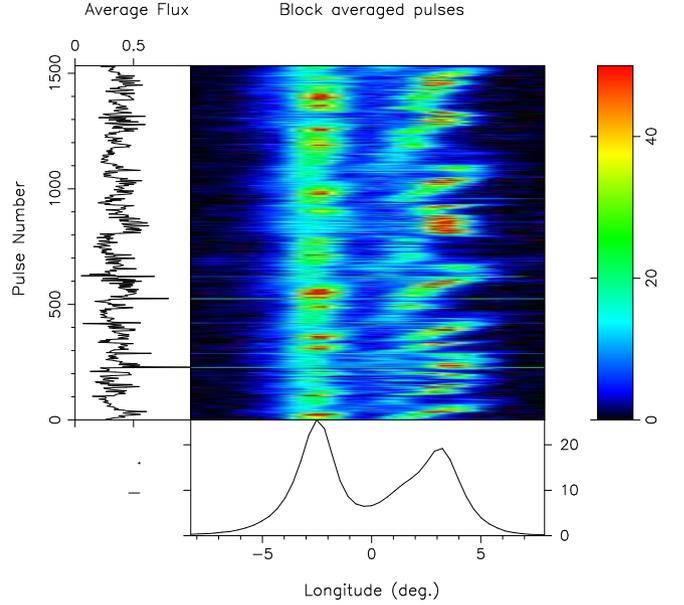}
\caption{Pulsar B2053+21 exhibits an unusual subpulse modulation pattern. Individual subpulses in the first component have a fixed longitude; whereas, in the second component they often show ``drift''.  The second component has two parts, evident in the average component as well, and the drifts sometimes connect them and sometimes not.}
\label{figA112}
\end{center}
\end{figure}
\noindent\textit{\textbf{B2053+21}}: This pulsar has two well resolved components with a weak intermediate third increasingly clear at lower frequencies. The  \citet{KTSD23} profiles permit width measurements down to 50 MHz.  Its subpulse modulation in the trailing region contrasts strongly with that in the leading feature as shown in Fig~\ref{figA112}.  We model it using an inner cone \textbf{D} configuration but also see some width growth with wavelength.  Here $\alpha$ cannot increase enough to support an outer conal configuration. It remains unclear whether the middle feature is a core or conal component.  
\vskip 0.1in

\noindent\textit{\textbf{B2053+36}}: PSR B2053+36 shows the usual evolution of a core-single \textbf{S$_t$} profile with evidence of inner conal outriders at the highest frequencies.  Fluctuation spectra show no strong periodicites; however, the subpulses do seem to fall in three different longitude intervals, supporting a three-part profile structure.  Substantial scattering at 327 MHz and below with MM10's profile perhaps so fully scattered out that its width decreases.  
\vskip 0.1in

\noindent\textit{\textbf{B2113+14}}:  This pulsar seems to have an outer conal \textbf{S$_d$} geometry \citep[see also][]{JKMG2008}.  \cite{WSE07} do not detect drift modulation, but the 8\degr\ $\beta$ indicates that the sightline would miss any core emission.  We use the 102-MHz KL99 profile that may have been corrected for the scattering \citep{kuzmin_LL2007} apparently seen in the \citet{IMS89} profile.  
\vskip 0.1in

\noindent\textit{\textbf{B2122+13}}: This pulsar has an outer conal double \textbf{D} geometry.  Our fluctuation spectra (not shown) indicate irregular modulation with a cycle of about 5 rotation periods.
\vskip 0.1in

\begin{figure}
\begin{center}
\includegraphics[width=75mm,angle=-90.]{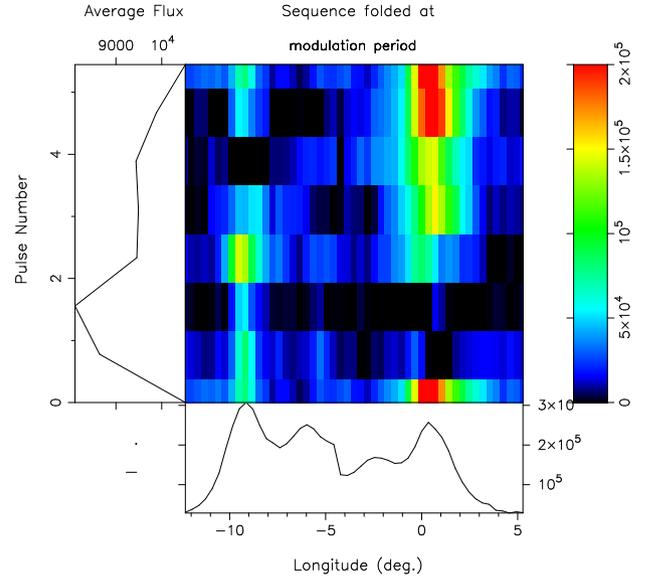}
\caption{B2210+29: The classic five-component profile of this pulsar is modulated on a 5.45-$P$ cycle in its conal components and a 6-4-$P$ one in the central core region.  Here the core region of the profile is omitted in the display, and only the fluctuating part of the power above a ``base'' is plotted.}
\label{figA113}
\end{center}
\end{figure}

\noindent\textit{\textbf{B2210+29}}:  This is a \textbf{class} \textbf{M} pulsar where unusually the inner cones can only be discerned at high frequency.  The mostly amplitude modulation of the inner and outer conal components in Fig.~\ref{figA113} strongly support this identification. Core widths are measurable in the LOFAR profiles, but not very accurately.  Scattering an issue in broadening at low frequency.

\label{lastpage}

\end{document}